\documentclass{aa}
\usepackage{amsmath}
\usepackage[varg]{txfonts}
\usepackage{pgfplotstable}
\usepackage{colortbl}
\usepackage{supertabular}
\usepackage{comment}
\usepackage{perpage}
\usepackage{subfigure}
\usepackage{enumitem} 
\usepackage{caption}
\MakePerPage{footnote}
\usepackage{booktabs}
\usepackage{float} 
\usepackage{appendix}
\usepackage{afterpage}
\usepackage[utf8]{inputenc}
\usepackage[colorlinks=true, linkcolor=blue, citecolor=blue, urlcolor=blue]{hyperref}
\usepackage{xcolor}
\usepackage{threeparttable}
\usepackage{xcolor}

\begin{document}
\titlerunning{Polarimetric maps AGB stars}
\authorrunning{Badolo et al.}

\title{A catalogue of high angular resolution and contrast polarimetric maps of 45 nearby AGB stars with SPHERE/ZIMPOL}
\author{Nekolgne Aymard Badolo\inst{1,2}
\and Eric Lagadec\inst{1} \and Mamadou N'Diaye\inst{1} \and Sié Zacharie Kam\inst{2} \and Iain McDonald\inst{3} \and Alexis Matter\inst{1} \and Jean Koulidiati\inst{2} \and Lyu Abe\inst{1} \and Marcel Carbillet \inst{1} \and Thierry Fusco\inst{4}
}

\institute{Université Côte d'Azur, Observatoire de la Côte d'Azur, CNRS, Laboratoire Lagrange, Nice, France
\and  Laboratoire de Physique et de Chimie de l'Environnement, Université Joseph Ki-Zerbo, Ouagadougou, Burkina Faso
\and Jodrell Bank Centre for Astrophysics, Alan Turing Building, Manchester M13 9PL, UK0
\and DOTA, ONERA, Université Paris Saclay, F-91123, Palaiseau, France
}

\date{Accepted 1 June 2026}
\abstract
{}
{We present the largest catalogue of asymptotic giant branch (AGB) stars, 45 targets in total, observed in polarized light at high angular resolution ($\sim$ 20 milliarcsec). The main goal of the study is to detect and characterize dust shells in the close environment of nearby AGB stars. This work also aims to systematically classify the AGB star circumstellar morphologies obtained with the SPHERE instrument installed at the  Very Large Telescope (VLT), thanks to its Zurich Imaging Polarimeter (ZIMPOL).}
{We extracted and analyzed polarized intensity maps for 45 AGB stars, constructed from polarimetric observation data obtained with the SPHERE/ZIMPOL instrument. An ellipse fitting method was applied to characterize the circumstellar envelopes. Stellar parameters (luminosity, effective temperature, surface gravity, extinction, metallicity) were compiled and recalculated when necessary from spectral energy distribution (SED) fitting using the Python SED fitting tool (PySSED) software. These data were then used to train a random forest machine learning model to determine the most discriminating variables for a resolved envelope around a given star.}
{We  constructed polarization maps for all stars in the sample, revealing a wide diversity of circumstellar morphologies. We detected 16 dusty circumstellar envelopes, including three never observed before. They display a wide range of morphologies, all of them showing a clear departure from spherical symmetry, indicating interaction with a companion or asymmetric mass ejections. The random forest model identified optimal thresholds for several physical parameters, thus providing robust criteria to anticipate SPHERE's ability to resolve dust envelopes around AGB stars. These results facilitate the selection of targets for future observations and contribute to a better understanding of the evolution mechanisms of circumstellar envelopes.}
{}
\maketitle 

\section{Introduction}

During the late stages of their evolution, low- and intermediate-mass stars (stars with an initial mass ranging between 0.8 and 8~M$_{\odot}$) evolve through the asymptotic giant branch (AGB)~\citep{Bressan2012}. This phase of stellar evolution is characterized by an intense mass loss, with rates up to 10$^{-4}$~M$_{\odot}$.yr$^{-1}$~\citep{Beck2010}. This mass loss is an important contributor to the chemical enrichment of galaxies, since $\sim$95\% of all stars will pass through this phase, bringing newly formed elements to the interstellar medium (ISM)~\citep{McDonald&Trabucchi2019}. The mechanisms at play to trigger this mass loss are not fully understood, yet a combination of pulsation and convection is proposed to extend the atmosphere and trigger dust formation via shocks. Radiation pressure on newly formed dust grains triggers mass loss and gas is carried along via friction
\citep{Hofner2018}. After the AGB phase, the star shrinks and warms. The envelope formed during the mass-loss phase becomes detached and then ionized by the central star. A planetary nebula (PNe) is formed. For more than three  decades, observations of PNe have shown that departures from spherical symmetry are common \citep{balick1987}. It is now widely accepted that the bipolar and multipolar morphologies of PNe are due to the presence of binary companions \citep{Miszalski2009}. The excess of linear momentum observed after the AGB phase \citep{bujarrabal2001} cannot be explained by a single star \citep{Soker2006,Nordhaus2006}.

Recent observations at high angular resolution of the gas in the millimetre range with the ALMA array have revealed a wealth of morphologies for the inner circumstellar environments of a sample of 17 AGB stars \citep{Decin2020}. The leading interpretation is that this is due to the influence of (sub)stellar companions that are shaping the inner winds of these giant stars. Observations of the dust distribution in the inner part of the envelopes of AGB stars should thus give us clues about its distribution around AGB stars and the envelopes' alteration by binary companions. To better understand this mass loss process, high angular and high-contrast observations of such mass-losing stars are thus crucial.

The  Zurich Imaging Polarimeter (ZIMPOL) camera \citep{Schmid2018} onboard the SPHERE extreme adaptive optics instrument \citep{Beuzit2019} on the Very Large Telescope (VLT) is one of the instruments best suited for this. The combination of an extreme adaptive optics system with an 8.2m-diameter telescope offers images with an angular resolution of $\sim$20~mas in the optical range. The polarimetric capabilities of the instrument enable us to study faint dusty structures around bright stars: the light from the central star is not polarized, while light scattered by dust becomes polarized. 

In this work, we present the largest catalogue to date of AGB stars observed in polarized light at high angular resolution with SPHERE/ZIMPOL, with 45 targets. This catalogue provides a unique resource for the study of circumstellar envelopes and mass loss mechanisms in these evolved stars. We propose a systematic classification of circumstellar morphologies according to their resolved or unresolved character. We then use these results to train a machine learning model based on the random forest approach \citep{Ho1995,Breiman2001}. This aims to identify the most discriminating physical parameters to predict the resolved character of AGB circumstellar envelopes observed with SPHERE. Based on stellar properties, this work allows us to define simple and robust criteria to anticipate SPHERE's ability to resolve dust envelopes around AGB stars. In the long term, this work will help us optimize the selection of targets for future observation campaigns dedicated to studying the evolution of circumstellar envelopes.

The paper is structured as follows. Section~\ref{sec:2} presents the observations and data reduction of the 45 stars observed with SPHERE/ZIMPOL. In Section~\ref{sec:3}, we detail our methodology to produce the polarized intensity maps of our targets. We then assess the resolved nature of their circumstellar envelopes to enable a classification of their morphology. We finally present a first random forest model to predict the resolved nature of our targets. The results of this analysis are discussed in Section~\ref{sec:4}. We present our conclusions and prospects for future campaigns of AGB stars in Section~\ref{sec:5}.

\section{Observations and data reduction}\label{sec:2}
\subsection{Target selection}

This catalogue is mostly based on observations of nearby AGB stars from the SPHERE 'Other Science' guaranteed time observations program (GTO).  Some of the observations were taken as fillers for the program. Other available ZIMPOL observations of nearby AGB stars were also retrieved from the archive. The goal was to produce the most homogeneous sample possible, despite the multiple observing constraints (e.g. stars' observability, the weather and so on). 

The sample of stars is mostly selected from the 'fundamental parameters and infrared (IR) excesses catalogue  of Tycho-Gaia stars' \citep{McDonald2012,McDonald2017}. It contains fundamental parameters for 1475 921 Tycho-2 and 107 145 Hipparcos stars, based on distances from Gaia Data Release 1. It contains a subsample  of 558 stars classified as giants, all located within a radius of 300~pc around the Sun (hereinafter McDonald's catalogue).

For our Other Science program, we observed 19 stars from McDonald's catalogue. We selected  bright stars ($V<8$~mag, when possible) with a large IR excess ($\mathrm{E_{IR}} \geq 1.2$, as we expect them to have observable circumstellar envelopes)  to ensure optimal correction by adaptive optics. The targets were chosen according to their observability during the given observing nights. To this set, we added 18 targets observed by ZIMPOL for other programs, and present in McDonald's catalogue.
The Other Science program  included three nearby AGB stars not present in McDonald's catalogue ($\rm L_2 \ Puppis$, V~Hya, and Y~Pav). Finally, to increase our sample of resolved envelopes around AGB stars by ZIMPOL, we added previously published data from five AGB stars (U~Del, W~Aql, V~Psa, R~Aql, and GY~Aql).
This selection is therefore not random, but intentionally biased towards bright AGB stars with significant IR excess, and is representative of the subset of objects in the McDonald's catalogue that are accessible to visible polarimetric imaging (see Appendix~\ref{an:sample_comparison} for complementary details).
Table~\ref{tab:morpho_param} presents the properties of the 45 targets we observed.

\subsection{Observations}
All observations presented in this study were acquired between September 2015 and January 2021 using the SPHERE / ZIMPOL instrument installed on the VLT \citep{Schmid2018}. The log of the observations is presented in Table~\ref{tab:log}. The data were obtained with the following filters: I$\_$PRIM, CntHa, N$\_$R, R$\_$PRIM, VBB, Cnt820, V, Cnt748, N$\_$Ha, and B$\_$Ha, the characteristics of which are given in Table~\ref{tab:filters}.

We observed 26 stars as part of the GTO, to which we added 19 previously published datasets, which we reduced in a uniform way.  Part of the observations were carried out within the framework of the ALMA Tracing the Origins of Molecules In dUst-forming oxygen-rich M-type stars (ATOMIUM) project. ATOMIUM focused on 17 evolved stars observed with ALMA for molecular mapping~\citep{Gottlieb2022}, and also combines high-spatial-resolution observations with SPHERE/ZIMPOL for polarized dust imaging on 14 of the 17 sources~\citep{Montarges2023}.

This diversity of origin of the observations made it possible to cover a wide range of instrumental configurations and seeing conditions, reinforcing the representativeness of the analyzed sample. Depending on the case, some stars were observed with alongside observations of a dedicated point-spread function (PSF) calibrator, while others were observed without.

\begin{table}[!ht]
\caption{Characteristics of the ZIMPOL filters \citep{Schmid2018} used for our observations.}
    \label{tab:filters}
    \centering
    \begin{tabular}{ccc}
    \toprule
    \hline
        Filter name & $\lambda_c$ ($nm$)& $\Delta \lambda ($nm$)$ \\
   \midrule
        VBB & 735 & 290 \\
        R\_PRIM & 626 & 149 \\
        I\_PRIM & 790 & 153 \\
        V & 554 & 81 \\
        N\_R & 646 & 57 \\
        Cnt748 & 747.4 & 20.6\\
        Cnt820 & 817.3 & 19.8 \\
        CntHa & 644.9 & 4.1 \\
        B\_Ha & 655.6 & 5.5 \\   
        N\_Ha & 656.9 & 1.0 \\
   \bottomrule
    \end{tabular}
\end{table}

\subsection{Data reduction}
The raw observational data of all our catalogue were reduced using the SPHERE Data Centre (SPHERE-DC), a data processing service developed by the French scientific community for users of the SPHERE instrument \citep{Delorme2017}. This centre uses a reduction pipeline enriched with specific routines to optimize the reduction of SPHERE data.
The raw data were calibrated using bias and flat-field frames. We then produced aligned images at different polarization modulator positions, recombination of polarized beams ($I^+$, $I^-$), and differential subtraction to extract the Stokes polarization maps ($Q$, $U$) necessary for the construction of intensity maps \citep{Schmid2018,Boer2020}.

%\subsection{Retrieval of polarimetric observables}
In addition to data reduction, the SPHERE-DC pipeline also provides polarimetric observables such as the Stokes intensity (I), the polarized intensity (PIL), the degree of linear polarization (DoLP), and the angle of linear polarization (AoLP). These observables are used to shape the dust distribution around the structures. We described the polarized intensity and the degree of polarization using the following relations
\begin{equation}
\begin{split}
\rm PIL & = \rm \sqrt{U^2 + Q^2}\\
\rm DoLP & = \rm \frac{PIL}{I}\\
\rm AoLP & = \rm \frac{1}{2}\arctan(U/Q)\,.
\end{split}
\label{eq:pol_bright_stars}
\end{equation} 

\section{Results}\label{sec:3}

\subsection{Classification criteria for our 45 stars sample}
We classified our sources envelopes into two main groups: resolved objects (clearly resolved and marginally resolved) and unresolved objects. For the classification, we used a resolution criterion based on both a study of the 2D Gaussian profile of the envelopes and the calculation of their mean radial profile of the data in polarized light.

To optimize the envelope characterization and improve the stellar classification, we applied automatic thresholding of the star intensity in polarized light to all 45 stars studied. Unlike the conventional method based on the estimation of the full width at half maximum (FWHM), we determine the width $w_h$ at a normalized intensity height $h$ which is much lower than 50\% of the intensity peak for each star. The selection of $h$ is automatically adapted to the observed morphology of the considered star.

For each image, we explore a range of $h$ that is linearly distributed between 0.5\% and 10\% of the maximum intensity. At each threshold h, a set A is defined as

\begin{equation}\label{eq:ellips_a}
    \rm {A(x, y)} = 
\begin{cases}
1 & \rm if \ PIL(x, y) > h \times PIL_{\max} \\
0 & \rm otherwise \ ,
\end{cases}
\end{equation}
where PIL$_{\max}$ is the maximum linear polarized intensity of the image and (x,y) denote the coordinates of a given pixel. This region therefore defines the set of pixels above a given threshold $h$.
To determine how close A is to an ellipse region, we set a function ellipse with semi-major axis a, semi-minor axis b, its centre coordinates (x$_0$, y$_0$), and rotation angle $\theta$ from the positive horizontal axis to the ellipse’s major axis with

\begin{equation}
\rm{Ellipse(X,Y)} = \frac{X^2}{a^2}+\frac{Y^2}{b^2}-1\,,
\end{equation}
with the coordinates (X, Y) given by 
\begin{equation}
\begin{cases}
\rm X &= \rm{ (x-x_0) \cos \theta + (y-y_0) \sin\theta}\\
\rm Y &= \rm{-(x-x_0)\sin\theta  + (y-y_0)\cos\theta\,}.
\end{cases}
\end{equation}

We define a set B with all pixels of coordinates (x,y) located inside the area determined by the ellipse function. It is written as
\begin{equation}\label{eq:ellip_b}
    \rm{ {B(x, y)} =} 
\begin{cases}
\rm 1 &  \rm{if \ Ellipse(x, y) < 0} \\
\rm 0 & \rm otherwise \,.
\end{cases}
\end{equation}

For each image, we define the Dice coefficient~\citep{Dice1945}, which is a standard estimate of similarity between binary sets defined as
\begin{equation}\label{eq:Dice}
    \rm {Dice_h(A, B)} = \rm{\frac{2|A \cap B|}{|A| + |B|}} \ .
\end{equation}
Based on this Dice coefficient, we then minimize a function cost such as 

\begin{equation}\label{eq:cost}
   \rm {Cost} = \rm 1 - \rm{Dice_h(A, B) }\ .
\end{equation}

For a given $h$, the ellipse parameters are optimized to minimize the cost function, ensuring the best overlap between $A$ and $B$. 

For each $h$, we obtain a corresponding Dice$_h$. The optimal height $\hat{h}$ is defined as the value of $h$ that maximizes Dice$_h$. With $\hat{h}$, we keep the associated ellipse parameters $a$ and $b$ for the remainder of the paper for a given star. 

The choice of $h$ and the ellipse parameters corresponds to a trade-off between maximizing the spatial extent and S/N ratio of the polarized signal in the detected region $A$. The ellipse is used here as a simple geometrical proxy, and therefore the optimal value $\hat{h}$ can be interpreted as a characteristic intensity level that captures a significant fraction of the circumstellar polarized emission.

We then calculated the mean radial PIL profile from the brightest pixel of total intensity, which we take as the origin, and computed the width $\hat{w}_h$ at a given normalized intensity level $\hat{h}$. Here, $\hat{w}_h$ is defined as the characteristic radial extent, derived from the azimuthally averaged radial profile, at which the polarized intensity exceeds the threshold $\hat{h}$.
The same operation is performed for the star and its PSF. The ratio $\eta$ is then calculated as the fraction $w_h^{\star}/w_h^{\mathrm{PSF}}$, where $w_h^{\star}$ and $w_h^{\mathrm{PSF}}$ are the widths computed for the star and the PSF, respectively.

For targets with multiple ZIMPOL observations obtained in different filters, we verified that the global envelope morphology and resolved or unresolved nature remain consistent across filters within the observational uncertainties, although moderate variations in polarized intensity and apparent extent are expected from the wavelength dependence of dust scattering and polarization~\citep{Khouri2020,Montarges2023}.
However, a detailed, systematic analysis of the wavelength dependence is beyond the scope of this work, as the available observations were not designed to provide homogeneous multi-filter coverage for each source.

Therefore, we retained a single observation per star to ensure a homogeneous and conservative classification. Specifically, we selected the filter that maximizes the measured stellar width $w_h^\star$ (and thus $\eta$), so that the envelope extent is not underestimated when several  observations are available for the same star.

If the star has no dedicated PSF observation (this was the case for 21 of the 45 stars in our sample), the average value of $w_h$ measured from the PSFs of other stars observed in the same filter is assigned as $w_h^{\mathrm{PSF}}$, allowing the computation of $\eta$ for this object.  
This average PSF width is systematically accompanied by its standard deviation, which quantifies the intrinsic dispersion of the PSF for a given filter and is propagated as an uncertainty when a mean PSF is used instead of a dedicated reference. The corresponding statistics (mean, median, standard deviation, and minimum--maximum range of $w_h$ for each filter) are reported in Table~\ref{tab:fwhm_stats}.

\begin{figure}[!ht]
    \centering
    \includegraphics[width=\linewidth]{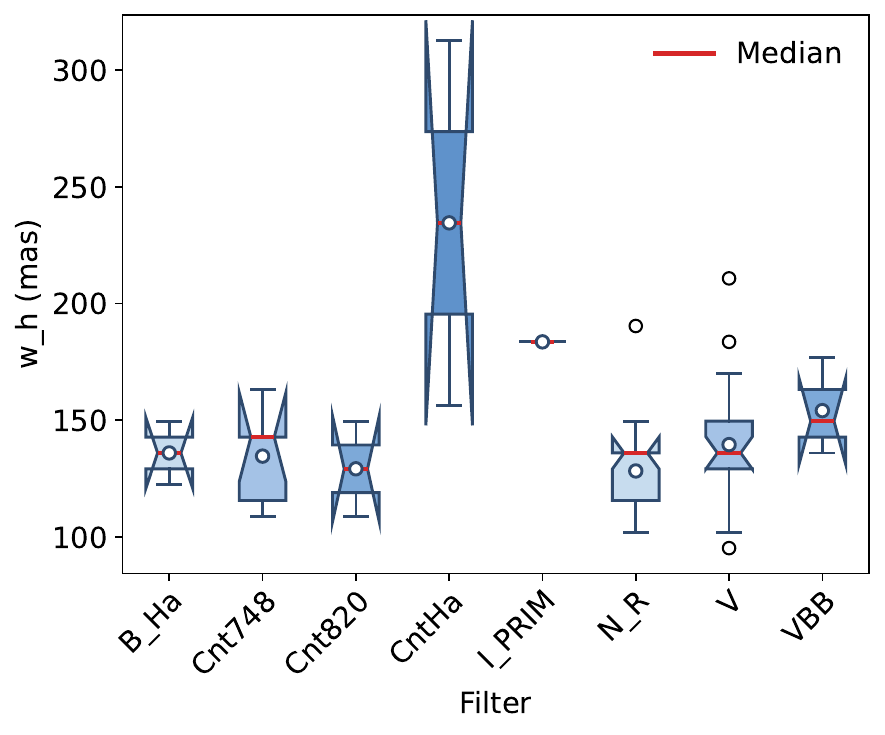}
    \caption{Distribution of the PSF width $w_h$ measured in polarized intensity for each SPHERE/ZIMPOL filter. 
    The boxplots show the median (horizontal red line), the interquartile range (boxes), and the overall spread of the measurements, 
    while the notches indicate the confidence interval on the median and the symbols mark the mean values. 
    This figure illustrates the intrinsic variability of the PSF width as a function of the filter and provides a visual estimate of the uncertainty 
    associated with the use of an average PSF width when no dedicated PSF observation is available.}
    \label{fig:fwhm_boxplot_by_filter}
\end{figure}

The dispersion of the PSF widths as a function of the filter is further illustrated in Figure~\ref{fig:fwhm_boxplot_by_filter}, where notched boxplots (with red lines marking the medians and symbols indicating the mean values) provide a visual assessment of the PSF variability available in each band. This representation allows us to directly evaluate the level of uncertainty introduced when adopting an average PSF width in the absence of a dedicated PSF observation.

After this, we set ourselves a resolution ratio threshold $\eta$ of 1.35,  and proceeded to rank stars on the basis of this threshold. This value was chosen because it corresponds to the point at which clear morphological features start to become visible in the images. Stars for which $\eta$ is larger than 1.35 are considered to be resolved; those for which $\eta$ ranges between 1 and 1.35 are considered to be marginally resolved; and those for which $\eta$ is smaller than 1 are considered to be unresolved (we include $\eta = 1$ in this category to take into account any errors in calculation, reading, and assessment). 

The classification criterion relies on the normalized width ratio $\eta = w_h^\star / w_h^{\mathrm{PSF}}$, such that variations in atmospheric conditions (as illustrated by the range of seeing values reported in Table~\ref{tab:log}) affect both the target and the PSF in a similar way and therefore are largely cancelled out. Although variations in seeing can influence the quality of the adaptive optics (AO) correction, they do not set the final angular resolution delivered by SPHERE, as the AO system corrects a large fraction of the atmospheric turbulence. In addition, our analysis is performed in polarized light, for which the impact of residual seeing-related effects is further reduced. In addition, although the absolute degree of linear polarization may vary with wavelength and filter width, the large-scale morphology and spatial extent of the polarized emission are expected to be robust across the visible range probed by ZIMPOL. For stars observed in several filters, we therefore retained a single dataset per target by selecting the filter that yields the largest measured $w_h^\star$, corresponding to the most extended polarized signal at the adopted threshold. This choice provides a conservative estimate of the envelope extent and prevents filter-dependent underestimation. As our analysis relies primarily on relative and morphological criteria through the ratio $\eta$, neither seeing variations nor filter-dependent effects significantly impact the resolved/unresolved classification results.

\begin{table*}[t]
\centering
\caption{Summary statistics of the PSF width $w_h$ (in polarized intensity) for each SPHERE/ZIMPOL filter.}
\label{tab:fwhm_stats}
\footnotesize
\setlength{\tabcolsep}{7pt}
\renewcommand{\arraystretch}{1.2}
\begin{tabular}{lccccccccc}
\toprule\hline
Filter & $N_{\rm obs}$ & Mean & Median & Std & Min & Q1 & Q3 & Max & CV \\
       &               & (mas) & (mas) & (mas) & (mas) & (mas) & (mas) & (mas) & (\%) \\
\midrule
B\_Ha   & 2  & 136.00 & 136.00 & 13.60 & 122.40 & 129.20 & 142.80 & 149.60 & 10.00 \\
Cnt748  & 5  & 134.64 & 142.80 & 19.90 & 108.80 & 115.60 & 142.80 & 163.20 & 14.78 \\
Cnt820  & 2  & 129.20 & 129.20 & 20.40 & 108.80 & 119.00 & 139.40 & 149.60 & 15.79 \\
CntHa   & 2  & 234.60 & 234.60 & 78.20 & 156.40 & 195.50 & 273.70 & 312.80 & 33.33 \\
I\_PRIM & 1  & 183.60 & 183.60 &  0.00 & 183.60 & 183.60 & 183.60 & 183.60 &  0.00 \\
N\_R    & 22 & 128.27 & 136.00 & 19.16 & 102.00 & 115.60 & 136.00 & 190.40 & 14.93 \\
V       & 21 & 139.56 & 136.00 & 27.01 &  95.20 & 129.20 & 149.60 & 210.80 & 19.35 \\
VBB     & 3  & 154.13 & 149.60 & 16.96 & 136.00 & 142.80 & 163.20 & 176.80 & 11.00 \\
\bottomrule
\end{tabular}
\tablefoot{$N_{\rm obs}$ denotes the number of available PSF observations per filter. The coefficient of variation (CV) is defined as the ratio between the standard deviation (Std) and the mean (Mean) value of $w_h$, and provides a normalized measure of the PSF dispersion
within each filter.}
\end{table*}

\begin{figure*}[!ht]
\centering
\includegraphics[width=1.82\columnwidth]{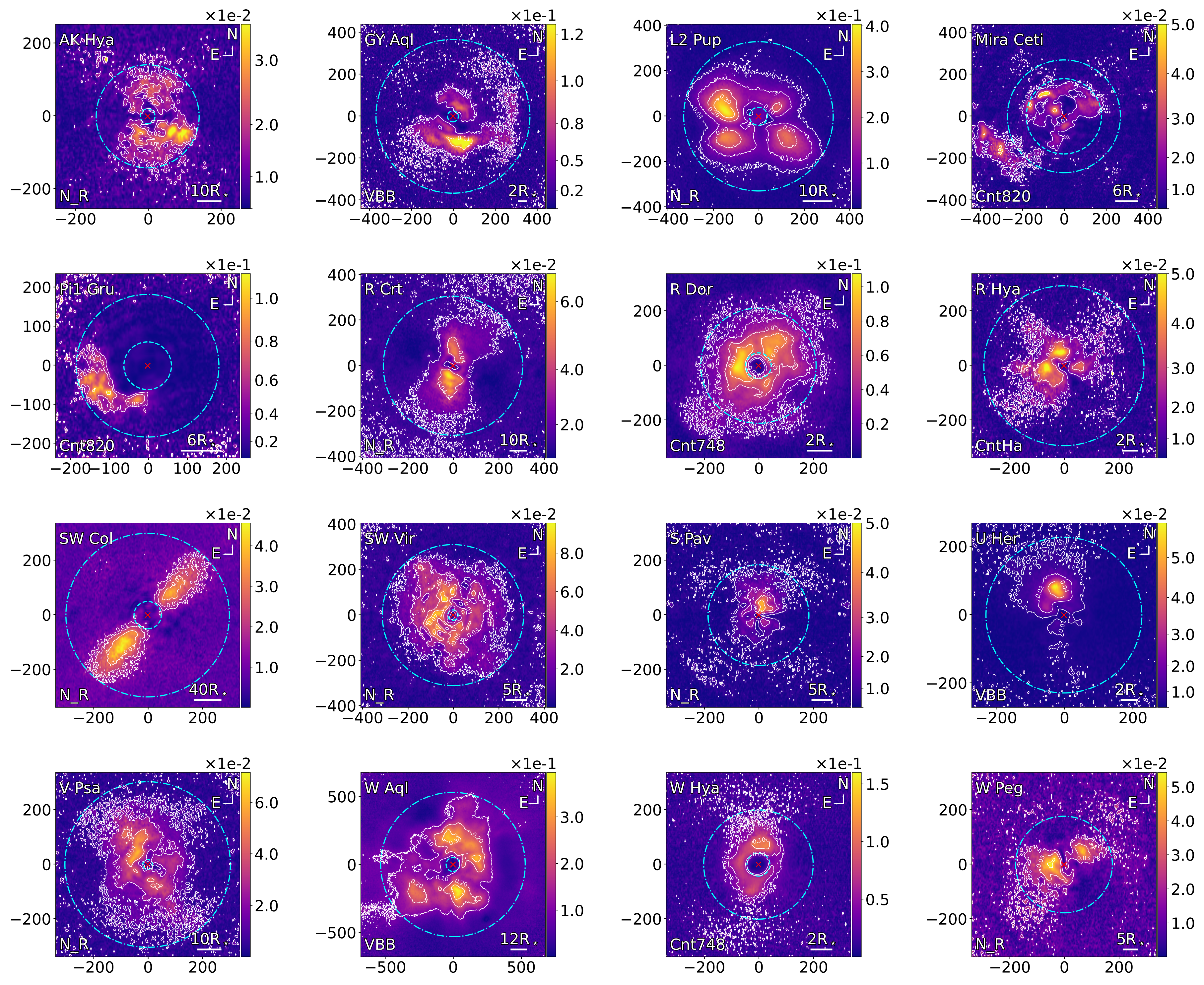}
\caption{Linear polarization degree maps of the 16 clearly resolved objects in our sample.
White contours, customized for each star, highlight the morphological structure of the circumstellar envelope. Cyan circles, also customized for each star, delineate regions containing strong polarized signals, indicating the presence of dust.
The horizontal axis corresponds to relative right ascension (RA) mas, while the vertical axis corresponds to relative declination (Dec) in mas. The scale bar at the bottom right is expressed in units of stellar radius (R$_\star$).}

\label{fig:18_dolp_panel}
\end{figure*}

\subsection{Previously published resolved targets}

In this section, we briefly present the properties of the objects in our catalogue observed with SPHERE/ZIMPOL that have already been published. We also discuss, based on DoLP maps, the morphologies of envelopes already mapped with SPHERE. The DoLP maps, on which the morphological analysis of the individual stars presented in this section and the following one is based, are grouped in Figure~\ref{fig:18_dolp_panel}.  Table~\ref{tab:overview_stellar_props} summarizes the main properties of the 16 objects with clearly resolved circumstellar envelopes, including their stellar characteristics, variability, mass-loss rates, known companions reported in the literature, and the main polarimetric properties derived in this work.

\subsubsection{S Pav}
$\rm S~Pav$  is an oxygen-rich AGB star of type M7IIe, classified as a semi-regular variable (SRa). Its distance was estimated at 470\,pc by~\citep{Heras2005}, who also observed a relatively low mass loss rate of $1.0 \times 10^{-7}$ M$_\odot.\mathrm{yr}^{-1}$. Several studies suggest that S~Pav could be a binary system, a hypothesis supported by the asymmetry of its circumstellar envelope revealed by high resolution imaging with ALMA~\citep{Decin2020}. The polarimetric study of \citet{Montarges2023} highlighted a maximum DoLP of 6\% in N$\_$R filter for S~Pav, revealing a relatively spherical material distribution extending up to a radius of 20 R$_\star$. 
Our observations show a maximum DoLP of $\sim \rm{5}\%$. Around S~Pav, an asymmetry is observed with a well-organized structure, extending up to 12R$_\star$(184.7mas), with material continuing beyond this radius in the form of small clusters.

\subsubsection{V PsA}
V~PsA is an M-type semi-regular variable (SRb) AGB star with a pulsation period of 148 days \citep{Aringer1999}. The available observations of the star do not reveal the presence of a stellar companion. \cite{Wallstrom2024} estimated its distance at 278~pc, its effective temperature at 2400~K, and its mass loss rate at $4.8 \times 10^{-7}$  M$_\odot.\mathrm{yr}^{-1}$. For V~PsA, polarization revealed an elongated structure extending beyond 20 R$_\star$ around the star, with a maximum DoLP of 6\% in N$\_$R~\citep{Montarges2023} indicating an extended and significant dust distribution. 
Our observations reveal a slightly higher maximum DoLP of $\sim \rm{7}\%$ and confirm a strongly asymmetric morphology, with a dominant structure oriented at a position angle of $\sim45^\circ$. Polarized emission is detected from $\sim2$~R$_\star$ ($\sim$20~mas) and extends outward up to $\sim34$~R$_\star$ ($\sim$305~mas), indicating that dust is distributed over a wide range of spatial scales around the star.

\subsubsection{R Hya}
R~Hya is an oxygen-rich AGB star of spectral type M6--9 and one of the best studied Mira variables \citep{Wenger2000}. Its pulsation period is 385 days, with a high amplitude in the visible, and its distance is estimated at 165~pc based on the parallax and the light curve \citep{Zijlstra2002}. Its stellar diameter is about 34~mas at $\sim$0.90~$\mu$m \citep{Haniff1995}, while near-IR measurements yield smaller values of $\sim$24~mas \citep{Richichi2005}, giving a radius of about 2.6~AU. Its effective temperature is 2830~K \citep{Feast1996}, and its mass loss rate is of the order $3 \times 10^{-7}$ M$_\odot.\mathrm{yr}^{-1}$, with significant historical variations \citep{Zijlstra2002}. Infrared observations from the MIPS InfraRed Imaging of AGB Dust shells (MIRIAD) program with Spitzer have revealed a nebulous extent of about 400" around R Hya, interpreted as a bow shock at the interface between the stellar wind and the interstellar medium \citep{Ueta2006}. No companion has been detected \citep{Haniff1995,Richichi2005}, but the circumstellar envelope shows asymmetries and complex structures, with the presence of amorphous silicate dust (olivine, pyroxene) and oxides typical of oxygen-rich AGB stars \citep{Begemann1997}. The polarimetric study of~\cite{Montarges2023} revealed structures in the form of isolated clumps in the environment of R~Hya, extending up to 10~R$_\star$, with a maximum DoLP of $ \sim \rm{5}$\%. Our observations confirmed this structure with a global radius of 10~R$_\star$ ($\sim$293~mas).

\subsubsection{W Aql}
W Aquilae (W~Aql) is an evolved intermediate-mass S star on the AGB branch, with a C/O ratio of 0.98 \citep{Brunner2018} and Mira variability (period of 490 days). Its mass loss rate is $3.0 \times 10^{-6}\mathrm{M}_{\odot}.\mathrm{yr}^{-1}$, its expansion velocity is $16.5\mathrm{kms}^{-1}$, and its distance is 395~pc~\citep{Ramstedt2018, Danilovich2014}. W~Aql has a F8–G0 companion at 0.46" (180 AU), inducing spiral structures and arcs in the circumstellar envelope \citep{Ramstedt2017}, with an overall smooth envelope that exhibits local asymmetries and a southwest-polarized extension. The W~Aql object exhibited the strongest and most extensive DoLP, 
with a maximum of $39$\%, and material localized in places around the star, mainly beyond 20 R$_\star$~\citep{Montarges2023}. This pronounced polarized DoLP indicates a particularly efficient dust production and the presence of dust grains with sizes and compositions that efficiently scatter stellar light and produce a high degree of linear polarization in the visible. For this object, our DoLP map  shows an extremely large asymmetrical morphology around the star with an inner radius of 5~R$_\star$ (51.1~mas), reaching 53~R$_\star$ (530.7 mas). Inside the large envelope, we can see mainly asymmetrical clumps of dust with a DoLP of $\sim 39$\%. This study confirms the results of~\citep{Montarges2023} and suggests an important mass loss activity in the star. They are also in agreement with previously reported structures observed with  ALMA~\citep{Ramstedt2017}.

\subsubsection{GY Aql}
Gamma Aquilae (GY~Aql) is an evolved M6–M10 star identified as a Mira-type AGB star that stands out as one of the most powerful SiO emitters among AGB stars~\citep{Pardo2004}. Initially considered semi-regular, studies such as~\cite{Alcolea1999} revealed a remarkable regularity in SiO variability and a contrast of 4-5, typical of Mira stars.
The optical period of GY~Aql, set at $461.5 \pm 76$ days, agrees perfectly with the period of the SiO maser determined by time series analysis \citep{Pardo2004}. The SiO maser emission has a main peak at $34 \pm 0.5$ km.s$^{-1}$, reflecting the velocity of the circumstellar gas, with a typical width of 1 to 6 km.s$^{-1}$~\citep{Alcolea1999}. To date, no interpretation in favour of the presence of a binary companion has been proposed for GY Aql, and the line profiles do not show signatures of binary interactions or complex circumstellar structures. The linear polarization of GY~Aql, as observed by~\cite{Montarges2023}, appears as two lobes: the smaller lobe is located within 2~R$_\star$, while the larger one is situated outside 2~R$_\star$, with a maximum DoLP of $\sim 18\%$. This analysis coincides with the onset of spiral arms in ALMA gas maps, suggesting that dust forms or grows in these structures. For GY~Aql, our DoLP map shows a maximum DoLP of $\sim \rm{13}\%$ and two asymmetrical clumps at 1.2~R$_\star$ (25.9~mas) in the north and $\sim20$~ AU in the south of the star. The matter extends clearly as clumpy structures and reaches a radius of$\sim$17~R$_\star$ ($\sim$337~mas) around the star.

\subsubsection{$\rm \pi ^1$ Gruis}
$\rm \pi ^1$ Gruis ($\pi^1$~Gru)  is an AGB star of type S5.7, located at about 163~pc, at the transition between oxygen-rich and carbon-rich stars~\citep{Mayer2014}. Its initial mass is about 2~M$_\odot$, for a current mass close to 1.5~M$_\odot$ \citep{Mayer2014}. The star exhibits semi-regular variability and a mass-loss rate estimated between $7.7 \times 10^{-7}$~M$_\odot$.yr$^{-1}$ \citep{Doan2017} and $2.7 \times 10^{-6}$~M$_\odot$.yr$^{-1}$ \citep{Winters2003}. It has an effective temperature of 3100~K \citep{VanEck1998}, a luminosity of 7240~L$_\odot$ \citep{Mayer2014}, a gas-to-dust mass ratio of 380 \citep{Groenewegen1998}, and a wind expansion velocity of 14.5~km.s$^{-1}$~\citep{Guandalini2008}. ALMA observations have revealed a complex circumstellar envelope around $\pi^1$~Gru, dominated by an extended and inclined equatorial disk, whose expansion varies with latitude \citep{Nhung2016, Doan2017}. This disk is associated with a rapid bipolar outflow, exceeding 100~km.s$^{-1}$, well above typical AGB values~\citep{Doan2017, Doan2020}, probably due to an episode of enhanced mass loss a century ago. The gas morphology also reveals a spiral structure, detected in CO, HCN, and SiO emissions \citep{Doan2020, Homan2020}.
The observed dynamics cannot be explained by the distant G0V-type companion located at a separation $\sim$460~AU~\citep{Mayer2014}, but rather by a hidden stellar companion, recently revealed by ALMA and SPHERE at 6.05$\pm$0.55~AU~\citep{Montarges2025}, responsible for the formation of the equatorial disk and the spiral structure \citep{Homan2020}. Recent studies further constrained the nature of this close companion. Using multi-epoch ALMA observations, \cite{Esseldeurs2026} showed that it follows a Keplerian circular orbit and is likely a main-sequence star slightly more massive than $\rm \pi ^1$ Gru itself. In addition, VLTI/MATISSE imaging by \cite{Drevon2026} revealed ongoing mass transfer through a WRLOF scenario, together with a circumcompanion disk-like structure and plume features linked to the onset of large-scale spiral morphology. The study of the $\pi^1$~Gru DoLP by~\cite{Montarges2023} presents a polarized signal localized on one side of the star, between 10 and 20~R$_\star$, with a maximum DoLP of $20\%$. This is probably related to the interaction between the stellar wind and a nearby companion, forming a polarized dust trail. For $\pi^1$~Gru, our observations reveal a  DoLP maximum of $\sim \rm{11}$ \% and show clumpy polarized structures organized in an arc-like morphology, evocative of a spiral pattern, extending up to  a radius of 18~R$_\star$ (313.3~mas). This interpretation is supported by ALMA observations, which reveal spiral patterns and an equatorial disk in the gas component, suggesting a close coupling between the gas dynamics and the dust distribution.

\subsubsection{U Herculis}
U~Herculis (U Her) is a Mira oxygen-rich AGB star,  exhibiting intense maser activity in the H$_2$O, SiO, and OH lines~\citep{vanLangevelde2000}. Interferometric observations using very long baseline interferometry (VLBI) have allowed the precise localization of the maser regions. The brightest H$_2$O maser spot is identified as the amplified stellar image \citep{Vlemmings2002}, and for the OH maser at 1667 MHz, the amplified stellar image model is confirmed: VLBI measurements show that the most blueshifted maser spot traces the stellar position, $\mu_{\delta} = -10.9 \pm 1.4$ mas.yr$^{-1}$)~\citep{vanLangevelde2000}. The SiO masers, studied by \citep{Cotton2010}, form rings about twice as large as the stellar disk in the near-IR, comparable to the size of the molecular envelope, without marked asymmetry but with radial velocity gradients.

U~Her was observed with a coronagraph, and its polarization map reveals a relatively extensive structure around the central star. The overall DoLP reaches relatively high values in the circumstellar environment of the star, with a maximum of $\sim18$\%. The morphology of the scattering material reveals a highly asymmetric structure, preferentially extended towards the north-east, reaching up to 9 R$_\star$ ($\sim$200 mas). This complex structure is consistent with the rings and the stellar disk reported by~\cite{Cotton2010}, and suggests the presence of a binary interaction.

\subsubsection{SW Vir}
SW Virginis (SW Vir)   is an M7III AGB star located at 143$\pm20$~pc  according to \citet{vanLeeuwen2007}. It exhibits semi-regular variability (SRb) with a period of 154 days and a visual amplitude of approximately 1.5~mag \citep{Jura1992}. The presence of technetium indicates that it is a thermally pulsating AGB \citep{Lebzelter2003}.
Its uniform angular diameter is 16.8~mas and 16.7~mas in H- and K-band \citep{Ridgway1982,Schmidtke1986}, giving a physical radius of approximately 1.2~AU. SW~Vir has a luminosity of 4500~$\pm$~1100~L$_\odot$ and an effective temperature of 2990 $\pm$ 50K according to~\citep{Ohnaka2019}, for an estimated mass between 1 and 1.25~M$_\odot$. The Astronomical Multi-BEam combineR installed on Very Large Telescope Interferometer (VLTI/AMBER) observations revealed an extended outer atmosphere, with a diameter increasing from 16.2~mas in the continuum to 22$-$24~mas in the CO lines, indicating a MOLsphere up to $\sim2$~R$_\star$ and a H$_2$O column of 10$^{19}$–$10^{20}$cm$^{-2}$~ \citep{Ohnaka2019}. According to~\cite{Olofsson2002} and~\cite{Khouri2020} the SW~Vir mass loss rate is estimated between $1.7 \times 10^{-7}$ and $9.8 \times 10^{-7}$~M$_\odot$.yr$^{-1}$ with an expansion velocity of $7.5$–$8.5$~km.s$^{-1}$. \cite{Khouri2020} estimated a less resolved polarized envelope in the visible, but with polarization detected at distances shorter than 2 $R_\star$ (R$_\star \sim$ 0 .9$-$1.1 AU). The maximum DoLP is about $\rm{9.5}$~\%, and the polarized region remains close to the star, mainly in the eastern hemisphere. With our observations, the DoLP map  shows an inner radius of 1.2~R$_\star$ (22.6~mas) and an outer radius of 16.5~R$_\star$ (310.7~mas).
This morphology is consistent with previous studies that report the presence of an extended MOLsphere and dust formation that occurs already in the close vicinity of the star \citep{Ohnaka2019,Khouri2020}.

\subsubsection{R~Crt}
R Crateris (R~Crt)  is a semi-regular variable AGB star of type SRb, with a period of about 160 days and optical brightness varying between 9.8 and 11.2  mag, i.e. an amplitude of $\sim$1.4 mag~\citep{Kholopov1987, Szymczak1999}. Its spectral classification varies from M4 to M7/M8III depending on the studies \citep{Szymczak1999}. The O-rich star R Crt exhibits OH (1612, 1665, 1667 MHz), H$_2$O, and SiO maser emissions \citep{Jewell1991, LeSqueren1979}.
The OH maser emission, imaged using polarimetric data with MERLIN, reveals a compact structure (<0.2 arcsec) and an envelope modelled as a thin shell of radius $0.14 \pm 0.04$ arcsec (42–66 AU at 300 pc) \citep{Szymczak1999}. The OH envelope expansion velocity is 7.9 km.s$^{-1}$, smaller than that of the H$_2$O masers (12.8 km.s$^{-1}$) and CO (11~km.s$^{-1}$), suggesting an axisymmetric structure or regional velocity variations. The estimated mass loss rate  of the star via OH is $0.7$–$1\times10^{-7}$ M$_\odot$.yr$^{-1}$, while estimates via CO are ten times higher~\citep{Kahane1994}. The study of~\cite{Khouri2020} on R~Crt showed an envelope with a well-marked bipolar morphology in visible polarization, with a polarized cone extending up to $\sim$ 6$-$12 AU from the centre (i.e. 5 R$_\star$, with  R$_\star \sim $ 1.25$-$2.5 AU in the visible). The maximum DoLP reaches about $\sim \rm{6.5}$\%, with the south-eastern cone being the brightest in polarization. Our DoLP map (the same data as~\citep{Khouri2020})   confirms the large bipolar morphology around the star which reaches 40~R$_\star$ (306.5 mas). In this large envelope, two asymmetrical clumps, consistent with the observations of \citet{Kahane1994}, suggest high mass loss activity in the inner regions of the envelope.

\subsubsection{$\rm L_2 \ Pup$}
$\rm L_2 \ Puppis$  is an AGB star located at 64~pc, with a mass $\sim 2 M_\odot$ \citep{Kervella_2014}, a prototype for the study of the latest evolutionary phases~\citep{Kervella_2014}. Observations with the NAOS-CONICA adaptive optics infrared camera at the VLT (NACO) in 2013 and 2015 revealed an opaque dust lane in the J band, interpreted as a disk inclined at 84°~\citep{Kervella_2014,Lykou_2015}. In 2015, SPHERE/ZIMPOL  confirmed the disk and discovered a companion at 2~AU~\citep{Kervella_2015}. Modelling with RADMC-3D allowed \cite{Kervella2015} to locate the inner edge of the disk at 6 AU and the dense outer edge at 13 AU from the centre of the star. The DoLP reaches a maximum of 42\% in the V filter and 58\% in the NR filter, indicating efficient scattering by dust grains on the disk. Their analysis suggested an early stage of bipolar planetary nebula formation. A recent ALMA study \citep{Van_de_Sande_2024} modelled the disk, estimated its age at $\lesssim 10^5$ years, and showed carbon-rich chemistry despite an oxygenated disk, paving the way for chemical exploration of disks around AGBs. For $\rm L_2 \ Pup$, our DoLP map shows an elongated structure with a P.A of $\sim50°$ and a size of 25~R$_\star$ (308.8~mas). Inside this envelope are located mainly four asymmetrical clumps with different levels of polarization, as observed by \citep{Kervella2015}, and explained as the polarimetric signature of an edge-on disk.

\subsubsection{R Doradus}
R Doradus is an oxygen-rich AGB star with low mass loss and a complex circumstellar envelope. The abundance profiles of SiO and HCN in the stellar wind show high initial abundances and specific radial declines, illustrating the chemical dynamics of the circumstellar medium \citep{VanDeSande2018}. ALMA observations have modelled gas density, temperature, and velocity, revealing complex motions and local overdensities in the extended atmosphere \citep{Khouri2024}. Interaction with the ISM is manifested by a bubble detected in UV, interpreted as the result of the stellar wind on the galactic environment \citep{Ortiz2023}. The VLTI/AMBER infrared interferometry on this object revealed an expanding outer atmosphere, linked to dust formation \citep{Ohnaka2019}. Visible studies carried out  with SPHERE/ZIMPOL by \citet {Khouri2016} revealed a resolved stellar disk around R~Doradus, with a FWHM ranging from 58 to 71~mas depending on the filter and the epoch. The most representative size of the visible continuum is about 59~mas (i.e. $\sim$3.4~AU at 59~pc), reflecting the influence of the variable molecular opacity in the star's atmosphere. Finally, ALMA revealed rotation, fast flows, and cold layers of SiO, revealing the morpho-kinematic complexity of the envelope \citep{Nhung2021}. For R~Doradus, our observations (ESO program~098.D-0731(A), PI: Theo Khouri), which had not yet been published at the time of this study and are therefore distinct from previous observations reported in the literature, in particular those of~\cite{Khouri2016,Schirmer2025},  show a high polarization with a maximum of $\sim \rm{10}\%$. Matter  is detected at a distance less than 2~R$_\star$ ($<$105~mas) with the highest DoLP close to the star. The envelope of this matter is clearly structured in a radius of 4.5~R$_\star$~($\sim$210~mas) and extends beyond the envelope as clumpy structures. Our observations are consistent with previous SPHERE/ZIMPOL studies of R Doradus \citep{Khouri2016,Schirmer2025}, confirming an asymmetric dusty environment close to the star. The prominent polarized clumpy structures and high DoLP observed within a few stellar radii further support a highly inhomogeneous and time-variable dust formation process.

\subsubsection{Omi Ceti}
Mira~Ceti (or Omi Ceti) is the prototype star of the Mira variables, an AGB giant of type M5-9IIIe located at 92±10~pc \citep{vanLeeuwen2007}. It forms a binary system with a white dwarf (Mira B) at $\sim$70-80 AU, with an orbital period of approximately 500 years~\citep{Prieur2002, Planesas2016}. Mira has a pulsation period of 333 days and a visual amplitude of approximately 8 mag \citep{Templeton2009}, with a temperature of 2900$-$3200~K, a radius of 21~mas, and a mass-loss rate of 2–3×10$^{-7}$~M$_\odot$.yr$^{-1}$~\citep{Khouri2018, Ryde2001}. Its circumstellar envelope shows complex dynamics, pulsational shocks, and inhomogeneous dust formation \citep{Wong2016, Khouri2018}. ALMA observations of Mira~Ceti reveal fragmented morphology and preferential outflows \citep{Ramstedt2014, Hoai2020}. The detection of neutral atomic carbon (CI) emission around Mira, shifted by $\sim$4 km.s$^{-1}$, suggests an origin related to photodissociation by Mira B~\citep{Saberi_2018}. The visible morphology of Mira A, observed with SPHERE/ZIMPOL, revealed a marginally resolved stellar disk with a radius of about 21~mas corresponding to 2.15~AU at 102~pc (the distance estimated by~\citet{vanLeeuwen2007}), as well as polarized dust structures around the star and its companion~\citep{Khouri2018}. For this object, our observations (program ESO 100.D-0737(B), PI: T. Khouri, Cnt820 filter) show an asymmetrical structure located at a radius of 17~R$_\star$~($\sim$305~mas) and 1.8~R$_\star$ ($\sim$32~mas). 

The recent study by \cite{Khouri2026}, based on the same SPHERE/ZIMPOL observing program but using V-band data, shows a broadly similar polarized morphology. Differences in wavelength sensitivity lead to variations in contrast and apparent spatial extent, but both datasets are consistent with a highly structured and asymmetric dust environment around Mira A.

This asymmetric distribution, characterized by arcs and localized condensations, is consistent with a perturbation of the stellar wind induced by a gravitational interaction with the companion. Such structures are in line with ALMA interferometric observations, which reveal pulsation-driven shocks and strongly inhomogeneous dust formation in the close environment of Mira~Ceti~\citep{Wong2016,Khouri2018}.

\subsubsection{W Hya}
W Hydrae (W Hya) is an AGB star of type M7IIIe, located at 78$\pm$7~pc \citep{Vlemmings2003}, a semi-regular variable (SRa) with a period of 382 days and a visual amplitude of 6 to 10~mag \citep{Lebzelter2003}. The star exhibits SiO, H$_2$O, and OH maser emissions (1612, 1665, and 1667 MHz), but the OH maser emission at 1612 and 1665 MHz is weak, typical of SRa~\citep{Etoka2003}.
Analysis with the Infrared Space Observatory Short Wavelength Spectrometer (ISO-SWS) reveals a rich molecular inventory (CO, H$_2$O, OH, CO$_2$, SiO, SO$_2$) around the star, with excitation temperatures of 300 to 3000~K \citep{Justtanont2004}. Spectral modelling indicates the presence of amorphous silicates, Al$_2$O$_3$ and Mg$_{0.1}$Fe$_{0.9}$O, with a dust mass loss rate of 3×10$^{-10}$M$_\odot$.yr$^{-1}$ and a gaseous mass loss rate of (3.5$-$8)×10$^{-8}$M$_\odot$.yr$^{-1}$. VLTI/MIDI observations reveal an asymmetric structure of W~Hya with a photospheric diameter of 42$\pm$2~mas, correlated with a detached shell and a north-easterly jet observed by the Herschel Space Observatory’s Photodetector Array Camera and Spectrometer (Herschel-PACS). The masers trace different asymmetric regions of the envelope, suggesting non-spherical mass loss~\citep{Zhao-Geisler2015}. The star is listed as an optical double, but no physical companion has been confirmed.
The dust envelope traced by visible polarization extends to about 30~AU from the centre~\citep{Khouri2020}, with a DoLP reaching a maximum of about 2.5\%, observed very close to or within the visible photosphere. Similar structures have also been reported in previous studies~\citep{Ohnaka2016,Ohnaka2017,Ohnaka2024}. These multi-epoch and multi-wavelength studies of W~Hya \citep{Ohnaka2016,Ohnaka2017,Ohnaka2024,Khouri2020} reveal a highly time-variable and clumpy dust environment within a few stellar radii. Although the overall asymmetric and inhomogeneous nature of the envelope is consistently observed, the detailed morphology (e.g. location, number, and contrast of clumps, as well as apparent large-scale asymmetries) varies significantly with time and wavelength, reflecting both dynamical evolution and phase-dependent dust formation. Our DoLP map, based on the same dataset as~\citep{Khouri2020} (ESO program~096.D-0930(B), PI: Theo Khouri), reveals a generally bipolar morphology, oriented at a position angle of~$\sim$~0°. Within this bipolar envelope appear four main clusters, distributed asymmetrically around the star and organized into two perpendicular structures, one of which is aligned with the global axis. These clusters exhibit high degrees of local polarization (up to $\sim \rm{34}$ \%). The entire structure is contained within a radius of approximately 5R$_\star$ ($\sim$ 200~mas), with matter being detected as low as $\sim$ 1R$_\star$ ($\sim$ 40 mas).

\subsection{Newly resolved objects}
Among the objects classified as clearly resolved,  AK Hya, SW Col, and W~Peg are presented here for the first time in high-resolution polarized visible light. In this section, we describe these sources. 

\subsubsection{AK Hya}
AK Hydrae (AK~Hya) is an M6III-type AGB star~\citep{percy2001}. It is classified as a semi-regular variable (SRb), with a pulsation period of 50–75 days and a photometric amplitude of 1.16 mag in V~\citep{percy2001}. Its mass loss is moderate, probably less than $10^{-7}$~M$_\odot$.yr$^{-1}$~\citep{kerschbaum1999}, and the gas expansion velocity is 4.6 km.s$^{-1}$, typical of AGBs with low expansion velocities. The CO observations of AK~Hya show detections in CO (1$-$0), CO (2$-$1), and CO (3$-$2) transitions with integrated intensities of 0.43, 2.79, and 6.11~K.km.s$^{-1}$ \citep{kerschbaum1999}. The parabolic line profile and expansion velocity indicate an optically thick envelope and a moderate mass loss rate, typical of red SRVs with periods $<$ 200 days. The American Association of Variable Star Observers (AAVSO) observations over more than 3100 days confirm a complex multi-periodic variability of this star, characteristic of AGBs in the thermal pulsation phase~\citep{percy2001}.

The AK~Hya polarized intensity maps reveal a bipolar circumstellar  structure around the central source. The degree of linear polarization remains relatively low with a maximum of $\sim \rm{3.5}$\%, indicating an object with observed scattering activity that is less intense than other objects in the sample. The inner circle represents 5 R$_\star$ (33.7 mas) and the large one 20 R$_\star$. The circumstellar material we detect starts as close as 5 R$_\star$ ($\sim$\,35 mas). The structure is elongated along the north-south direction, with a P.A.$\sim$5$^o$. Weak polarized structures are revealed at the top and bottom of this structure and are concentrated toward its edges. This is more clearly seen in the northern part ($\sim$ 135 mas) of the bipolar shell. The bipolar structure is twice as narrow along the equator.

\subsubsection{SW Col}
SW Columbae (SW Col) is an M1III giant star on the AGB with an estimated distance of 92~pc by~\cite{Dumm1998}. 
SW~Col is a variable AGB star with a luminosity of 921~L$_\odot$ and an effective temperature of 3660~K~\citep{McDonald2012}. It has a visual amplitude of 0.33~mag in the V band and a main pulsation period of 78.7~days~\citep{Tabur2009}. Characterized by multi-periodic pulsations and moderate amplitudes, this behaviour is typical of SRVs~\citep{Tabur2009}. Although the majority of the local M~giants are variable, SW Col does not exhibit extreme variability~\citep{Adelman2000}. At the circumstellar level, SW~Col shows a moderate IR excess, indicative of dust production and an active stellar wind~\citep{McDonald2012}.

Due to its remarkable morphology, SW~Col constitutes one of the most fascinating objects in our sample. The DoLP map reveals relatively moderate values of the degree of linear polarization, reaching $\sim \rm{4.5}$\%.  The revealed structure is highly bipolar and extends up to 120 R$_*$ (300~mas)  in both directions. The circumstellar material we detect starts as close as 20 R$_*$ ($\sim$ 50 mas).

\subsubsection{W Peg}
W~Pegasi (W~Peg) is a Mira variable with a photometric amplitude of five magnitudes (8$-$13 mag) and a period of $344.9$ days~\citep{Wyatt1983}. Its average spectral type is M7e at maximum light, characterized by TiO bands, Balmer emission lines, and weak metallic lines~\citep{Keenan1974}. Its position and radial velocity associate it with the Galactic disk. In addition, IR measurements ($J=1.1$, $H=0.2$, $K=0.1$) confirm a cold atmosphere of this star ($T_{\text{eff}} \sim 2800$ K)~\citep{Cutri2003}. W~Peg has estimated angular diameters of $8.2 \pm 0.1$ mas and $9.6 \pm 0.1$ mas in the H and K bands ~\citep{Millan2005}, reflecting the effects of molecular opacity. It has an estimated distance of $310 \pm 31$~pc, which corresponds to a linear diameter of about 1.2~AU in K~\citep{Wyatt1983}. To date, no binary companion has been detected for W~Peg.

 Its polarization map reveals an extended structure around the central star. The structure is oriented along a P.A$\sim$120$^o$ and is larger in the south-east than in the north-west and extends up to 16 R$_\star$ ($\sim$ 180 mas). The overall DoLP remains relatively low in the circumstellar environment of the star, reaching a maximum of $\sim \rm{5.3}$\%. The degree of polarization is very low perpendicularly to the extend structure, as seen in other bipolar structures such as SW~Col.

\begin{table*}[h]
\centering
\caption{Overview of stellar, variability, mass loss, and polarimetric properties for the 18 resolved AGB stars observed with SPHERE/ZIMPOL.}
\label{tab:overview_stellar_props}
\footnotesize
\renewcommand{\arraystretch}{1.2}
\setlength{\tabcolsep}{4pt}

\begin{tabular}{lccccccccccc}
\toprule \hline
Star &
Stellar type &
Distance &
Var &
$P$ &
$T_{\rm eff}$ &
$L$ &
Mass-loss rate &
Max. &
$R_{\rm out}$ &
Companion \\
&
&
(pc) &
&
(d) &
(K) &
($L_\odot$) &
($M_\odot\,{\rm yr}^{-1}$) &
 DoLP(\%) &
($R_\star$) &
(sep.) \\
\midrule
S~Pav     & AGB O-rich & 470 (1) & SRa  & --        & --        & --        & $1.0\times10^{-7}$ (1) & $\sim$5  & 12  & suspected \\
V~PsA     & AGB O-rich & 278 (2) & SRb  & 148 (3)   & 2400 (2)  & --        & $4.8\times10^{-7}$ (2) & $\sim$7  & 34  & -- \\
R~Hya     & AGB O-rich & 165 (4) & Mira & 385 (4)   & 2830 (6)  & --        & $\sim3\times10^{-7}$ (4) & $\sim$5  & 10  & 3500~AU (23) \\
W~Aql     & AGB S-type & 395 (7) & Mira & 490 (7)   & --        & --        & $3.0\times10^{-6}$ (7) & $\sim$39 & 53  &  0.46\arcsec (24) \\
GY~Aql    & AGB O-rich & --      & Mira & $461.5$ (8) & -- & -- & -- & $\sim$30 & 17 & -- \\
$\pi^1$~Gru & AGB S-type & 163 (9) & SR & -- & 3100 (9) & 7240 (9) & $(7.7\times10^{-7}$) (9) & $\sim$11 & 20 &  6.05 AU (25) \\
U~Her    & AGB O-rich & -- & Mira & -- & -- & -- & -- & $\sim$18 & 8 & -- \\
SW~Vir   & AGB O-rich & $143$ (10) & SRb & 154 (10) & 2990 (11) & $4500$ (11) & $(1.7$--$9.8)\times10^{-7}$ (10) & $\sim$9.5 & 16.5 & -- \\
R~Crt    & AGB O-rich & --     & SRb  & $\sim$160 (12) & -- & -- & $(0.7$--$1)\times10^{-7}$ (12) & $\sim$6.5 & 40 & -- \\
L$_2$~Pup & AGB O-rich & 64 (13) & Mira & -- & -- & -- & -- & $\sim$40 & 25 &  $\sim$2 AU (26) \\
R~Dor    & AGB O-rich & 59 (14) & -- & -- & -- & -- & -- & $\sim$10 & 4.5 & -- \\
Mira~Ceti   & AGB O-rich & $92$ (10) & Mira & 333 (15) & 2900--3200 (15) & -- & $(2$--$3)\times10^{-7}$ (15) & $\sim 5$ & 17 &  $\sim$70 AU (27) \\
W~Hya    & AGB O-rich & $78$ (16) & SRa & 382 (16) & -- & -- & $(3.5$--$8)\times10^{-8}$ (16) & $\sim$34 & 5 & -- \\
AK~Hya   & AGB O-rich & -- & SRb & 50--75 (17) & -- & -- & $<10^{-7}$ (17) & $\sim$3.5& 20 & -- \\
BK~Vir   & AGB O-rich & 180 (18) & -- & -- & 2920 (18) & $2700$ (18) & $\sim2\times10^{-7}$ (18) & $\sim$2 & 35 & -- \\
RT~Vir   & AGB O-rich & $226$ (19) & SRb & 153--320 (19) & 2902 (19) & $\sim6000$ (19) & $10^{-7}$--$10^{-6}$ (19) & $\sim$7 & 23 & -- \\
SW~Col   & AGB O-rich & 92 (20) & SRV & 78.7 (20) & 3660 (20) & 921 (20) & -- & $\sim 4.5$ & 120 & -- \\
W~Peg    & AGB O-rich & $310$ (21) & Mira & 344.9 (21) & $\sim2800$ (21) & -- & -- & $\sim 5.3$ & 16 & -- \\
\bottomrule
\end{tabular}

\tablefoot{
$R_{\rm out}$ denotes the maximum detected radius of polarized dust expressed in stellar radii.
All objects listed in this table are resolved in polarized visible light.\\
\textbf{References:}
(1) \citep{Heras2005};
(2) \citep{Wallstrom2024};
(3) \citep{Aringer1999};
(4) \citep{Zijlstra2002};
(5) \citep{Wenger2000};
(6) \citep{Feast1996};
(7) \citep{Ramstedt2018,Danilovich2014};
(8) \citep{Pardo2004};
(9) \citep{Mayer2014,Doan2017,Winters2003};
(10) \citep{vanLeeuwen2007};
(11) \citep{Ohnaka2019};
(12) \citep{Kholopov1987,Szymczak1999,Kahane1994};
(13) \citep{Kervella_2014};
(14) \citep{VanDeSande2018};
(15) \citep{Templeton2009,Khouri2018};
(16) \citep{Vlemmings2003,Lebzelter2003,Justtanont2004};
(17) \citep{percy2001,kerschbaum1999};
(18) \citep{Perrin1998,Ohnaka2012,Winters2003};
(19) \citep{Zhang2017,Kudashkina2022,Brand2020};
(20) \citep{Dumm1998,McDonald2012,Tabur2009};
(21) \citep{Wyatt1983,Keenan1974,Cutri2003};
(22) \citep{vanLangevelde2000,Vlemmings2002};
(23) \citep{Homan2021}
(24) \citep{Ramstedt2014};
(25) \citep{Montarges2025};
(26) \citep{Kervella2015};
(27) \citep{Karovska_1997}.
}
\end{table*}

\subsection{Representativeness of the sample}

In order to identify possible discriminating parameters between our sample of 45 stars studied and the McDonald’s catalogue, we applied  two methods. We used first principal component analysis (PCA) to explore the data structure. Thus, we applied the Kolmogorov-Smirnov test (KS-test) to quantify the difference in the distribution of each parameter between the two groups. We conducted these two methods on a set of 11 stellar parameters: the distance (D), the luminosity (L), the effective temperature ($\rm T_{eff}$), the stellar radius (R$_\star$), the interstellar extinction (Av), the  IR excess (E\_IR), the fraction of reprocessed light in IR ($\mathrm{LIR/L_\star}$), the magnitude in the V, J, and K bands ($\rm M_V, \ M_J, \ M_K$), and the colour index ($\rm{J-K}$).  More details on these two methods and the definitions of the relevant parameters are given in Section~\ref{sec:comparison_method}. \\

\begin{figure}[!ht]
\centering
\resizebox{\hsize}{!}
{\includegraphics{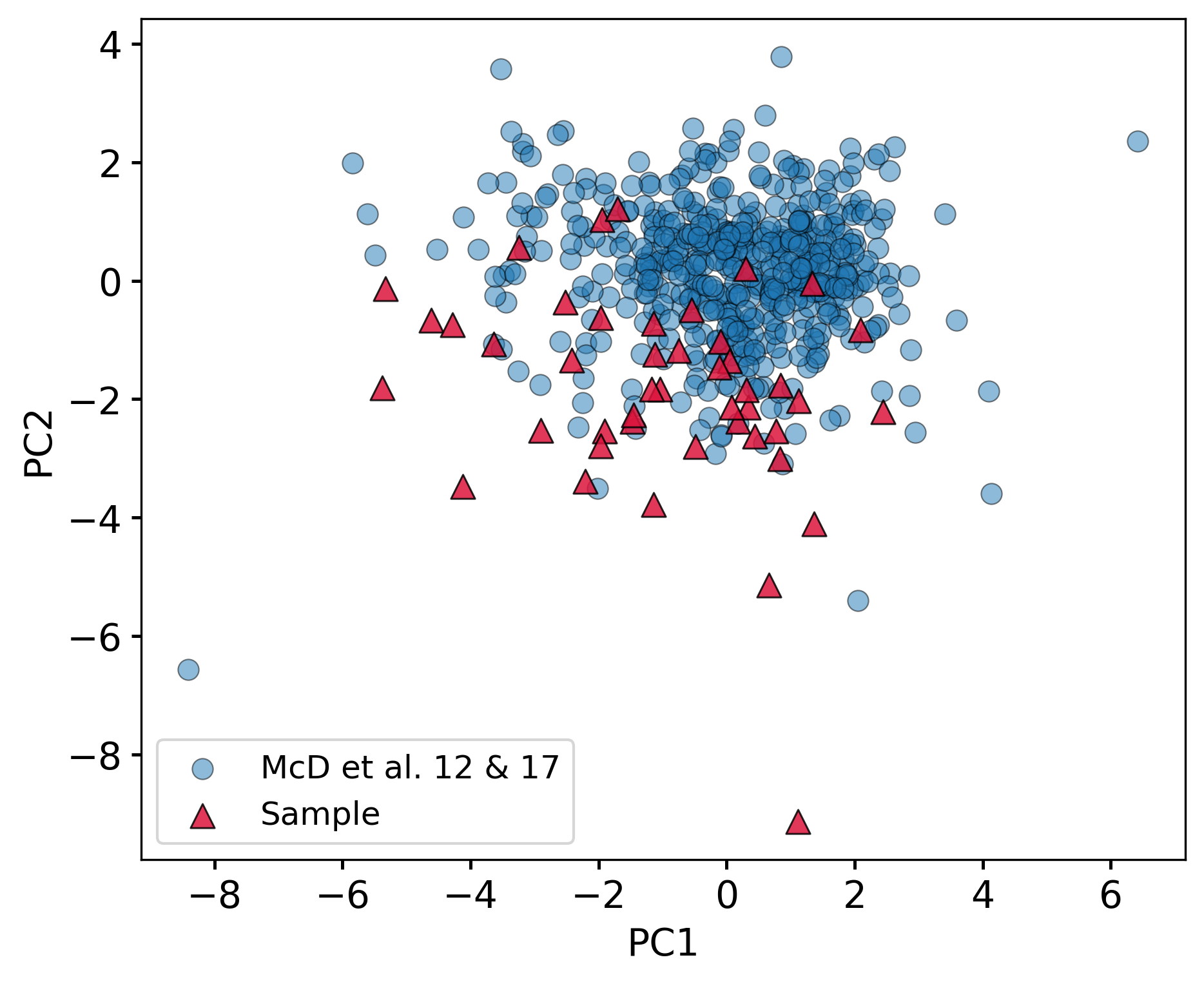}}
\caption{
Bi-dimensional projection of the PCA of the full sample \citep{McDonald2012,McDonald2017} in blue, and the selected subsample in red. The axes correspond to the first two principal components (PC1, PC2), which explain most of the variance in the data. This visualization highlights the distribution and separation of the selected stars in parameter space.}
\label{fig:PCA2D}
\end{figure}

The results of the PCA  analysis reveal the ability of these methods to distinguish our sample of 45 stars studied from the rest of the red giants. The 2D PCA (Figure~\ref{fig:PCA2D}) projection shows that our sample of 45 stars is not representative of the large sample of red giants. Most of the stars in the subsample appear offset to the same side relative to the majority of the stars in the large sample in the PCA projected space. This suggests the existence of discriminating physical parameters between the two samples. This result confirms the specificity of our sample in the physical parameter space.

\begin{figure*}[]
\resizebox{\hsize}{!}{\includegraphics{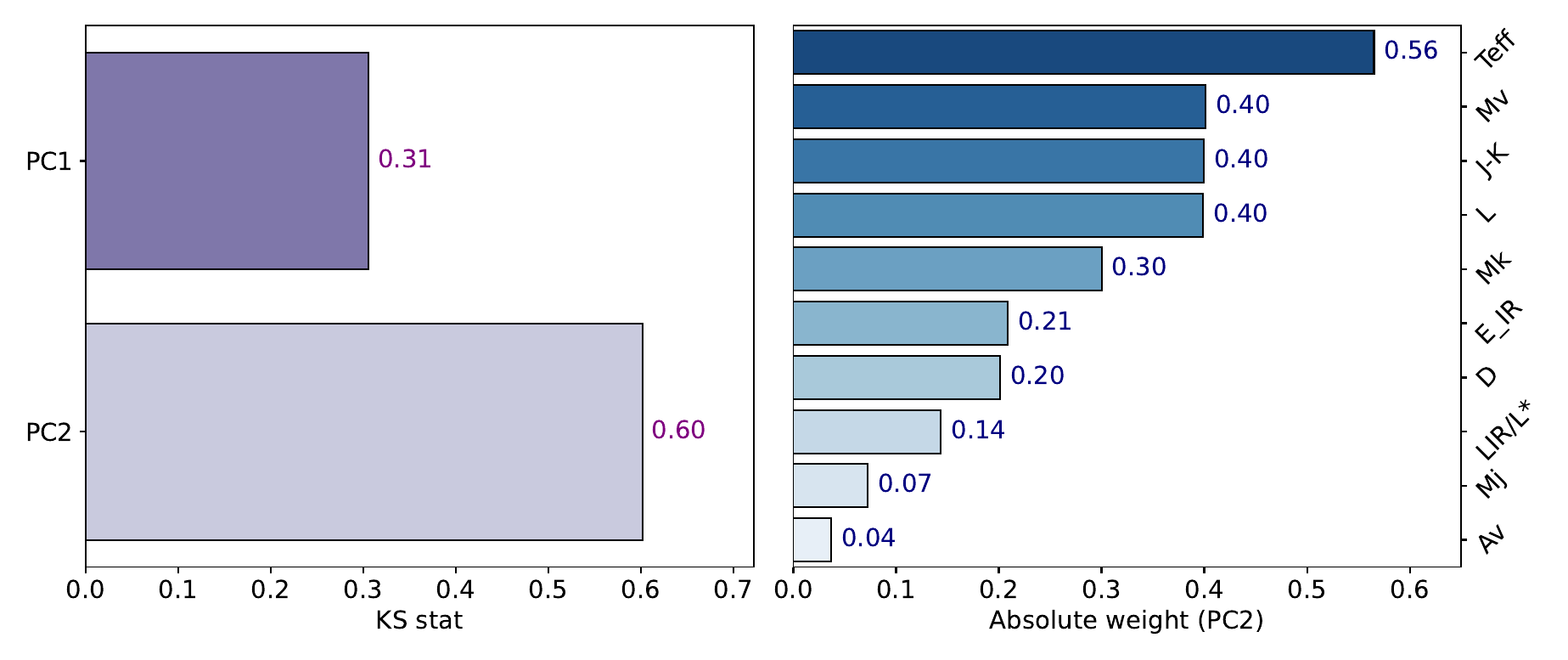}}
\caption{Variables most discriminative between the full sample and the selected subsample, identified via the KS test in the 2D PCA space. Left: KS test comparing PC1 and PC2; the KS statistics are 0.60 for PC2 and 0.31 for PC1, showing that variables loading heavily on PC2 drive the separation between the samples. Right: Absolute variable weights in PC2.}
\label{fig:KS_stat}
\end{figure*}

\subsection{Prediction of the resolved nature of star envelopes}
To propose some physical criteria for predicting the resolved nature of circumstellar envelopes by SPHERE in visible wavelengths, we applied a multi-step approach, combining machine learning and optimal threshold analysis on our sample of 45 stars. The main objective of this study is to identify, from the observed and/or calculated parameters, the most discriminating factors for resolved nature, and to propose simple and interpretable rules for the prediction of whether the dust envelope of a star will be resolved by SPHERE or not. The following methodology is structured into three steps: global modelling, identification of discriminating parameters, and determination of optimal thresholds.

Examination of the most discriminating principal components of the PCA, as well as analysis of the most important variables according to the KS test, indicate that luminosity, IR excess ($\mathrm{E_{IR}}$), effective temperature, and magnitude in the K band are the most important contributors to the differentiation of our sample (see Figure~\ref{fig:KS_stat}). These results confirm that the subsample selection allows us to isolate a subset of stars with distinct physical characteristics, thus justifying the adopted multivariate approach.  In addition, the visual magnitude $M_V$ and the near-infrared colour $J\!-\!K$ also appear as significant parameters separating the two samples.

These results show that, beyond the parameters that initially guided the selection of our 45 stars studied (luminosity, $M_V$, and $\mathrm{E_{IR}}$), other stellar properties such as $J\!-\!K$, $M_K$, and $T_{\mathrm{eff}}$ contribute significantly to distinguishing our sample of 45 AGB stars from the larger population of red giant stars. From a physical perspective, this reflects differences in both stellar temperature and circumstellar dust content, as redder colours and brighter near-IR magnitudes are indicative of cooler, more evolved stars surrounded by dust-rich environments.

\section{Discussion}\label{sec:4}

\subsection{Resolved envelope properties and classification criteria}

 All  45 nearby AGB stars observed in our sample display a clear departure from spherical symmetry, with the presence at large scale (more than 10 R$_\star$) of arcs, bipolar structures, or clumps. Since convection and pulsation can produce asymmetries in the immediate vicinity of single stars~\citep{Hofner2019,Freytag2023,Wiegert2024}, our observations alone do not always allow us to distinguish unambiguously between intrinsic stellar processes and shaping by a companion. In particular, although large-scale arcs, clumps, or bipolar structures are often interpreted as signatures of binarity, similar non-spherical morphologies can also arise from asymmetric mass ejection in single stars.

The high fraction of large-scale asymmetric structures among the resolved envelopes indicates that additional shaping mechanisms are at play beyond a purely spherical steady wind. Although such morphologies are often associated with the influence of a (sub-)stellar companion,  our data do not allow us to unambiguously establish binarity for all resolved systems. 
In several cases, the observed large-scale arcs, clumps, or bipolar-like features may therefore result from 
either binary interactions or intrinsically asymmetric mass ejection.  Moreover, certain envelopes present a pronounced clumpiness, confirming  that 
dust formation and mass loss processes are neither spatially nor temporally homogeneous.

Based on these morphological characteristics, we defined three observational classes: clearly resolved envelopes, marginally resolved envelopes, and unresolved objects. This classification allows us not only to distinguish between envelopes that are spatially resolved and those that remain unresolved with SPHERE but also to separate resolved envelopes according to the degree of morphological complexity. Of the 45 stars in our sample, 16 exhibit clearly resolved circumstellar envelopes, while 11 show marginally resolved structures, and 18 remain unresolved with SPHERE. For the purpose of predicting the resolved nature of circumstellar envelopes, the clearly resolved and marginally resolved objects were grouped into a single class of resolved envelopes. 

\begin{figure}[!ht]
\centering
\includegraphics[width=\columnwidth]{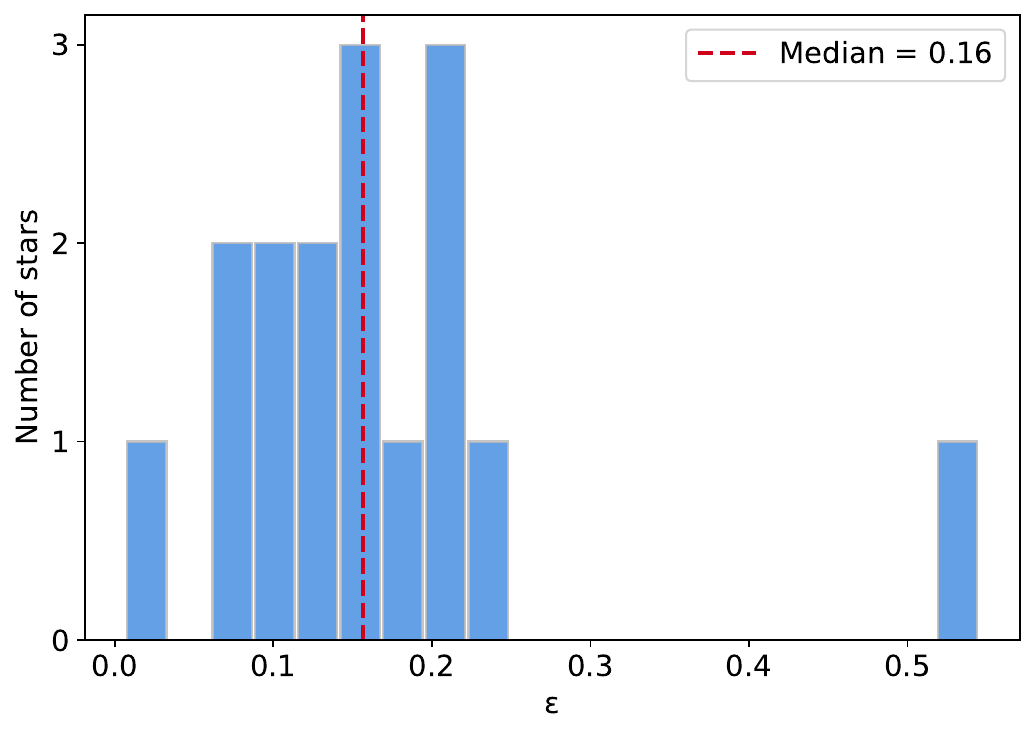}
\caption{Histogram of ellipticities $\varepsilon$ of the 16 clearly resolved circumstellar envelopes, calculated for each star from the filter with the largest apparent diameter. The dotted red line indicates the median of the distribution. The ellipticity values reflect the level of elongation or flattening of the observed envelopes.
}
\label{fig:histo_elip}
\end{figure}

Moreover, the analysis of circumstellar envelopes, illustrated by the ellipticity histogram (see Figure~\ref{fig:histo_elip}), reveals that all 16 clearly resolved envelopes analyzed exhibit a strictly non-zero ellipticity. This result indicates that none of these  envelopes is compatible with a perfectly spherical geometry but instead displays systematic departures from sphericity, even when these deviations remain moderate. From a physical point of view, high ellipticity can result from several mechanisms: the presence of a stellar companion favours the formation of elongated structures by gravitational interaction or mass transfer; binary systems can induce preferential matter flows and circumbinary disks; finally, magnetic fields or asymmetric pulsations can channel mass loss in certain directions. These factors contribute to the observed morphological diversity and testify to the complexity of mass loss processes in the last phases of stellar evolution.
This classification scheme also forms the basis for the random forest analysis presented in the following sections. 

\subsection{Impact of stellar physical properties on the morphology and extent of the dust shell}

Our random forest model, trained on the physical parameters of AGB stars, identifies favourable conditions for observing a resolved envelope (for stars located within a radius of 300\,pc) with the SPHERE instrument (see Table~\ref{tab:predict}).
 
Using the set of discriminating parameters, the optimal thresholds indicate that stellar luminosity plays a primary role in envelope detectability. The decision rule shows that the probability of resolving a circumstellar envelope increases when the luminosity exceeds the optimal threshold ($L$ > 4681\,$L_\odot$), a trend that is consistent with the higher median luminosity measured for resolved objects. Physically, more luminous AGB stars are expected to sustain stronger mass loss rates, leading to the development of more extended and brighter dust shells.

The apparent angular size, corresponding to the angular radius of the central star and quantified by $R_{\star}$, is another strong discriminant. The envelopes are more likely to be resolved when $R_{\star}$ exceeds the threshold value ($>9$\,mas). Larger values correspond to more evolved and intrinsically larger AGB stars, which are statistically more likely to host extended circumstellar dust shells detectable with SPHERE.

Visual extinction ($A_v$) appears as a secondary discriminant. According to the decision rule, the probability of resolution is higher for values below the threshold ($A_v < 0.02$), while the median values remain very similar for resolved and unresolved objects. This suggests that $A_v$ mainly traces line-of-sight effects rather than the global spatial extent of the circumstellar envelope and therefore plays a limited role in determining resolution.
Overall, these results indicate that the most favourable targets for resolved observations are luminous AGB stars surrounded by extended dust shells, whose large apparent sizes enhance their detectability with SPHERE.

\subsection{Observational criteria to select resolved targets}

Our analysis also shows that several observational parameters play a determining role in the detection and characterization of the circumstellar envelopes of AGB stars with SPHERE (see Table~\ref{tab:predict}). A significant fraction of stellar luminosity reprocessed in the IR (LIR/L$_\star$) is found to be a strong discriminant: the probability of resolution increases when LIR/L$_\star$ exceeds the optimal threshold ($> 0.007$). This reflects the presence of substantial amounts of circumstellar dust, which enhances both thermal emission and scattered-light signals from extended envelopes.

The IR excess $E_\mathrm{{IR}}$ also appears as a relevant discriminating parameter. According to the decision rule, the probability of resolving a circumstellar envelope increases for values above the threshold ($\rm{E_{IR} > 1.42}$), consistent with the slightly higher median value measured for resolved objects. A larger IR excess traces the presence of warm circumstellar dust and is therefore indicative of more developed dust envelopes that are more likely to be spatially resolved.

The contrast parameter ($\rm{C_{tr}}$), which quantifies the brightness contrast between the circumstellar envelope and the central star, plays an important role in envelope detectability. In this work, $\rm{C_{tr}}$ is directly taken as the optimal height $\hat{h}$, determined from the radial profile of the polarized intensity map ($\rm{PIL}$). This height corresponds to the relative polarized signal associated with the circumstellar envelope with respect to the peak polarized intensity in the image.  The probability of resolution increases when the contrast exceeds the optimal threshold ($\rm{C_{tr} > 0.04}$). Higher contrast values improve the visibility of circumstellar emission relative to the stellar PSF, facilitating the detection of extended structures in polarimetric imaging.

Finally, the axis ratio $\rm{b/a}$ provides insight into the morphology of the circumstellar envelope. In this case, the decision rule indicates a higher probability of resolution for values below the threshold ($\rm{b/a < 0.97}$). We emphasize that this threshold should be regarded as an empirical criterion derived from our classification procedure rather than as a strict physical boundary. This suggests that departures from spherical symmetry, such as elongated or flattened dust distributions, produce stronger spatial signatures and are therefore more readily detected with high angular resolution imaging. However, we note that asymmetries may also be present in envelopes that remain unresolved, and that part of the measured deviation from circular symmetry can be influenced by residual PSF anisotropies. Therefore, the axis ratio alone should not be interpreted as a definitive indicator of intrinsic envelope morphology.
Together, these observational criteria indicate that SPHERE preferentially resolves circumstellar envelopes that are extended, dust-rich, and often morphologically asymmetric, where both angular size and contrast play a decisive role.

\section{Conclusion}\label{sec:5}
In this study, we presented the largest catalogue (45) of AGB stars observed in polarized light at high angular resolution. Polarization maps enabled us to resolve 16 dusty circumstellar envelopes, including three for the first time. All of these envelopes clearly show a departure from spherical symmetry. Such asymmetries may arise either from intrinsically asymmetric mass loss processes in single stars or from shaping by (sub)stellar companions. 
While binarity provides a natural explanation for some of the most pronounced morphologies (e.g. large-scale arcs or bipolar structures), our data do not allow us to determine in all cases which mechanism dominates.
Moreover, certain envelopes present a pronounced clumpiness, confirming that dust formation and mass loss processes are neither spatially nor temporally homogeneous.
Finally, using a random forest model applied to our sample, we determine the most important stellar parameters to use to be able to predict which star should have a resolved envelope or not. This should be useful for future surveys that aim to characterize dust shells around AGB stars.

 \section*{Data availability}

(i) The supplementary polarimetric maps associated with this article, including the degree of linear polarization (DoLP) and polarized intensity (PIL) maps for all 45 AGB stars in the sample, are publicly available on Zenodo at \href{https://doi.org/10.5281/zenodo.20607881}{DOI: 10.5281/zenodo.20607881}.

\noindent (ii) The reduced SPHERE/ZIMPOL polarimetric images (DoLP and PIL maps) for the 45 AGB stars presented in this catalogue are available at the CDS via  \href{http://cdsweb.u-strasbg.fr/cgi-bin/qcat?J/A+A/}{http://cdsweb.u-strasbg.fr/cgi-bin/qcat?J/A+A/}.

\begin{acknowledgements}
This work has made use of the the SPHERE Data Centre, jointly operated by OSUG/IPAG (Grenoble), PYTHEAS/LAM/CESAM (Marseille), OCA/Lagrange (Nice), Observatoire de Paris/LESIA (Paris), and Observatoire de Lyon. 

We thank the referee for the very thorough work, which significantly helped to improve the paper.
\end{acknowledgements}

\bibliographystyle{aa}
\bibliography{catalog}

@ARTICLE{Hofner2018,
       author = {{H{\"o}fner}, Susanne and {Olofsson}, Hans},
        title = "{Mass loss of stars on the asymptotic giant branch. Mechanisms, models and measurements}",
      journal = {\aapr},
         year = 2018,
        month = jan,
       volume = {26},
       number = {1},
          eid = {1},
        pages = {1},
          doi = {10.1007/s00159-017-0106-5},
       adsurl = {https://ui.adsabs.harvard.edu/abs/2018A&ARv..26....1H}
}

@ARTICLE{Schmid2018,
       author = {{Schmid}, H.~M. and {Bazzon}, A. and {Roelfsema}, R. and {Mouillet}, D. and {Milli}, J. and {Menard}, F. and {Gisler}, D. and {Hunziker}, S. and {Pragt}, J. and {Dominik}, C. and {Boccaletti}, A. and {Ginski}, C. and {Abe}, L. and {Antoniucci}, S. and {Avenhaus}, H. and {Baruffolo}, A. and {Baudoz}, P. and {Beuzit}, J.~L. and {Carbillet}, M. and {Chauvin}, G. and {Claudi}, R. and {Costille}, A. and {Daban}, J. -B. and {de Haan}, M. and {Desidera}, S. and {Dohlen}, K. and {Downing}, M. and {Elswijk}, E. and {Engler}, N. and {Feldt}, M. and {Fusco}, T. and {Girard}, J.~H. and {Gratton}, R. and {Hanenburg}, H. and {Henning}, Th. and {Hubin}, N. and {Joos}, F. and {Kasper}, M. and {Keller}, C.~U. and {Langlois}, M. and {Lagadec}, E. and {Martinez}, P. and {Mulder}, E. and {Pavlov}, A. and {Podio}, L. and {Puget}, P. and {Quanz}, S.~P. and {Rigal}, F. and {Salasnich}, B. and {Sauvage}, J. -F. and {Schuil}, M. and {Siebenmorgen}, R. and {Sissa}, E. and {Snik}, F. and {Suarez}, M. and {Thalmann}, Ch. and {Turatto}, M. and {Udry}, S. and {van Duin}, A. and {van Holstein}, R.~G. and {Vigan}, A. and {Wildi}, F.},
        title = "{SPHERE/ZIMPOL high resolution polarimetric imager. I. System overview, PSF parameters, coronagraphy, and polarimetry}",
      journal = {\aap},
         year = 2018,
        month = nov,
       volume = {619},
          eid = {A9},
        pages = {A9},
          doi = {10.1051/0004-6361/201833620},
archivePrefix = {arXiv},
       eprint = {1808.05008},
 primaryClass = {astro-ph.IM},
       adsurl = {https://ui.adsabs.harvard.edu/abs/2018A&A...619A...9S}
}

@ARTICLE{McDonald2017,
       author = {{McDonald}, I. and {Zijlstra}, A.~A. and {Watson}, R.~A.},
        title = "{Fundamental parameters and infrared excesses of Tycho-Gaia stars}",
      journal = {\mnras},
         year = 2017,
        month = oct,
       volume = {471},
       number = {1},
        pages = {770-791},
          doi = {10.1093/mnras/stx1433},
archivePrefix = {arXiv},
       eprint = {1706.02208},
 primaryClass = {astro-ph.SR},
       adsurl = {https://ui.adsabs.harvard.edu/abs/2017MNRAS.471..770M}
}

@ARTICLE{McDonald2012,
       author = {{McDonald}, I. and {Zijlstra}, A.~A. and {Boyer}, M.~L.},
        title = "{Fundamental parameters and infrared excesses of Hipparcos stars}",
      journal = {\mnras},
         year = 2012,
        month = nov,
       volume = {427},
       number = {1},
        pages = {343-357},
          doi = {10.1111/j.1365-2966.2012.21873.x},
archivePrefix = {arXiv},
       eprint = {1208.2037},
 primaryClass = {astro-ph.SR},
       adsurl = {https://ui.adsabs.harvard.edu/abs/2012MNRAS.427..343M}
}

@ARTICLE{Olofsson2002,
       author = {{Olofsson}, H. and {Gonz{\'a}lez Delgado}, D. and {Kerschbaum}, F. and {Sch{\"o}ier}, F.~L.},
        title = "{Mass loss rates of a sample of irregular and semiregular M-type AGB-variables}",
      journal = {\aap},
         year = 2002,
        month = sep,
       volume = {391},
        pages = {1053-1067},
          doi = {10.1051/0004-6361:20020841},
archivePrefix = {arXiv},
       eprint = {astro-ph/0206172},
 primaryClass = {astro-ph},
       adsurl = {https://ui.adsabs.harvard.edu/abs/2002A&A...391.1053O}
}

@ARTICLE{Lebzelter2003,
       author = {{Lebzelter}, T. and {Hron}, J.},
        title = "{Technetium and the third dredge up in AGB stars. I. Field stars}",
      journal = {\aap},
         year = 2003,
        month = dec,
       volume = {411},
        pages = {533-542},
          doi = {10.1051/0004-6361:20031458},
archivePrefix = {arXiv},
       eprint = {astro-ph/0310018},
 primaryClass = {astro-ph},
       adsurl = {https://ui.adsabs.harvard.edu/abs/2003A&A...411..533L}
}

@ARTICLE{Heras2005,
       author = {{Heras}, A.~M. and {Hony}, S.},
        title = "{Oxygen-rich AGB stars with optically thin dust envelopes}",
      journal = {\aap},
         year = 2005,
        month = aug,
       volume = {439},
       number = {1},
        pages = {171-182},
          doi = {10.1051/0004-6361:20042296},
       adsurl = {https://ui.adsabs.harvard.edu/abs/2005A&A...439..171H}
}

@ARTICLE{Decin2020,
       author = {{Decin}, L. and {Montarg{\`e}s}, M. and {Richards}, A.~M.~S. and {Gottlieb}, C.~A. and {Homan}, W. and {McDonald}, I. and {El Mellah}, I. and {Danilovich}, T. and {Wallstr{\"o}m}, S.~H.~J. and {Zijlstra}, A. and {Baudry}, A. and {Bolte}, J. and {Cannon}, E. and {De Beck}, E. and {De Ceuster}, F. and {de Koter}, A. and {De Ridder}, J. and {Etoka}, S. and {Gobrecht}, D. and {Gray}, M. and {Herpin}, F. and {Jeste}, M. and {Lagadec}, E. and {Kervella}, P. and {Khouri}, T. and {Menten}, K. and {Millar}, T.~J. and {M{\"u}ller}, H.~S.~P. and {Plane}, J.~M.~C. and {Sahai}, R. and {Sana}, H. and {Van de Sande}, M. and {Waters}, L.~B.~F.~M. and {Wong}, K.~T. and {Yates}, J.},
        title = "{(Sub)stellar companions shape the winds of evolved stars}",
      journal = {Science},
         year = 2020,
        month = sep,
       volume = {369},
       number = {6510},
        pages = {1497-1500},
          doi = {10.1126/science.abb1229},
archivePrefix = {arXiv},
       eprint = {2009.11694},
 primaryClass = {astro-ph.SR},
       adsurl = {https://ui.adsabs.harvard.edu/abs/2020Sci...369.1497D}
}

@ARTICLE{Boer2020,
       author = {{de Boer}, J. and {Langlois}, M. and {van Holstein}, R.~G. and {Girard}, J.~H. and {Mouillet}, D. and {Vigan}, A. and {Dohlen}, K. and {Snik}, F. and {Keller}, C.~U. and {Ginski}, C. and {Stam}, D.~M. and {Milli}, J. and {Wahhaj}, Z. and {Kasper}, M. and {Schmid}, H.~M. and {Rabou}, P. and {Gluck}, L. and {Hugot}, E. and {Perret}, D. and {Martinez}, P. and {Weber}, L. and {Pragt}, J. and {Sauvage}, J. -F. and {Boccaletti}, A. and {Le Coroller}, H. and {Dominik}, C. and {Henning}, T. and {Lagadec}, E. and {M{\'e}nard}, F. and {Turatto}, M. and {Udry}, S. and {Chauvin}, G. and {Feldt}, M. and {Beuzit}, J. -L.},
        title = "{Polarimetric imaging mode of VLT/SPHERE/IRDIS. I. Description, data reduction, and observing strategy}",
      journal = {\aap},
         year = 2020,
        month = jan,
       volume = {633},
          eid = {A63},
        pages = {A63},
          doi = {10.1051/0004-6361/201834989},
archivePrefix = {arXiv},
       eprint = {1909.13107},
 primaryClass = {astro-ph.IM},
       adsurl = {https://ui.adsabs.harvard.edu/abs/2020A&A...633A..63D}
}

@ARTICLE{Miszalski2009,
       author = {{Miszalski}, B. and {Acker}, A. and {Parker}, Q.~A. and {Moffat}, A.~F.~J.},
        title = "{Binary planetary nebulae nuclei towards the Galactic bulge. II. A penchant for bipolarity and low-ionisation structures}",
      journal = {\aap},
         year = 2009,
        month = oct,
       volume = {505},
       number = {1},
        pages = {249-263},
          doi = {10.1051/0004-6361/200912176},
archivePrefix = {arXiv},
       eprint = {0907.2463},
 primaryClass = {astro-ph.SR},
       adsurl = {https://ui.adsabs.harvard.edu/abs/2009A&A...505..249M}
}

@ARTICLE{balick1987,
       author = {{Balick}, Bruce},
        title = "{The Evolution of Planetary Nebulae. I. Structures, Ionizations, and Morphological Sequences}",
      journal = {\aj},
         year = 1987,
        month = sep,
       volume = {94},
        pages = {671},
          doi = {10.1086/114504},
       adsurl = {https://ui.adsabs.harvard.edu/abs/1987AJ.....94..671B}
}

@ARTICLE{Bujarrabal2001,
       author = {{Bujarrabal}, V. and {Castro-Carrizo}, A. and {Alcolea}, J. and {S{\'a}nchez Contreras}, C.},
        title = "{Mass, linear momentum and kinetic energy of bipolar flows in protoplanetary nebulae}",
      journal = {\aap},
         year = 2001,
        month = oct,
       volume = {377},
        pages = {868-897},
          doi = {10.1051/0004-6361:20011090},
       adsurl = {https://ui.adsabs.harvard.edu/abs/2001A&A...377..868B}
}

@ARTICLE{Soker2006,
       author = {{Soker}, Noam},
        title = "{Why Magnetic Fields Cannot Be the Main Agent Shaping Planetary Nebulae}",
      journal = {\pasp},
         year = 2006,
        month = feb,
       volume = {118},
       number = {840},
        pages = {260-269},
          doi = {10.1086/498829},
archivePrefix = {arXiv},
       eprint = {astro-ph/0501647},
 primaryClass = {astro-ph},
       adsurl = {https://ui.adsabs.harvard.edu/abs/2006PASP..118..260S}
}

@ARTICLE{Nordhaus2006,
       author = {{Nordhaus}, J. and {Blackman}, E.~G.},
        title = "{Low-mass binary-induced outflows from asymptotic giant branch stars}",
      journal = {\mnras},
         year = 2006,
        month = aug,
       volume = {370},
       number = {4},
        pages = {2004-2012},
          doi = {10.1111/j.1365-2966.2006.10625.x},
archivePrefix = {arXiv},
       eprint = {astro-ph/0604445},
 primaryClass = {astro-ph},
       adsurl = {https://ui.adsabs.harvard.edu/abs/2006MNRAS.370.2004N}
}

@ARTICLE{Beuzit2019,
       author = {{Beuzit}, J. -L. and {Vigan}, A. and {Mouillet}, D. and {Dohlen}, K. and {Gratton}, R. and {Boccaletti}, A. and {Sauvage}, J. -F. and {Schmid}, H.~M. and {Langlois}, M. and {Petit}, C. and {Baruffolo}, A. and {Feldt}, M. and {Milli}, J. and {Wahhaj}, Z. and {Abe}, L. and {Anselmi}, U. and {Antichi}, J. and {Barette}, R. and {Baudrand}, J. and {Baudoz}, P. and {Bazzon}, A. and {Bernardi}, P. and {Blanchard}, P. and {Brast}, R. and {Bruno}, P. and {Buey}, T. and {Carbillet}, M. and {Carle}, M. and {Cascone}, E. and {Chapron}, F. and {Charton}, J. and {Chauvin}, G. and {Claudi}, R. and {Costille}, A. and {De Caprio}, V. and {de Boer}, J. and {Delboulb{\'e}}, A. and {Desidera}, S. and {Dominik}, C. and {Downing}, M. and {Dupuis}, O. and {Fabron}, C. and {Fantinel}, D. and {Farisato}, G. and {Feautrier}, P. and {Fedrigo}, E. and {Fusco}, T. and {Gigan}, P. and {Ginski}, C. and {Girard}, J. and {Giro}, E. and {Gisler}, D. and {Gluck}, L. and {Gry}, C. and {Henning}, T. and {Hubin}, N. and {Hugot}, E. and {Incorvaia}, S. and {Jaquet}, M. and {Kasper}, M. and {Lagadec}, E. and {Lagrange}, A. -M. and {Le Coroller}, H. and {Le Mignant}, D. and {Le Ruyet}, B. and {Lessio}, G. and {Lizon}, J. -L. and {Llored}, M. and {Lundin}, L. and {Madec}, F. and {Magnard}, Y. and {Marteaud}, M. and {Martinez}, P. and {Maurel}, D. and {M{\'e}nard}, F. and {Mesa}, D. and {M{\"o}ller-Nilsson}, O. and {Moulin}, T. and {Moutou}, C. and {Orign{\'e}}, A. and {Parisot}, J. and {Pavlov}, A. and {Perret}, D. and {Pragt}, J. and {Puget}, P. and {Rabou}, P. and {Ramos}, J. and {Reess}, J. -M. and {Rigal}, F. and {Rochat}, S. and {Roelfsema}, R. and {Rousset}, G. and {Roux}, A. and {Saisse}, M. and {Salasnich}, B. and {Santambrogio}, E. and {Scuderi}, S. and {Segransan}, D. and {Sevin}, A. and {Siebenmorgen}, R. and {Soenke}, C. and {Stadler}, E. and {Suarez}, M. and {Tiph{\`e}ne}, D. and {Turatto}, M. and {Udry}, S. and {Vakili}, F. and {Waters}, L.~B.~F.~M. and {Weber}, L. and {Wildi}, F. and {Zins}, G. and {Zurlo}, A.},
        title = "{SPHERE: the exoplanet imager for the Very Large Telescope}",
      journal = {\aap},
         year = 2019,
        month = nov,
       volume = {631},
          eid = {A155},
        pages = {A155},
          doi = {10.1051/0004-6361/201935251},
archivePrefix = {arXiv},
       eprint = {1902.04080},
 primaryClass = {astro-ph.IM},
       adsurl = {https://ui.adsabs.harvard.edu/abs/2019A&A...631A.155B}
}

@INPROCEEDINGS{Delorme2017,
       author = {{Delorme}, P. and {Meunier}, N. and {Albert}, D. and {Lagadec}, E. and {Le Coroller}, H. and {Galicher}, R. and {Mouillet}, D. and {Boccaletti}, A. and {Mesa}, D. and {Meunier}, J. -C. and {Beuzit}, J. -L. and {Lagrange}, A. -M. and {Chauvin}, G. and {Sapone}, A. and {Langlois}, M. and {Maire}, A. -L. and {Montarg{\`e}s}, M. and {Gratton}, R. and {Vigan}, A. and {Surace}, C.},
        title = "{The SPHERE Data Center: a reference for high contrast imaging processing}",
    booktitle = {SF2A-2017},
         year = 2017,
       editor = {{Reyl{\'e}}, C. and {Di Matteo}, P. and {Herpin}, F. and {Lagadec}, E. and {Lan{\c{c}}on}, A. and {Meliani}, Z. and {Royer}, F.},
        month = dec,
        pages = {Di},
          doi = {10.48550/arXiv.1712.06948},
archivePrefix = {arXiv},
       eprint = {1712.06948},
 primaryClass = {astro-ph.IM},
       adsurl = {https://ui.adsabs.harvard.edu/abs/2017sf2a.conf..347D}
}

@ARTICLE{Ramstedt2018,
       author = {{Ramstedt}, S. and {Mohamed}, S. and {Olander}, T. and {Vlemmings}, W.~H.~T. and {Khouri}, T. and {Liljegren}, S.},
        title = "{CO envelope of the symbiotic star R Aquarii seen by ALMA}",
      journal = {\aap},
         year = 2018,
        month = aug,
       volume = {616},
          eid = {A61},
        pages = {A61},
          doi = {10.1051/0004-6361/201833394},
archivePrefix = {arXiv},
       eprint = {1806.04979},
 primaryClass = {astro-ph.SR},
       adsurl = {https://ui.adsabs.harvard.edu/abs/2018A&A...616A..61R}
}

@ARTICLE{Brunner2018,
       author = {{Brunner}, M. and {Danilovich}, T. and {Ramstedt}, S. and {Marti-Vidal}, I. and {De Beck}, E. and {Vlemmings}, W.~H.~T. and {Lindqvist}, M. and {Kerschbaum}, F.},
        title = "{Molecular line study of the S-type AGB star W Aquilae. ALMA observations of CS, SiS, SiO and HCN}",
      journal = {\aap},
         year = 2018,
        month = sep,
       volume = {617},
          eid = {A23},
        pages = {A23},
          doi = {10.1051/0004-6361/201832724},
archivePrefix = {arXiv},
       eprint = {1806.01622},
 primaryClass = {astro-ph.SR},
       adsurl = {https://ui.adsabs.harvard.edu/abs/2018A&A...617A..23B}
}

@article{Ortiz2023,
    author = {Ortiz, R and Guerrero, M A},
    title = {A large bubble around the AGB star R Dor detected in the UV},
    journal = {\mnras},
    volume = {522},
    number = {1},
    pages = {811-818},
    year = {2023},
    month = {04},
    issn = {0035-8711},
    doi = {10.1093/mnras/stad984},
    url = {https://doi.org/10.1093/mnras/stad984},
    eprint = {https://academic.oup.com/mnras/article-pdf/522/1/811/49994417/stad984.pdf},
}

@ARTICLE{Ohnaka2019,
       author = {{Ohnaka}, K. and {Hadjara}, M. and {Maluenda Berna}, M.~Y.~L.},
        title = "{Spatially resolving the atmosphere of the non-Mira-type AGB star SW Vir in near-infrared molecular and atomic lines with VLTI/AMBER}",
      journal = {\aap},
         year = 2019,
        month = jan,
       volume = {621},
          eid = {A6},
        pages = {A6},
          doi = {10.1051/0004-6361/201834171},
archivePrefix = {arXiv},
       eprint = {1811.05989},
 primaryClass = {astro-ph.SR},
       adsurl = {https://ui.adsabs.harvard.edu/abs/2019A&A...621A...6O}
}

@ARTICLE{Brand2020,
       author = {{Brand}, J. and {Engels}, D. and {Winnberg}, A.},
        title = "{Water vapour masers in long-period variable stars. II. The semi-regular variables R Crt and RT Vir}",
      journal = {\aap},
         year = 2020,
        month = dec,
       volume = {644},
          eid = {A45},
        pages = {A45},
          doi = {10.1051/0004-6361/202039157},
archivePrefix = {arXiv},
       eprint = {2011.00294},
 primaryClass = {astro-ph.GA},
       adsurl = {https://ui.adsabs.harvard.edu/abs/2020A&A...644A..45B}
}

@ARTICLE{Alcolea1999,
       author = {{Alcolea}, J. and {Pardo}, J.~R. and {Bujarrabal}, V. and {Bachiller}, R. and {Barcia}, A. and {Colomer}, F. and {Gallego}, J.~D. and {G{\'o}mez-Gonz{\'a}lez}, J. and {del Pino Cisneros}, A. and {Planesas}, P. and {del R{\'\i}o}, S. and {Rodr{\'\i}guez-Franco}, A. and {del Romero}, A. and {Tafalla}, M. and {de Vicente}, P.},
        title = "{Six years of short-spaced monitoring of the v=1 and v=2, J=1-0 $^{28}$SiO maser emission in evolved stars}",
      journal = {\aaps},
         year = 1999,
        month = nov,
       volume = {139},
        pages = {461-482},
          doi = {10.1051/aas:1999112},
       adsurl = {https://ui.adsabs.harvard.edu/abs/1999A&AS..139..461A}
}

@ARTICLE{Pardo2004,
       author = {{Pardo}, J.~R. and {Alcolea}, J. and {Bujarrabal}, V. and {Colomer}, F. and {del Romero}, A. and {de Vicente}, P.},
        title = "{$^{28}$SiO v = 1 and v = 2, J = 1-0 maser variability in evolved stars. Eleven years of short spaced monitoring}",
      journal = {\aap},
         year = 2004,
        month = sep,
       volume = {424},
        pages = {145-156},
          doi = {10.1051/0004-6361:20040309},
       adsurl = {https://ui.adsabs.harvard.edu/abs/2004A&A...424..145P}
}

@ARTICLE{Kervella_2014,
       author = {{Kervella}, P. and {Montarg{\`e}s}, M. and {Ridgway}, S.~T. and {Perrin}, G. and {Chesneau}, O. and {Lacour}, S. and {Chiavassa}, A. and {Haubois}, X. and {Gallenne}, A.},
        title = "{An edge-on translucent dust disk around the nearest AGB star, L$_{2}$ Puppis. VLT/NACO spectro-imaging from 1.04 to 4.05 {\ensuremath{\mu}}m and VLTI interferometry}",
      journal = {\aap},
         year = 2014,
        month = apr,
       volume = {564},
          eid = {A88},
        pages = {A88},
          doi = {10.1051/0004-6361/201323273},
archivePrefix = {arXiv},
       eprint = {1404.3189},
 primaryClass = {astro-ph.SR},
       adsurl = {https://ui.adsabs.harvard.edu/abs/2014A&A...564A..88K}
}

@ARTICLE{Lykou_2015,
       author = {{Lykou}, F. and {Klotz}, D. and {Paladini}, C. and {Hron}, J. and {Zijlstra}, A.~A. and {Kluska}, J. and {Norris}, B.~R.~M. and {Tuthill}, P.~G. and {Ramstedt}, S. and {Lagadec}, E. and {Wittkowski}, M. and {Maercker}, M. and {Mayer}, A.},
        title = "{Dissecting the AGB star L$_{2}$ Puppis: a torus in the making}",
      journal = {\aap},
         year = 2015,
        month = apr,
       volume = {576},
          eid = {A46},
        pages = {A46},
          doi = {10.1051/0004-6361/201322828},
archivePrefix = {arXiv},
       eprint = {1503.05031},
 primaryClass = {astro-ph.SR},
       adsurl = {https://ui.adsabs.harvard.edu/abs/2015A&A...576A..46L}
}

@ARTICLE{Kervella_2015,
       author = {{Kervella}, P. and {Montarg{\`e}s}, M. and {Lagadec}, E. and {Ridgway}, S.~T. and {Haubois}, X. and {Girard}, J.~H. and {Ohnaka}, K. and {Perrin}, G. and {Gallenne}, A.},
        title = "{The dust disk and companion of the nearby AGB star L$_{2}$ Puppis. SPHERE/ZIMPOL polarimetric imaging at visible wavelengths}",
      journal = {\aap},
         year = 2015,
        month = jun,
       volume = {578},
          eid = {A77},
        pages = {A77},
          doi = {10.1051/0004-6361/201526194},
archivePrefix = {arXiv},
       eprint = {1511.04448},
 primaryClass = {astro-ph.SR},
       adsurl = {https://ui.adsabs.harvard.edu/abs/2015A&A...578A..77K}
}

@article{Van_de_Sande_2024,
    author = {Van de Sande, M and Walsh, C and Danilovich, T and De Ceuster, F and Ceulemans, T},
    title = {Modelling predicts a molecule-rich disc around the AGB star L2 Puppis},
    journal = {\mnras},
    volume = {532},
    number = {1},
    pages = {734-754},
    year = {2024},
    month = {06},
    issn = {0035-8711},
    doi = {10.1093/mnras/stae1553},
    url = {https://doi.org/10.1093/mnras/stae1553},
    eprint = {https://academic.oup.com/mnras/article-pdf/532/1/734/58366107/stae1553.pdf},
}

@ARTICLE{Aringer1999,
       author = {{Aringer}, B. and {H{\"o}fner}, S. and {Wiedemann}, G. and {Hron}, J. and {J{\o}rgensen}, U.~G. and {K{\"a}ufl}, H.~U. and {Windsteig}, W.},
        title = "{SiO rotation-vibration bands in cool giants II. The behaviour of SiO bands in AGB stars}",
      journal = {\aap},
         year = 1999,
        month = feb,
       volume = {342},
        pages = {799-808},
       adsurl = {https://ui.adsabs.harvard.edu/abs/1999A&A...342..799A}
}

@ARTICLE{Wallstrom2024,
       author = {{Wallstr{\"o}m}, S.~H.~J. and {Danilovich}, T. and {M{\"u}ller}, H.~S.~P. and {Gottlieb}, C.~A. and {Maes}, S. and {Van de Sande}, M. and {Decin}, L. and {Richards}, A.~M.~S. and {Baudry}, A. and {Bolte}, J. and {Ceulemans}, T. and {De Ceuster}, F. and {de Koter}, A. and {El Mellah}, I. and {Esseldeurs}, M. and {Etoka}, S. and {Gobrecht}, D. and {Gottlieb}, E. and {Gray}, M. and {Herpin}, F. and {Jeste}, M. and {Kee}, D. and {Kervella}, P. and {Khouri}, T. and {Lagadec}, E. and {Malfait}, J. and {Marinho}, L. and {McDonald}, I. and {Menten}, K.~M. and {Millar}, T.~J. and {Montarg{\`e}s}, M. and {Nuth}, J.~A. and {Plane}, J.~M.~C. and {Sahai}, R. and {Waters}, L.~B.~F.~M. and {Wong}, K.~T. and {Yates}, J. and {Zijlstra}, A.},
        title = "{ATOMIUM: Molecular inventory of 17 oxygen-rich evolved stars observed with ALMA}",
      journal = {\aap},
         year = 2024,
        month = jan,
       volume = {681},
          eid = {A50},
        pages = {A50},
          doi = {10.1051/0004-6361/202347632},
archivePrefix = {arXiv},
       eprint = {2312.03467},
 primaryClass = {astro-ph.SR},
       adsurl = {https://ui.adsabs.harvard.edu/abs/2024A&A...681A..50W}
}

@ARTICLE{Montarges2023,
       author = {{Montarg{\`e}s}, M. and {Cannon}, E. and {de Koter}, A. and {Khouri}, T. and {Lagadec}, E. and {Kervella}, P. and {Decin}, L. and {McDonald}, I. and {Homan}, W. and {Waters}, L.~B.~F.~M. and {Sahai}, R. and {Gottlieb}, C.~A. and {Malfait}, J. and {Maes}, S. and {Pimpanuwat}, B. and {Jeste}, M. and {Danilovich}, T. and {De Ceuster}, F. and {Van de Sande}, M. and {Gobrecht}, D. and {Wallstr{\"o}m}, S.~H.~J. and {Wong}, K.~T. and {El Mellah}, I. and {Bolte}, J. and {Herpin}, F. and {Richards}, A.~M.~S. and {Baudry}, A. and {Etoka}, S. and {Gray}, M.~D. and {Millar}, T.~J. and {Menten}, K.~M. and {M{\"u}ller}, H.~S.~P. and {Plane}, J.~M.~C. and {Yates}, J. and {Zijlstra}, A.},
        title = "{The VLT/SPHERE view of the ATOMIUM cool evolved star sample. I. Overview: Sample characterization through polarization analysis}",
      journal = {\aap},
         year = 2023,
        month = mar,
       volume = {671},
          eid = {A96},
        pages = {A96},
          doi = {10.1051/0004-6361/202245398},
archivePrefix = {arXiv},
       eprint = {2301.02081},
 primaryClass = {astro-ph.SR},
       adsurl = {https://ui.adsabs.harvard.edu/abs/2023A&A...671A..96M}
}

@ARTICLE{Karovska_1997,
       author = {{Karovska}, Margarita and {Hack}, Warren and {Raymond}, John and {Guinan}, Edward},
        title = "{First Hubble Space Telescope Observations of Mira AB Wind-accreting Binary Systems}",
      journal = {\apjl},
         year = 1997,
        month = jun,
       volume = {482},
       number = {2},
        pages = {L175-L178},
          doi = {10.1086/310704},
       adsurl = {https://ui.adsabs.harvard.edu/abs/1997ApJ...482L.175K}
}

@ARTICLE{Keenan1974,
       author = {{Keenan}, Philip C. and {Garrison}, Robert F. and {Deutsch}, Armin J.},
        title = "{Revised Catalog of Spectra of Mira Variables of Types ME and Se}",
      journal = {\apjs},
         year = 1974,
        month = nov,
       volume = {28},
        pages = {271},
          doi = {10.1086/190318},
       adsurl = {https://ui.adsabs.harvard.edu/abs/1974ApJS...28..271K}
}

@book{Hastie2009,
  author    = {Trevor Hastie and Robert Tibshirani and Jerome Friedman},
  title     = {The Elements of Statistical Learning: Data Mining, Inference, and Prediction},
  edition   = {2nd},
  publisher = {Springer},
  year      = {2009},
  isbn      = {978-0-387-84857-0},
  url       = {https://web.stanford.edu/~hastie/ElemStatLearn/}
}

@ARTICLE{Beguin2024,
       author = {{B{\'e}guin}, E. and {Chiavassa}, A. and {Ahmad}, A. and {Freytag}, B. and {Uttenthaler}, S.},
        title = "{Retrieving stellar parameters and dynamics of AGB stars with Gaia parallax measurements and CO$^{5}$BOLD RHD simulations}",
      journal = {\aap},
         year = 2024,
        month = oct,
       volume = {690},
          eid = {A125},
        pages = {A125},
          doi = {10.1051/0004-6361/202450245},
archivePrefix = {arXiv},
       eprint = {2409.03422},
 primaryClass = {astro-ph.SR},
       adsurl = {https://ui.adsabs.harvard.edu/abs/2024A&A...690A.125B}
}

@article{Khouri2024,
  author = {Khouri, T. and Olofsson, H. and Vlemmings, W. H. T. and Schirmer, T. and Tafoya, D. and Maercker, M. and De Beck, E. and Nyman, L.-A. and Saberi, M.},
  title = {An empirical view of the extended atmosphere and inner envelope of the asymptotic giant branch star R Doradus - I. Physical model based on CO lines},
  journal = {\aap},
  year = {2024},
  volume = {685},
  pages = {A11},
  doi = {10.1051/0004-6361/202348382},
  url = {https://doi.org/10.1051/0004-6361/202348382}
}

@article{Cotton2010,
  author = {Cotton, W. D. and Ragland, S. and Pluzhnik, E. A. and Danchi, W. C. and Traub, W. A. and Willson, L. A. and Lacasse, M. G.},
  title = {SiO MASERS IN ASYMMETRIC MIRAS. IV. χ CYGNI, R AQUILAE, R LEO MINORIS, RU HERCULIS, U HERCULIS, AND U ORIONIS},
  journal = {\apjs},
  year = {2010},
  volume = {188},
  number = {2},
  pages = {506},
  doi = {10.1088/0067-0049/188/2/506},
  url = {https://dx.doi.org/10.1088/0067-0049/188/2/506}
}

@article{Vlemmings2002,
  author = {Vlemmings, W. H. T. and van Langevelde, H. J. and Diamond, P. J.},
  title = {Astrometry of the stellar image of U Her amplified by the circumstellar 22 GHz water masers},
  journal = {\aap},
  year = {2002},
  volume = {393},
  number = {2},
  pages = {L33-L36},
  doi = {10.1051/0004-6361:20021169},
  url = {https://doi.org/10.1051/0004-6361:20021169}
}

@ARTICLE{vanLangevelde2000,
  author = {van Langevelde, H.~J. and Vlemmings, W. and Diamond, P.~J. and Baudry, A. and Beasley, A.~J.},
  title = {VLBI astrometry of the stellar image of U Herculis, amplified by the 1667 MHz OH maser},
  journal = {\aap},
  year = {2000},
  volume = {357},
  pages = {945-950},
  doi = {10.48550/arXiv.astro-ph/0003300},
  url = {https://ui.adsabs.harvard.edu/abs/2000A&A...357..945V}
}

@ARTICLE{Kervella2015,
       author = {{Kervella}, P. and {Montarg{\`e}s}, M. and {Lagadec}, E. and {Ridgway}, S.~T. and {Haubois}, X. and {Girard}, J.~H. and {Ohnaka}, K. and {Perrin}, G. and {Gallenne}, A.},
        title = "{The dust disk and companion of the nearby AGB star L$_{2}$ Puppis. SPHERE/ZIMPOL polarimetric imaging at visible wavelengths}",
      journal = {\aap},
         year = 2015,
        month = jun,
       volume = {578},
          eid = {A77},
        pages = {A77},
          doi = {10.1051/0004-6361/201526194},
archivePrefix = {arXiv},
       eprint = {1511.04448},
 primaryClass = {astro-ph.SR},
       adsurl = {https://ui.adsabs.harvard.edu/abs/2015A&A...578A..77K}
}

@ARTICLE{McDonald2024,
       author = {{McDonald}, Iain and {Zijlstra}, Albert A. and {Cox}, Nick L.~J. and {Alexander}, Emma L. and {Csukai}, Alexander and {Ramkumar}, Ria and {Hollings}, Alexander},
        title = "{PYSSED: an automated method of collating and fitting stellar spectral energy distributions}",
      journal = {RASTI},
         year = 2024,
        month = jan,
       volume = {3},
       number = {1},
        pages = {89-107},
          doi = {10.1093/rasti/rzae005},
archivePrefix = {arXiv},
       eprint = {2402.12496},
 primaryClass = {astro-ph.IM},
       adsurl = {https://ui.adsabs.harvard.edu/abs/2024RASTI...3...89M}
}

@ARTICLE{Gottlieb2022,
       author = {{Gottlieb}, C.~A. and {Decin}, L. and {Richards}, A.~M.~S. and {De Ceuster}, F. and {Homan}, W. and {Wallstr{\"o}m}, S.~H.~J. and {Danilovich}, T. and {Millar}, T.~J. and {Montarg{\`e}s}, M. and {Wong}, K.~T. and {McDonald}, I. and {Baudry}, A. and {Bolte}, J. and {Cannon}, E. and {De Beck}, E. and {de Koter}, A. and {El Mellah}, I. and {Etoka}, S. and {Gobrecht}, D. and {Gray}, M. and {Herpin}, F. and {Jeste}, M. and {Kervella}, P. and {Khouri}, T. and {Lagadec}, E. and {Maes}, S. and {Malfait}, J. and {Menten}, K.~M. and {M{\"u}ller}, H.~S.~P. and {Pimpanuwat}, B. and {Plane}, J.~M.~C. and {Sahai}, R. and {Van de Sande}, M. and {Waters}, L.~B.~F.~M. and {Yates}, J. and {Zijlstra}, A.},
        title = "{ATOMIUM: ALMA tracing the origins of molecules in dust forming oxygen rich M-type stars. Motivation, sample, calibration, and initial results}",
      journal = {\aap},
         year = 2022,
        month = apr,
       volume = {660},
          eid = {A94},
        pages = {A94},
          doi = {10.1051/0004-6361/202140431},
archivePrefix = {arXiv},
       eprint = {2112.04399},
 primaryClass = {astro-ph.SR},
       adsurl = {https://ui.adsabs.harvard.edu/abs/2022A&A...660A..94G}
}

@ARTICLE{Saberi_2018,
       author = {{Saberi}, M. and {Vlemmings}, W.~H.~T. and {De Beck}, E. and {Montez}, R. and {Ramstedt}, S.},
        title = "{Detection of CI line emission towards the oxygen-rich AGB star omi Ceti}",
      journal = {\aap},
         year = 2018,
        month = apr,
       volume = {612},
          eid = {L11},
        pages = {L11},
          doi = {10.1051/0004-6361/201833080},
archivePrefix = {arXiv},
       eprint = {1804.09958},
 primaryClass = {astro-ph.SR},
       adsurl = {https://ui.adsabs.harvard.edu/abs/2018A&A...612L..11S}
}

@ARTICLE{Vlemmings2003,
       author = {{Vlemmings}, W.~H.~T. and {van Langevelde}, H.~J. and {Diamond}, P.~J. and {Habing}, H.~J. and {Schilizzi}, R.~T.},
        title = "{VLBI astrometry of circumstellar OH masers: Proper motions and parallaxes of four AGB stars}",
      journal = {\aap},
         year = 2003,
        month = aug,
       volume = {407},
        pages = {213-224},
          doi = {10.1051/0004-6361:20030766},
archivePrefix = {arXiv},
       eprint = {astro-ph/0305405},
 primaryClass = {astro-ph},
       adsurl = {https://ui.adsabs.harvard.edu/abs/2003A&A...407..213V}
}

@article{Nhung2021,
  author = {Nhung, P. T. and Hoai, D. T. and Tuan-Anh, P. and Darriulat, P. and Diep, P. N. and Ngoc, N. B. and Phuong, N. T. and Thai, T. T.},
  title = {Morpho-kinematics of the circumstellar envelope of the AGB star R Dor: a global view},
  journal = {\mnras},
  year = {2021},
  volume = {504},
  number = {2},
  pages = {2687-2706},
  month = {05},
  doi = {10.1093/mnras/stab954},
  url = {https://doi.org/10.1093/mnras/stab954}
}

@article{Zijlstra2002,
  author = {Zijlstra, A. A. and Bedding, T. R. and Mattei, J. A.},
  title = {The evolution of the Mira variable R Hydrae},
  journal = {\mnras},
  volume = {334},
  number = {3},
  pages = {498-510},
  year = {2002},
  doi = {10.1046/j.1365-8711.2002.05467.x}
}

@article{Haniff1995,
  author = {Haniff, C. A. and Scholz, M. and Tuthill, P. G.},
  title = {Angular diameter measurements of Mira variables at 902 nm},
  journal = {\mnras},
  volume = {276},
  pages = {640-654},
  year = {1995}
}

@article{Richichi2005,
  author = {Richichi, A. and Percheron, I. and Khristoforova, M.},
  title = {CHARM2: An updated Catalog of High Angular Resolution Measurements},
  journal = {\aap},
  volume = {431},
  pages = {773-779},
  year = {2005}
}

@article{Feast1996,
  author = {Feast, M. W.},
  title = {Mira variables in the Large Magellanic Cloud},
  journal = {\mnras},
  volume = {278},
  pages = {11-26},
  year = {1996}
}

@article{Ueta2006,
  author = {Ueta, T. and Speck, A. K. and Stencel, R. E. and Herwig, F. and Gehrz, R. D. and Szczerba, R. and Izumiura, H. and Zijlstra, A. A. and Latter, W. B. and Matsuura, M. and Meixner, M. and Steffen, M. and Elitzur, M.},
  title = {Detection of a Far-Infrared Bow Shock Nebula around R Hya: The First MIRIAD Results},
  journal = {\apj},
  volume = {648},
  pages = {L39-L42},
  year = {2006},
  doi = {10.1086/507017}
}

@article{Begemann1997,
  author = {Begemann, B. and Dorschner, J. and Henning, Th. and Mutschke, H. and Guertler, J. and Koempe, C. and Nass, R.},
  title = {Aluminum oxide and the opacity of oxygen-rich circumstellar dust in the 12–17 micron range},
  journal = {\apj},
  volume = {476},
  pages = {199-208},
  year = {1997},
  doi = {10.1086/303609}
}

@ARTICLE{Danilovich2014,
       author = {{Danilovich}, T. and {Bergman}, P. and {Justtanont}, K. and {Lombaert}, R. and {Maercker}, M. and {Olofsson}, H. and {Ramstedt}, S. and {Royer}, P.},
        title = "{Detailed modelling of the circumstellar molecular line emission of the S-type AGB star W Aquilae}",
      journal = {\aap},
         year = 2014,
        month = sep,
       volume = {569},
          eid = {A76},
        pages = {A76},
          doi = {10.1051/0004-6361/201322807},
archivePrefix = {arXiv},
       eprint = {1408.1825},
 primaryClass = {astro-ph.SR},
       adsurl = {https://ui.adsabs.harvard.edu/abs/2014A&A...569A..76D}
}

@ARTICLE{Ramstedt2017,
       author = {{Ramstedt}, S. and {Mohamed}, S. and {Vlemmings}, W.~H.~T. and {Danilovich}, T. and {Brunner}, M. and {De Beck}, E. and {Humphreys}, E.~M.~L. and {Lindqvist}, M. and {Maercker}, M. and {Olofsson}, H. and {Kerschbaum}, F. and {Quintana-Lacaci}, G.},
        title = "{The circumstellar envelope around the S-type AGB star W Aql. Effects of an eccentric binary orbit}",
      journal = {\aap},
         year = 2017,
        month = sep,
       volume = {605},
          eid = {A126},
        pages = {A126},
          doi = {10.1051/0004-6361/201730934},
archivePrefix = {arXiv},
       eprint = {1709.07327},
 primaryClass = {astro-ph.SR},
       adsurl = {https://ui.adsabs.harvard.edu/abs/2017A&A...605A.126R}
}

@article{Mayer2014,
  author = {A. Mayer and A. Jorissen and C. Paladini and F. Kerschbaum and D. Pourbaix and C. Siopis and R. Ottensamer and M. Mečina and N. L. J. Cox and M. A. T. Groenewegen and D. Klotz and G. Sadowski and A. Spang and P. Cruzalèbes and C. Waelkens},
  title = {Large-scale environments of binary AGB stars probed by Herschel II. Two companions interacting with the wind of π1 Gruis},
  journal = {\aap},
  volume = {570},
  pages = {A113},
  year = {2014},
  doi = {10.1051/0004-6361/201424465}
}

@article{Winters2003,
  author = {J. M. Winters and T. Le Bertre and K. S. Jeong and L.-A. Nyman and N. Epchtein},
  title = {CO observations of S stars: mass-loss rates and gas expansion velocities},
  journal = {\aap},
  volume = {409},
  pages = {715},
  year = {2003},
  doi = {10.1051/0004-6361:20031146}
}

@article{Guandalini2008,
  author = {R. Guandalini and M. Busso},
  title = {Infrared properties of Galactic S stars},
  journal = {\aap},
  volume = {488},
  pages = {675},
  year = {2008},
  doi = {10.1051/0004-6361:200809736}
}

@article{Groenewegen1998,
  author = {M. A. T. Groenewegen and T. de Jong},
  title = {Dust and gas mass loss rates of AGB stars},
  journal = {\aap},
  volume = {337},
  pages = {797},
  year = {1998}
}

@article{VanEck1998,
  author = {S. Van Eck and A. Jorissen and S. Udry and M. Mayor and B. Pernier},
  title = {The Hertzsprung-Russell diagram of S stars from Hipparcos data},
  journal = {\aap},
  volume = {329},
  pages = {971},
  year = {1998}
}

@article{Doan2017,
  author = {L. Doan and S. Ramstedt and W. H. T. Vlemmings and S. Höfner and E. De Beck and F. Kerschbaum and M. Lindqvist and M. Maercker and S. Mohamed and C. Paladini and M. Wittkowski},
  title = {The extended molecular envelope of the asymptotic giant branch star π1 Gruis as seen by ALMA I. Large-scale kinematic structure and CO excitation properties},
  journal = {\aap},
  volume = {605},
  pages = {A28},
  year = {2017},
  doi = {10.1051/0004-6361/201730703}
}

@article{Doan2020,
  author = {L. Doan and S. Ramstedt and W. H. T. Vlemmings and S. Mohamed and S. Höfner and E. De Beck and F. Kerschbaum and M. Lindqvist and M. Maercker and C. Paladini and M. Wittkowski},
  title = {The extended molecular envelope of the asymptotic giant branch star π1 Gruis as seen by ALMA II. The spiral-outflow observed at high-angular resolution},
  journal = {\aap},
  volume = {633},
  pages = {A13},
  year = {2020},
  doi = {10.1051/0004-6361/201935245}
}

@article{Homan2020,
  author = {Ward Homan and Miguel Montargès and Bannawit Pimpanuwat and Anita M. S. Richards and Sofia H. J. Wallström and Pierre Kervella and Leen Decin and Albert Zijlstra and Taissa Danilovich and Alex de Koter and Karl Menten and Raghvendra Sahai and John Plane and Kelvin Lee and Rens Waters and Alain Baudry and Ka Tat Wong and Tom J. Millar and Marie Van de Sande and Eric Lagadec and David Gobrecht and Jeremy Yates and Daniel Price and Emily Cannon and Jan Bolte and Frederik De Ceuster and Fabrice Herpin and Joe Nuth and Jan Philip Sindel and Dylan Kee and Malcolm D. Grey and Sandra Etoka and Manali Jeste and Carl A. Gottlieb and Elaine Gottlieb and Iain McDonald and Ileyk El Mellah and Holger S. P. Müller},
  title = {ATOMIUM: A high-resolution view on the highly asymmetric wind of the AGB star π1Gruis I. First detection of a new companion and its effect on the inner wind},
  journal = {\aap},
  volume = {644},
  pages = {A61},
  year = {2020},
  doi = {10.1051/0004-6361/202039185}
}

@article{Dumm1998,
  author = {Dumm, T. and Schild, H.},
  title = {Fundamental parameters of M giants in the solar neighbourhood},
  journal = {NewA},
  volume = {3},
  number = {2},
  pages = {137--161},
  year = {1998},
  doi = {10.1016/S1384-1076(98)00016-2}
}

@article{Tabur2009,
  author = {Tabur, V. and Kiss, L. L. and Bedding, T. R. and Moon, T. T. and Szeidl, B. and Kjeldsen, H.},
  title = {Long-term photometry and periods for 261 nearby pulsating M giants},
  journal = {\mnras},
  volume = {400},
  number = {4},
  pages = {1945--1961},
  year = {2009},
  doi = {10.1111/j.1365-2966.2009.15570.x}
}

@article{Adelman2000,
  author = {Adelman, S. J.},
  title = {Variability of M Giants in the Hipparcos Catalogue},
  journal = {IBVS},
  number = {4870},
  year = {2000},
  url = {http://www.konkoly.hu/cgi-bin/IBVS?4870}
}

@ARTICLE{Khouri2020,
       author = {{Khouri}, T. and {Vlemmings}, W.~H.~T. and {Paladini}, C. and {Ginski}, C. and {Lagadec}, E. and {Maercker}, M. and {Kervella}, P. and {De Beck}, E. and {Decin}, L. and {de Koter}, A. and {Waters}, L.~B.~F.~M.},
        title = "{Inner dusty envelope of the AGB stars W Hydrae, SW Virginis, and R Crateris using SPHERE/ZIMPOL}",
      journal = {\aap},
         year = 2020,
        month = mar,
       volume = {635},
          eid = {A200},
        pages = {A200},
          doi = {10.1051/0004-6361/201834618},
archivePrefix = {arXiv},
       eprint = {2003.06195},
 primaryClass = {astro-ph.SR},
       adsurl = {https://ui.adsabs.harvard.edu/abs/2020A&A...635A.200K}
}

@ARTICLE{Ridgway1982,
       author = {{Ridgway}, S.~T. and {Jacoby}, G.~H. and {Joyce}, R.~R. and {Siegel}, M.~J. and {Wells}, D.~C.},
        title = "{Angular diameters by the lunar occultation technique. IV - Alpha Leo and the Cepheid Zeta GEM}",
      journal = {\aj},
         year = 1982,
        month = apr,
       volume = {87},
        pages = {680-684},
          doi = {10.1086/113144},
       adsurl = {https://ui.adsabs.harvard.edu/abs/1982AJ.....87..680R}
}

@article{Schmidtke1986,
  author = {Schmidtke, P. C. and Africano, J. L. and Jacoby, G. H. and Joyce, R. R. and Ridgway, S. T.},
  title = {Angular diameters of stars from lunar occultations. II},
  journal = {\aj},
  volume = {91},
  pages = {961--967},
  year = {1986},
  doi = {10.1086/114073}
}

@article{Jura1992,
  author = {Jura, M. and Kleinmann, S. G.},
  title = {Infrared observations of mass loss from oxygen-rich stars},
  journal = {\aaps},
  volume = {83},
  pages = {329--352},
  year = {1992},
  doi = {10.1086/191741}
}

@ARTICLE{Szymczak1999,
       author = {{Szymczak}, M. and {Cohen}, R.~J. and {Richards}, A.~M.~S.},
        title = "{MERLIN polarimetry of the OH maser emission from R Crateris}",
      journal = {\mnras},
         year = 1999,
        month = apr,
       volume = {304},
       number = {4},
        pages = {877-882},
          doi = {10.1046/j.1365-8711.1999.02365.x},
       adsurl = {https://ui.adsabs.harvard.edu/abs/1999MNRAS.304..877S}
}

@article{Kholopov1987,
author = {Kholopov, P. and Samus, N. and Kazarovets, E. and Kireeva, N.},
year = {1987},
month = {07},
pages = {1},
title = {The 68th Name-List of Variable Stars},
volume = {3058},
journal = {IBVS}
}

@article{Jewell1991,
  title={Observational properties of V= 1, J= 1-0 SiO masers},
  author={Jewell, PR and Snyder, LE and Walmsley, CM and Wilson, TL and Gensheimer, PD},
  journal= {\aap},
  volume={242},
  pages={211--234},
  year={1991}
}

@article{LeSqueren1979,
  title={New OH sources in CRL objects and late type stars-On the correlation of OH velocity pattern and stellar period},
  author={Le Squeren, AM and Baudry, A and Brillet, J and Darchy, B},
  journal={\aap},
  volume={72},
  pages={39--44},
  year={1979}
}

@article{Kahane1994,
  title={Circumstellar CO around bright oxygen-rich semi-regulars.},
  author={Kahane, C and Jura, M},
  journal={\aap},
  volume={290},
  pages={183--197},
  year={1994}
}

@article{kerschbaum1999,
  title={Oxygen-rich semiregular and irregular variables-A catalogue of circumstellar CO observations},
  author={Kerschbaum, Franz and Olofsson, Hans},
  journal={\aap},
  volume={138},
  number={2},
  pages={299--322},
  year={1999},
  publisher={EDP Sciences}
}

@article{percy2001,
  author = {Percy, John R. and Dunlop, Heather and Kassim, Lola and Thompson, Raymond R.},
  title = {Periods of 25 Pulsating Red Giants},
  journal = {IBVS},
  year = {2001},
  volume = {5041},
  pages = {1-3},
  publisher = {Konkoly Observatory}
}

@ARTICLE{VanDeSande2018,
       author = {{Van de Sande}, M. and {Decin}, L. and {Lombaert}, R. and {Khouri}, T. and {de Koter}, A. and {Wyrowski}, F. and {De Nutte}, R. and {Homan}, W.},
        title = "{Chemical content of the circumstellar envelope of the oxygen-rich AGB star R Doradus. Non-LTE abundance analysis of CO, SiO, and HCN}",
      journal = {\aap},
         year = 2018,
        month = jan,
       volume = {609},
          eid = {A63},
        pages = {A63},
          doi = {10.1051/0004-6361/201731298},
archivePrefix = {arXiv},
       eprint = {1708.09190},
 primaryClass = {astro-ph.SR},
       adsurl = {https://ui.adsabs.harvard.edu/abs/2018A&A...609A..63V}
}

@ARTICLE{Ohnaka2012,
       author = {{Ohnaka}, K. and {Hofmann}, K. -H. and {Schertl}, D. and {Weigelt}, G. and {Malbet}, F. and {Massi}, F. and {Meilland}, A. and {Stee}, Ph.},
        title = "{Spatially resolving the outer atmosphere of the M giant BK Virginis in the CO first overtone lines with VLTI/AMBER}",
      journal = {\aap},
         year = 2012,
        month = jan,
       volume = {537},
          eid = {A53},
        pages = {A53},
          doi = {10.1051/0004-6361/201118128},
       adsurl = {https://ui.adsabs.harvard.edu/abs/2012A&A...537A..53O}
}

@ARTICLE{vanLeeuwen2007,
       author = {{van Leeuwen}, F.},
        title = "{Validation of the new Hipparcos reduction}",
      journal = {\aap},
         year = 2007,
        month = nov,
       volume = {474},
       number = {2},
        pages = {653-664},
          doi = {10.1051/0004-6361:20078357},
       adsurl = {https://ui.adsabs.harvard.edu/abs/2007A&A...474..653V}
}

@ARTICLE{Perrin1998,
       author = {{Perrin}, G. and {Coud{\'e} du Foresto}, V. and {Ridgway}, S.~T. and {Mariotti}, J. -M. and {Traub}, W.~A. and {Carleton}, N.~P. and {Lacasse}, M.~G.},
        title = "{Extension of the effective temperature scale of giants to types later than M6}",
      journal = {\aap},
         year = 1998,
        month = mar,
       volume = {331},
        pages = {619-626},
       adsurl = {https://ui.adsabs.harvard.edu/abs/1998A&A...331..619P}
}

@ARTICLE{Kudashkina2022,
       author = {{Kudashkina}, L.~S. and {Andronov}, I.~L.},
        title = "{Updated Catalog of Variable Stars in the Field of the Globular Cluster M5}",
      journal = {Odessa Astron. Publ.},
         year = 2022,
        month = jan,
       volume = {35},
        pages = {142-149},
          doi = {10.18524/1810-4215.2022.35.268364},
       adsurl = {https://ui.adsabs.harvard.edu/abs/2022OAP....35..142K}
}

@ARTICLE{Zhang2017,
       author = {{Zhang}, Bo and {Zheng}, Xingwu and {Reid}, Mark J. and {Honma}, Mareki and {Menten}, Karl M. and {Brunthaler}, Andreas and {Kim}, Jaeheon},
        title = "{VLBA Trigonometric Parallax Measurement of the Semi-regular Variable RT Vir}",
      journal = {\apj},
         year = 2017,
        month = nov,
       volume = {849},
       number = {2},
          eid = {99},
        pages = {99},
          doi = {10.3847/1538-4357/aa8ee9},
       adsurl = {https://ui.adsabs.harvard.edu/abs/2017ApJ...849...99Z}
}

@article{Templeton2009,
doi = {10.1088/0004-637X/691/2/1470},
url = {https://dx.doi.org/10.1088/0004-637X/691/2/1470},
year = {2009},
month = {feb},
publisher = {The American Astronomical Society},
volume = {691},
number = {2},
pages = {1470},
author = {Templeton, Matthew R. and Karovska, Margarita},
title = {LONG-PERIOD VARIABILITY IN o CETI},
journal = {\apj}
}

@article{Prieur2002,
doi = {10.1086/338029},
url = {https://dx.doi.org/10.1086/338029},
year = {2002},
month = {mar},
publisher = {},
volume = {139},
number = {1},
pages = {249},
author = {Prieur, J. L. and Aristidi, E. and Lopez, B. and Scardia, M. and Mignard, F. and Carbillet, M.},
title = {High Angular Resolution Observations of Late-Type Stars},
journal = {\apjs}
}

@ARTICLE{Planesas2016,
       author = {{Planesas}, P. and {Alcolea}, J. and {Bachiller}, R.},
        title = "{The radio continuum spectrum of Mira A and Mira B up to submillimeter wavelengths}",
      journal = {\aap},
         year = 2016,
        month = feb,
       volume = {586},
          eid = {A69},
        pages = {A69},
          doi = {10.1051/0004-6361/201527833},
archivePrefix = {arXiv},
       eprint = {1512.08638},
 primaryClass = {astro-ph.SR},
       adsurl = {https://ui.adsabs.harvard.edu/abs/2016A&A...586A..69P}
}

@ARTICLE{Khouri2018,
       author = {{Khouri}, T. and {Vlemmings}, W.~H.~T. and {Olofsson}, H. and {Ginski}, C. and {De Beck}, E. and {Maercker}, M. and {Ramstedt}, S.},
        title = "{High-resolution observations of gas and dust around Mira using ALMA and SPHERE/ZIMPOL}",
      journal = {\aap},
         year = 2018,
        month = nov,
       volume = {620},
          eid = {A75},
        pages = {A75},
          doi = {10.1051/0004-6361/201833643},
archivePrefix = {arXiv},
       eprint = {1810.03886},
 primaryClass = {astro-ph.SR},
       adsurl = {https://ui.adsabs.harvard.edu/abs/2018A&A...620A..75K}
}

@article{Ryde2001,
doi = {10.1086/318341},
url = {https://dx.doi.org/10.1086/318341},
year = {2001},
month = {jan},
publisher = {},
volume = {547},
number = {1},
pages = {384},
author = {Ryde, N. and Schöier, F. L.},
title = {Modeling CO Emission from Mira’s Wind},
journal = {\apj}
}

@ARTICLE{Hoai2020,
       author = {{Hoai}, D.~T. and {Tuan-Anh}, P. and {Nhung}, P.~T. and {Darriulat}, P. and {Diep}, P.~N. and {Phuong}, N.~T. and {Thai}, T.~T.},
        title = "{Revealing new features of the millimetre emission of the circumbinary envelope of Mira Ceti}",
      journal = {\mnras},
         year = 2020,
        month = jun,
       volume = {495},
       number = {1},
        pages = {943-961},
          doi = {10.1093/mnras/staa1173},
archivePrefix = {arXiv},
       eprint = {2004.11518},
 primaryClass = {astro-ph.SR},
       adsurl = {https://ui.adsabs.harvard.edu/abs/2020MNRAS.495..943H}
}

@INPROCEEDINGS{Wong2016,
       author = {{Wong}, Ka Tat and {Kami{\'n}ski}, Tomasz and {Menten}, Karl M. and {Wyrowski}, Friedrich},
        title = "{Resolving The Extended Atmosphere and The Inner Wind of Mira (o Cet) With Long ALMA Baselines}",
    booktitle = {19th Cambridge Workshop on Cool Stars, Stellar Systems, and the Sun (CS19)},
         year = 2016,
        month = oct,
          eid = {38},
        pages = {38},
          doi = {10.5281/zenodo.163626},
       adsurl = {https://ui.adsabs.harvard.edu/abs/2016csss.confE..38W}
}

@ARTICLE{Ramstedt2014,
       author = {{Ramstedt}, S. and {Mohamed}, S. and {Vlemmings}, W.~H.~T. and {Maercker}, M. and {Montez}, R. and {Baudry}, A. and {De Beck}, E. and {Lindqvist}, M. and {Olofsson}, H. and {Humphreys}, E.~M.~L. and {Jorissen}, A. and {Kerschbaum}, F. and {Mayer}, A. and {Wittkowski}, M. and {Cox}, N.~L.~J. and {Lagadec}, E. and {Leal-Ferreira}, M.~L. and {Paladini}, C. and {P{\'e}rez-S{\'a}nchez}, A. and {Sacuto}, S.},
        title = "{The wonderful complexity of the Mira AB system}",
      journal = {\aap},
         year = 2014,
        month = oct,
       volume = {570},
          eid = {L14},
        pages = {L14},
          doi = {10.1051/0004-6361/201425029},
archivePrefix = {arXiv},
       eprint = {1410.1529},
 primaryClass = {astro-ph.SR},
       adsurl = {https://ui.adsabs.harvard.edu/abs/2014A&A...570L..14R}
}

@article{Nhung2016,
    author = {Nhung, P. T. and Hoai, D. T. and Diep, P. N. and Phuong, N. T. and Thao, N. T. and Tuan-Anh, P. and Darriulat, P.},
    title = {Morphology and kinematics of the gas envelope of Mira Ceti},
    journal = {\mnras},
    volume = {460},
    number = {1},
    pages = {673-688},
    year = {2016},
    month = {04},
    issn = {0035-8711},
    doi = {10.1093/mnras/stw996},
    url = {https://doi.org/10.1093/mnras/stw996},
    eprint = {https://academic.oup.com/mnras/article-pdf/460/1/673/8116377/stw996.pdf},
}

@ARTICLE{Etoka2003,
       author = {{Etoka}, S. and {Le Squeren}, A.~M. and {Gerard}, E.},
        title = "{Detection of 1612 MHz OH emission in the semiregular variable stars RT Vir, R Crt and W Hya}",
      journal = {\aap},
         year = 2003,
        month = may,
       volume = {403},
        pages = {L51-L54},
          doi = {10.1051/0004-6361:20030578},
       adsurl = {https://ui.adsabs.harvard.edu/abs/2003A&A...403L..51E}
}

@ARTICLE{Justtanont2004,
       author = {{Justtanont}, K. and {de Jong}, T. and {Tielens}, A.~G.~G.~M. and {Feuchtgruber}, H. and {Waters}, L.~B.~F.~M.},
        title = "{W Hya: Molecular inventory by ISO-SWS}",
      journal = {\aap},
         year = 2004,
        month = apr,
       volume = {417},
        pages = {625-635},
          doi = {10.1051/0004-6361:20031772},
archivePrefix = {arXiv},
       eprint = {astro-ph/0402068},
 primaryClass = {astro-ph},
       adsurl = {https://ui.adsabs.harvard.edu/abs/2004A&A...417..625J}
}

@article{Zhao-Geisler2015,
doi = {10.1086/682261},
url = {https://dx.doi.org/10.1086/682261},
year = {2015},
month = {aug},
publisher = {University of Chicago Press},
volume = {127},
number = {954},
pages = {732},
author = {Zhao-Geisler, R. and Köhler, R. and Kemper, F. and Kerschbaum, F. and Mayer, A. and Quirrenbach, A. and Lopez, B.},
title = {Spectro-Imaging of the Asymmetric Inner Molecular and Dust Shell Region of the Mira Variable W Hya with MIDI/VLTI1},
journal = {PASP}
}

@ARTICLE{Millan2005,
       author = {{Millan-Gabet}, R. and {Pedretti}, E. and {Monnier}, J.~D. and {Schloerb}, F.~P. and {Traub}, W.~A. and {Carleton}, N.~P. and {Lacasse}, M.~G. and {Segransan}, D.},
        title = "{Diameters of Mira Stars Measured Simultaneously in the J, H, and K' Near-Infrared Bands}",
      journal = {\apj},
         year = 2005,
        month = feb,
       volume = {620},
       number = {2},
        pages = {961-969},
          doi = {10.1086/427163},
archivePrefix = {arXiv},
       eprint = {astro-ph/0411073},
 primaryClass = {astro-ph},
       adsurl = {https://ui.adsabs.harvard.edu/abs/2005ApJ...620..961M}
}

@ARTICLE{Wyatt1983,
       author = {{Wyatt}, S.~P. and {Cahn}, J.~H.},
        title = "{Kinematics and ages of Mira variables in the greater solar neighborhood.}",
      journal = {\apj},
         year = 1983,
        month = dec,
       volume = {275},
        pages = {225-239},
          doi = {10.1086/161527},
       adsurl = {https://ui.adsabs.harvard.edu/abs/1983ApJ...275..225W}
}

@BOOK{Cutri2003,
       author = {{Cutri}, R.~M. and {Skrutskie}, M.~F. and {van Dyk}, S. and {Beichman}, C.~A. and {Carpenter}, J.~M. and {Chester}, T. and {Cambresy}, L. and {Evans}, T. and {Fowler}, J. and {Gizis}, J. and {Howard}, E. and {Huchra}, J. and {Jarrett}, T. and {Kopan}, E.~L. and {Kirkpatrick}, J.~D. and {Light}, R.~M. and {Marsh}, K.~A. and {McCallon}, H. and {Schneider}, S. and {Stiening}, R. and {Sykes}, M. and {Weinberg}, M. and {Wheaton}, W.~A. and {Wheelock}, S. and {Zacarias}, N.},
        title = "{2MASS All Sky Catalog of point sources.}",
         year = 2003,
       adsurl = {https://ui.adsabs.harvard.edu/abs/2003tmc..book.....C}
}

@ARTICLE{Wenger2000,
       author = {{Wenger}, M. and {Ochsenbein}, F. and {Egret}, D. and {Dubois}, P. and {Bonnarel}, F. and {Borde}, S. and {Genova}, F. and {Jasniewicz}, G. and {Lalo{\"e}}, S. and {Lesteven}, S. and {Monier}, R.},
        title = "{The SIMBAD astronomical database. The CDS reference database for astronomical objects}",
      journal = {\aaps},
         year = 2000,
        month = apr,
       volume = {143},
        pages = {9-22},
          doi = {10.1051/aas:2000332},
archivePrefix = {arXiv},
       eprint = {astro-ph/0002110},
 primaryClass = {astro-ph},
       adsurl = {https://ui.adsabs.harvard.edu/abs/2000A&AS..143....9W}
}

@PROCEEDINGS{bt_settl2011,
        title = "{16$^{th}$ Cambridge Workshop on Cool Stars, Stellar Systems, and the Sun}",
    booktitle = {16th Cambridge Workshop on Cool Stars, Stellar Systems, and the Sun},
         year = 2011,
       editor = {{Johns-Krull}, Christopher and {Browning}, Matthew K. and {West}, Andrew A.},
       series = {ASP Conf. Ser.},
       volume = {448},
        month = dec,
       adsurl = {https://ui.adsabs.harvard.edu/abs/2011ASPC..448.....J}
}

@INPROCEEDINGS{Ho1995,
  author={Tin Kam Ho},
  booktitle={Proceedings of 3rd International Conference on Document Analysis and Recognition}, 
  title={Random decision forests}, 
  year={1995},
  volume={1},
  number={},
  pages={278-282},
  doi={10.1109/ICDAR.1995.598994}
}

@ARTICLE{Khouri2016,
       author = {{Khouri}, T. and {Maercker}, M. and {Waters}, L.~B.~F.~M. and {Vlemmings}, W.~H.~T. and {Kervella}, P. and {de Koter}, A. and {Ginski}, C. and {De Beck}, E. and {Decin}, L. and {Min}, M. and {Dominik}, C. and {O'Gorman}, E. and {Schmid}, H. -M. and {Lombaert}, R. and {Lagadec}, E.},
        title = "{Study of the inner dust envelope and stellar photosphere of the AGB star R Doradus using SPHERE/ZIMPOL}",
      journal = {\aap},
         year = 2016,
        month = jun,
       volume = {591},
          eid = {A70},
        pages = {A70},
          doi = {10.1051/0004-6361/201628435},
archivePrefix = {arXiv},
       eprint = {1605.05504},
 primaryClass = {astro-ph.SR},
       adsurl = {https://ui.adsabs.harvard.edu/abs/2016A&A...591A..70K}
}

@ARTICLE{McDonald&Trabucchi2019,
       author = {{McDonald}, I. and {Trabucchi}, M.},
        title = "{The onset of the AGB wind tied to a transition between sequences in the period-luminosity diagram}",
      journal = {\mnras},
         year = 2019,
        month = apr,
       volume = {484},
       number = {4},
        pages = {4678-4682},
          doi = {10.1093/mnras/stz324},
archivePrefix = {arXiv},
       eprint = {1901.06325},
 primaryClass = {astro-ph.SR},
       adsurl = {https://ui.adsabs.harvard.edu/abs/2019MNRAS.484.4678M}
}

@ARTICLE{Bressan2012,
       author = {{Bressan}, Alessandro and {Marigo}, Paola and {Girardi}, L{\'e}o. and {Salasnich}, Bernardo and {Dal Cero}, Claudia and {Rubele}, Stefano and {Nanni}, Ambra},
        title = "{PARSEC: stellar tracks and isochrones with the PAdova and TRieste Stellar Evolution Code}",
      journal = {\mnras},
         year = 2012,
        month = nov,
       volume = {427},
       number = {1},
        pages = {127-145},
          doi = {10.1111/j.1365-2966.2012.21948.x},
archivePrefix = {arXiv},
       eprint = {1208.4498},
 primaryClass = {astro-ph.SR},
       adsurl = {https://ui.adsabs.harvard.edu/abs/2012MNRAS.427..127B}
}

@ARTICLE{Beck2010,
       author = {{De Beck}, E. and {Decin}, L. and {de Koter}, A. and {Justtanont}, K. and {Verhoelst}, T. and {Kemper}, F. and {Menten}, K.~M.},
        title = "{Probing the mass-loss history of AGB and red supergiant stars from CO rotational line profiles. II. CO line survey of evolved stars: derivation of mass-loss rate formulae}",
      journal = {\aap},
         year = 2010,
        month = nov,
       volume = {523},
          eid = {A18},
        pages = {A18},
          doi = {10.1051/0004-6361/200913771},
archivePrefix = {arXiv},
       eprint = {1008.1083},
 primaryClass = {astro-ph.SR},
       adsurl = {https://ui.adsabs.harvard.edu/abs/2010A&A...523A..18D}
}

@ARTICLE{Wiegert2024,
       author = {{Wiegert}, Joachim and {Freytag}, Bernd and {H{\"o}fner}, Susanne},
        title = "{Asymmetries in asymptotic giant branch stars and their winds: I. From 3D RHD models to synthetic observables}",
      journal = {\aap},
         year = 2024,
        month = oct,
       volume = {690},
          eid = {A162},
        pages = {A162},
          doi = {10.1051/0004-6361/202450077},
archivePrefix = {arXiv},
       eprint = {2407.17317},
 primaryClass = {astro-ph.SR},
       adsurl = {https://ui.adsabs.harvard.edu/abs/2024A&A...690A.162W}
}

@ARTICLE{Freytag2023,
       author = {{Freytag}, Bernd and {H{\"o}fner}, Susanne},
        title = "{Global 3D radiation-hydrodynamical models of AGB stars with dust-driven winds}",
      journal = {\aap},
         year = 2023,
        month = jan,
       volume = {669},
          eid = {A155},
        pages = {A155},
          doi = {10.1051/0004-6361/202244992},
archivePrefix = {arXiv},
       eprint = {2301.11836},
 primaryClass = {astro-ph.SR},
       adsurl = {https://ui.adsabs.harvard.edu/abs/2023A&A...669A.155F}
}

@ARTICLE{Hofner2019,
       author = {{H{\"o}fner}, Susanne and {Freytag}, Bernd},
        title = "{Exploring the origin of clumpy dust clouds around cool giants. A global 3D RHD model of a dust-forming M-type AGB star}",
      journal = {\aap},
         year = 2019,
        month = mar,
       volume = {623},
          eid = {A158},
        pages = {A158},
          doi = {10.1051/0004-6361/201834799},
archivePrefix = {arXiv},
       eprint = {1902.04074},
 primaryClass = {astro-ph.SR},
       adsurl = {https://ui.adsabs.harvard.edu/abs/2019A&A...623A.158H}
}

@ARTICLE{Breiman2001,
       author = {{Breiman}, Leo},
        title = "{Random Forests.}",
      journal = {Mach. Learn.},
         year = 2001,
        month = jan,
       volume = {45},
        pages = {5-32},
          doi = {10.1023/A:1010933404324},
       adsurl = {https://ui.adsabs.harvard.edu/abs/2001MachL..45....5B}
}

@article{Dice1945,
  author  = {Dice, L.~R.},
  title   = {Measures of the Amount of Ecologic Association Between Species},
  journal = {Ecology},
  volume  = {26},
  number  = {3},
  pages   = {297--302},
  year    = {1945},
  doi     = {10.2307/1932409}
}

@ARTICLE{Montarges2025,
       author = {{Montarg{\`e}s}, M. and {Malfait}, J. and {Esseldeurs}, M. and {de Koter}, A. and {Baron}, F. and {Kervella}, P. and {Danilovich}, T. and {Richards}, A.~M.~S. and {Sahai}, R. and {McDonald}, I. and {Khouri}, T. and {Shetye}, S. and {Zijlstra}, A. and {Van de Sande}, M. and {El Mellah}, I. and {Herpin}, F. and {Siess}, L. and {Etoka}, S. and {Gobrecht}, D. and {Marinho}, L. and {Wallstr{\"o}m}, S.~H.~J. and {Wong}, K.~T. and {Yates}, J.},
        title = "{An accreting dwarf star orbiting the S-type giant star {\ensuremath{\pi}}$^{1}$ Gru}",
      journal = {\aap},
         year = 2025,
        month = jul,
       volume = {699},
          eid = {A22},
        pages = {A22},
          doi = {10.1051/0004-6361/202452587},
archivePrefix = {arXiv},
       eprint = {2504.16845},
 primaryClass = {astro-ph.SR},
       adsurl = {https://ui.adsabs.harvard.edu/abs/2025A&A...699A..22M}
}

@ARTICLE{Homan2021,
       author = {{Homan}, Ward and {Pimpanuwat}, Bannawit and {Herpin}, Fabrice and {Danilovich}, Taissa and {McDonald}, Iain and {Wallstr{\"o}m}, Sofia H.~J. and {Richards}, Anita M.~S. and {Baudry}, Alain and {Sahai}, Raghvendra and {Millar}, Tom J. and {de Koter}, Alex and {Gottlieb}, C.~A. and {Kervella}, Pierre and {Montarg{\`e}s}, Miguel and {Van de Sande}, Marie and {Decin}, Leen and {Zijlstra}, Albert and {Etoka}, Sandra and {Jeste}, Manali and {M{\"u}ller}, Holger S.~P. and {Maes}, Silke and {Malfait}, Jolien and {Menten}, Karl and {Plane}, John and {Lee}, Kelvin and {Waters}, Rens and {Wong}, Ka Tat and {Lagadec}, Eric and {Gobrecht}, David and {Yates}, Jeremy and {Price}, Daniel and {Cannon}, Emily and {Bolte}, Jan and {De Ceuster}, Frederik and {Nuth}, Joe and {Philip Sindel}, Jan and {Kee}, Dylan and {Gray}, Malcolm D. and {El Mellah}, Ileyk},
        title = "{ATOMIUM: The astounding complexity of the near circumstellar environment of the M-type AGB star R Hydrae. I. Morpho-kinematical interpretation of CO and SiO emission}",
      journal = {\aap},
         year = 2021,
        month = jul,
       volume = {651},
          eid = {A82},
        pages = {A82},
          doi = {10.1051/0004-6361/202140512},
archivePrefix = {arXiv},
       eprint = {2104.07297},
 primaryClass = {astro-ph.SR},
       adsurl = {https://ui.adsabs.harvard.edu/abs/2021A&A...651A..82H}
}

@ARTICLE{Schirmer2025,
       author = {{Schirmer}, Thiebaut and {Khouri}, Theo and {Vlemmings}, Wouter and {Nyman}, Gunnar and {Maercker}, Matthias and {Unnikrishnan}, Ramlal and {Bojnordi Arbab}, Behzad and {Knudsen}, Kirsten K. and {Aalto}, Susanne},
        title = "{An empirical view of the extended atmosphere and inner envelope of the asymptotic giant branch star R Doradus: II. Constraining the dust properties with radiative transfer modelling}",
      journal = {\aap},
         year = 2025,
        month = nov,
       volume = {704},
          eid = {A4},
        pages = {A4},
          doi = {10.1051/0004-6361/202556884},
archivePrefix = {arXiv},
       eprint = {2511.20816},
 primaryClass = {astro-ph.SR},
       adsurl = {https://ui.adsabs.harvard.edu/abs/2025A&A...704A...4S}
}

@ARTICLE{Ohnaka2016,
       author = {{Ohnaka}, K. and {Weigelt}, G. and {Hofmann}, K.-H.},
        title = "{Clumpy dust clouds and extended atmosphere of the AGB star W Hydrae revealed with VLT/SPHERE-ZIMPOL and VLTI/AMBER}",
      journal = {\aap},
         year = 2016,
        month = may,
       volume = {589},
          eid = {A91},
        pages = {A91},
          doi = {10.1051/0004-6361/201628229},
archivePrefix = {arXiv},
       eprint = {1603.01197},
 primaryClass = {astro-ph.SR},
       adsurl = {https://ui.adsabs.harvard.edu/abs/2016A&A...589A..91O}
}

@ARTICLE{Ohnaka2017,
       author = {{Ohnaka}, K. and {Weigelt}, G. and {Hofmann}, K.-H.},
        title = "{Clumpy dust clouds and extended atmosphere of the AGB star W Hydrae revealed with VLT/SPHERE-ZIMPOL and VLTI/AMBER. II. Time variations between pre-maximum and minimum light}",
      journal = {\aap},
         year = 2017,
        month = jan,
       volume = {597},
          eid = {A20},
        pages = {A20},
          doi = {10.1051/0004-6361/201629761},
archivePrefix = {arXiv},
       eprint = {1611.04622},
 primaryClass = {astro-ph.SR},
       adsurl = {https://ui.adsabs.harvard.edu/abs/2017A&A...597A..20O}
}

@ARTICLE{Ohnaka2024,
       author = {{Ohnaka}, K. and {Wong}, K.~T. and {Weigelt}, G. and {Hofmann}, K.-H.},
        title = "{Contemporaneous high-angular-resolution imaging of the AGB star W Hya in vibrationally excited H$_{2}$O lines and visible polarized light with ALMA and VLT/SPHERE-ZIMPOL}",
      journal = {\aap},
         year = 2024,
        month = nov,
       volume = {691},
          eid = {L14},
        pages = {L14},
          doi = {10.1051/0004-6361/202451977},
archivePrefix = {arXiv},
       eprint = {2411.09759},
 primaryClass = {astro-ph.SR},
       adsurl = {https://ui.adsabs.harvard.edu/abs/2024A&A...691L..14O}
}

@ARTICLE{Esseldeurs2026,
       author = {{Esseldeurs}, Mats and {Decin}, Leen and {De Ridder}, Joris and {Mori}, Yoshiya and {Karakas}, Amanda I. and {Malfait}, Jolien and {Danilovich}, Ta{\'\i}ssa and {Mathis}, St{\'e}phane and {Richards}, Anita M.~S. and {Sahai}, Raghvendra and {Yates}, Jeremy and {Van de Sande}, Marie and {Baes}, Maarten and {Baudry}, Alain and {Bolte}, Jan and {Ceulemans}, Thomas and {De Ceuster}, Frederik and {El Mellah}, Ileyk and {Etoka}, Sandra and {Gottlieb}, Carl and {Herpin}, Fabrice and {Kervella}, Pierre and {Landri}, Camille and {Marinho}, Louise and {McDonald}, Iain and {Menten}, Karl and {Millar}, Tom and {Osborn}, Zara and {Pimpanuwat}, Bannawit and {Plane}, John and {Price}, Daniel J. and {Siess}, Lionel and {Vermeulen}, Owen and {Wong}, Ka Tat},
        title = "{Evidence for the Keplerian orbit of a close companion around a giant star}",
      journal = {Nat. Astron.},
         year = 2026,
        month = jan,
       volume = {10},
        pages = {124-143},
          doi = {10.1038/s41550-025-02697-2},
archivePrefix = {arXiv},
       eprint = {2511.11247},
 primaryClass = {astro-ph.SR},
       adsurl = {https://ui.adsabs.harvard.edu/abs/2026NatAs..10..124E}
}

@ARTICLE{Drevon2026,
       author = {{Drevon}, J. and {Paladini}, C. and {H{\"o}fner}, S. and {Planquart}, L. and {Siess}, L. and {Jorissen}, A. and {Montarg{\'e}s}, M. and {Vlemmings}, W. and {Khouri}, T. and {Olofsson}, H. and {Alonso-Hernandez}, J. and {De Beck}, E. and {Fonfria}, J.~P. and {Hron}, J. and {Matter}, A. and {Nardetto}, N. and {Ohnaka}, K. and {Sanchez-Contreras}, C. and {Weigelt}, G. and {Wittkowski}, M. and {Bojnordi Arbab}, B. and {Aringer}, B. and {Baron}, F. and {Chiavassa}, A. and {Cruzal{\'e}bes}, P. and {Danchi}, W.~C. and {Kerschbaum}, F. and {Leftley}, J. and {Lagadec}, E. and {Lopez}, B. and {Lykou}, F. and {Millour}, F. and {Rau}, G. and {Sanchez-Bermudez}, J. and {Th{\'e}venin}, F. and {Van Eck}, S. and {Velilla-Prieto}, L.},
        title = "{Direct imaging of mass transfer and circumcompanion structures in {\ensuremath{\pi}}$^{1}$ Gru with VLTI/MATISSE}",
      journal = {\aap},
         year = 2026,
        month = jan,
       volume = {706},
          eid = {L1},
        pages = {L1},
          doi = {10.1051/0004-6361/202558298},
       adsurl = {https://ui.adsabs.harvard.edu/abs/2026A&A...706L...1D}
}

@ARTICLE{Khouri2026,
       author = {{Khouri}, T. and {Vlemmings}, W.~H.~T. and {Raudales Oseguera}, D.~A. and {Tafoya}, D. and {Olofsson}, H. and {Paladini}, C. and {Maercker}, M. and {Saberi}, M. and {Gorai}, P. and {Danilovich}, T.},
        title = "{A surprisingly large asymmetric ejection from Mira A}",
      journal = {\aap},
         year = 2026,
        month = mar,
       volume = {707},
          eid = {A162},
        pages = {A162},
          doi = {10.1051/0004-6361/202557159},
archivePrefix = {arXiv},
       eprint = {2602.03159},
 primaryClass = {astro-ph.SR},
       adsurl = {https://ui.adsabs.harvard.edu/abs/2026A&A...707A.162K}
}

\begin{appendix}
\section{Samples comparison and star classification methods}\label{an:sample_comparison}

\subsection{Comparison between our catalogue and the parent sample}\label{sec:comparison_method}

In this section, we compare our catalogue of 45 AGB stars observed in polarized light with SPHERE/ZIMPOL to the parent catalogue of 558 nearby giant stars compiled by~\citep{McDonald2012,McDonald2017} and  introduced in Section~\ref{sec:2}. As explained in the Section~\ref{sec:2}, the selection of our catalogue from the parent sample was guided by four main criteria: infrared excess, luminosity, observability with SPHERE, and AGB type.
To evaluate how these criteria shape our catalogue relative to the parent population, we investigated whether additional stellar parameters could further discriminate between the two samples.

To this end, we conducted an in-depth analysis based on a set of 11 stellar parameters: the Distance (D), the luminosity (L), the effective temperature ($\rm T_{eff}$), the stellar radius (R$_\star$), the interstellar extinction (Av), the  infrared (IR) excess (E\_IR), the fraction of reprocessed light in IR ($\mathrm{LIR/L_\star}$), the magnitude in V, J, and K bands ($\rm {M_V, \ M_J, \ M_K}$), and the color index ($\rm{J-K}$).

Of these 11 parameters, 10 were determined using the Python Stellar Spectral Energy Distribution (PySSED) code \citep{McDonald2024}.  The code corrects for interstellar extinction affecting the stellar flux in observational data using the 3D GTOMO map of $A_V$. It then fits the corrected flux with BT-Settl stellar atmosphere models~\citep{bt_settl2011} to determine physical parameters such as T$_\mathrm{eff}$, L, and R$_\star$. The distance and extinction adopted for the fits correspond to those reported by the code, with the Hipparcos 2007 parallax being favored. The magnitudes $\rm \rm{M_V, \ M_J, \ M_K}$ are obtained from catalogue queries. 

The remaining parameters, $\mathrm{E_{IR}}$ and $\mathrm{L_{IR}/L\star}$, are determined from the following equations, 

\begin{equation}\label{eq:eir}
   \rm E_{IR} = \sum_{\lambda \succ 2.2 \mu m}\frac{F_{\lambda}^{obs}/F_{\lambda}^{model}}{n_{obs}},
\end{equation} 

\begin{equation}\label{eq_fract_ligh}
   \rm \frac{L_{IR}}{L_\star} = \frac{\displaystyle \int_{2.2 \mu m}^{\infty} \left( F_{\nu}^{obs}-F_{\nu}^{model} \right) \, \mathrm{d} \nu}{\displaystyle \int_{0}^{\infty} F_{\nu}^{obs} \, \mathrm{d} \nu},
\end{equation}
where $F_{\lambda}^{obs}$ and $F_{\lambda}^{model}$ (resp. $F_{\nu}^{obs}$ and $F_{\nu}^{model}$) represent the flux of the star from observation and its model derived with PySSED for a given wavelength (resp. a given $\mathrm{\mu =c/ \lambda}$) with $c$, the light velocity.
In order to identify possible discriminating parameters between our sample of 45 stars studied and the McDonald’s catalogue, we applied two multivariate analysis methods:\\

\begin{enumerate}[label=(\roman*)]
    \item Principal Component Analysis (PCA) to explore the data structure and visualize the distribution of stars in the principal parameter space;
    \item Kolmogorov-Smirnov test (KS-test) to quantify the difference in the distribution of each parameter between the two groups.
\end{enumerate}

\subsection{Global modelling}
\subsubsection{Dataset for the Random Forest}

To predict the resolved nature of the star envelope, we trained a \texttt{Random Forest}~\citep{Ho1995,Breiman2001} model on 14 available variables i.e. the 10 variables previously defined in Section~\ref{sec:comparison_method}, excluding the effective temperature (since its contribution is already largely captured by the luminosity), plus the envelope-to-star contrast (Ctr), the axis ratio (b/a), the linear ellipticity $\rm \varepsilon = 1 - b/a$, and the eccentricity  $e = \rm \left(1 - (b/a)^2\right)^{1/2}$, resulting in a total of 14 parameters). The \texttt{Random Forest} is a machine learning algorithm that builds an ensemble of decision trees using random samples of the data and random subsets of variables, in order to improve prediction accuracy and reduce overfitting. We chose it in this study because of its robustness to correlations between variables, its ability to handle moderately sized datasets as in our case, and its ability to provide a direct measure of the importance of each parameter. Specifically, the Random Forest computes variable importance by measuring the average decrease in impurity (Gini importance) across all trees: each time a variable is used to split a node, the reduction in classification error is recorded, and the total contribution of that variable is averaged over the ensemble. Variables with higher cumulative impurity decrease are deemed more discriminant.

Given the limited sample size ($N{=}45$), we did not adopt a standard train/test split approach. While such a split (e.g., 80/20) yields a single trained model that can be applied to new data, it risks overfitting or underfitting when the training subset is very small, and it provides only a single performance estimate that may be highly sensitive to the particular random partition. Instead, we focused on a repeated stratified cross-validation strategy, which offers a more robust assessment of model stability and generalization.

To ensure a reliable evaluation and avoid overfitting, we adopted the following approach:\\
-- \textit{ 5-fold cross-validation stratified, repeated 30 times}. To assess prediction stability and quantify performance variability, we repeated this entire 5 fold procedure 30 times with different random partitions, yielded 150 independent models (30 repetitions $\times$ 5 times), and each star was tested 30 times under varying training sets. Performance metrics (accuracy, precision, recall, F1-score) were averaged over all test predictions to obtain robust estimates, and their standard deviations quantified inter-split variability.

We also compared the cross-validation 3 times (test size 15 stars per fold, training size $\sim$30) with 5 times across multiple repetition counts (5, 10, 30, 50, and 100). Both strategies yielded nearly identical mean performance (accuracy and F1-score differed by less than 0.01 at 100 repetitions), and the set of systematically misclassified objects remained identical. We retained 5-fold as our primary approach because (i) it maximizes the size of the training set per fold (36 stars vs.\ 30 for 3-fold), which is beneficial for small samples; and (ii) it exhibited slightly lower performance variance (standard deviations $\sim$0.03 vs.\ $\sim$0.04 for 3-fold).

-- \textit{Performance evaluation metrics}. The performance of the model was evaluated using several complementary metrics, defined as follows:

(i) \textit{Confusion matrix} $M$: a cross-tabulation $2{\times}2$ of predicted versus true labels, which serves as the basis for all other metrics. The four elements are: (a) True Negatives (TN): unresolved stars correctly classified as unresolved; (b) False Positives (FP): unresolved stars incorrectly classified as resolved; (c) False Negatives (FN): resolved stars incorrectly classified as unresolved; and (d) True Positives (TP): resolved stars correctly classified as resolved. The matrix provides a detailed view of classification errors by class.

(ii) \textit{Accuracy}: the proportion of correctly classified stars (both resolved and unresolved) among all test predictions, derived from the confusion matrix as $\mathrm{Accuracy} = (\mathrm{TP} + \mathrm{TN}) / (\mathrm{TP} + \mathrm{TN} + \mathrm{FP} + \mathrm{FN})$.

(iii) \textit{Precision}: the fraction of predicted resolved stars that are truly resolved, $\mathrm{Precision} = \mathrm{TP} / (\mathrm{TP} + \mathrm{FP})$. High precision indicates few false alarms.

(iv) \textit{Recall} (or sensitivity): the fraction of truly resolved stars that are correctly identified, $\mathrm{Recall} = \mathrm{TP} / (\mathrm{TP} + \mathrm{FN})$. High recall indicates few missed detections.

(v) \textit{F1-score}: the harmonic mean of precision and recall, $\mathrm{F1} = 2 \, (\mathrm{Precision} \times \mathrm{Recall}) / (\mathrm{Precision} + \mathrm{Recall})$, which balances both metrics and is particularly useful for imbalanced classes.

(vi) \textit{Macro-average} and \textit{Weighted-average}: aggregate metrics across the two classes. Macro-average treats each class equally, computing the unweighted mean of per-class metrics; this is useful for assessing whether the model performs equally well on minority and majority classes. Weighted-average weights each class by its support (number of samples), providing a global performance estimate that reflects the class distribution in the data. Both are reported to ensure transparency on the model's behavior across classes.

The primary goal of this repeated cross-validation is to (1) assess the stability and robustness of the Random Forest classifier under different training/test partitions, (2) identify which variables are consistently discriminant (via variable-importance rankings averaged across all 150 models), and (3) detect objects that are systematically misclassified regardless of the training set, signaling intrinsic ambiguity in feature space.

\subsubsection{Identification of discriminating parameters}

To identify the physical parameters that contribute most to predicting the object's resolved nature, we employed the Random Forest variable importance rankings as a complementary validation step.

The Random Forest computes importance via Mean Decrease in Impurity (MDI), a Gini-based measure. For each variable $X$, we measured its contribution to reducing classification error by tracking the decrease in Gini impurity across all splits using $X$ in the ensemble. The Gini impurity is defined as

\begin{equation}
\mathrm{Gini}(D) = 1 - \sum_{k=1}^{K} p_k^2
\end{equation}

where $p_k$ is the proportion of class $k$ in a node; a lower Gini indicates a more homogeneous (pure) node. We summed $X$ contributions to Gini reduction across all 150 trained models (30 repetitions $\times$ 5 folds), then normalized to obtain a fractional importance score for each parameter.

While Random Forest importances indicate which variables the model uses most, they do not provide decision rules or thresholds for classifying new sources. Our core methodology for parameter selection and threshold determination therefore relies on a model-independent ROC/AUC approach.

\subsubsection{ROC/AUC approach}
As noted above, repeated 5-fold cross-validation does not produce a single final model suitable for prediction on new observations. We therefore adopt an alternative approach based on model-independent decision rules derived from ROC analysis and the Youden criterion, which allows us to establish explicit, interpretable thresholds directly applicable to future sources.

For each variable $X$ corresponding to a physical parameter, we calculated the optimal class-separation threshold using the ROC curve and the Youden criterion. This criterion consists of choosing  the threshold ($t_{\mathrm{opt}}$) that maximizes the difference between the true positive rate (TPR) and the false positive rate (FPR), that is, $J = \mathrm{TPR} - \mathrm{FPR}$. Only parameters whose ROC curves yield an area under the ROC curve (AUC) satisfying $\mathrm{AUC} > 0.5$ are retained for the rule. Finally, to determine on which side of the optimal threshold ($t_{\mathrm{opt}}$) a parameter must fall to likely support the object's resolved character, we compared the medians of the two classes ("resolved" class and "unresolved" class). Thus, for a given parameter X, if the class "resolved" has the smallest median (resp. the largest median), we consider that the parameter strongly supports the resolved character “if $X \leq t_{\mathrm{opt}}$” (resp. “if $X \geq t_{\mathrm{opt}}$”).

\subsection{Discussion of model results}
\paragraph{\textit{Model performance}.}
Using stratified 5-fold cross-validation repeated 30 times, the classifier achieves a mean accuracy of $0.76 \pm 0.03$, with a weighted F1-score of $0.75 \pm 0.03$.  The per-class metrics reported in Table~\ref{tab:classif_cv} indicate a better performance for the ``resolved'' class (precision 0.77, recall 0.85, F1-score 0.81; $n{=}27$) than for the ``non-resolved'' class (precision 0.73, recall 0.61, F1-score 0.67; $n{=}18$).  The averaged confusion matrix is consistent with these values, with $\mathrm{TN}{=}11$, $\mathrm{TP}{=}23$, $\mathrm{FP}{=}7$, and $\mathrm{FN}{=}4$.

These results show that the model efficiently captures discriminant patterns between resolved and non-resolved envelopes (see the feature-importance and AUC analyses), while naturally reflecting the limitations imposed by a moderate sample size and a mild class imbalance.
The higher recall and F1-score obtained for the resolved class are likely driven by both its larger representation in the dataset and the presence of more distinctive morphological and photometric signatures.
In contrast, the non-resolved class exhibits a stronger overlap in feature space, as reflected by the larger number of false positives (7) compared to false negatives (4).

Crucially, repeating the 5-fold procedure 30 times provides a robust estimate of performance variability and prediction stability at the object level. Global metrics vary modestly across repetitions (standard deviations $\sim$0.03), and the set of systematically misclassified objects remains stable, indicating that our conclusions do not depend on a particular random split. As emphasized by \citet{Hastie2009}, the assessment of a supervised classifier depends on context, task difficulty, and class distribution. In our case, metrics around $\sim$0.74 (accuracy and weighted F1) are consistent with the limited sample size and the intrinsic complexity of the problem, and they provide a sound basis for the rule-based predictions derived from AUC thresholds presented next. Nevertheless, caution is warranted when interpreting individual outcomes for the minority (unresolved) class.

\begin{table}[ht]
\caption{Classification metrics and confusion matrix for 5-fold cross-validation repeated 30 times. 
}
\label{tab:classif_cv}
\centering
\renewcommand{\arraystretch}{1.15}
\setlength{\tabcolsep}{7pt} 

\begin{tabular}{lcccc}
\toprule\hline
Class & Precision & Recall & F1-score & Support \\
\hline
Non-resolved & 0.73 & 0.61 & 0.67 & 18 \\
Resolved     & 0.77 & 0.85 & 0.81 & 27 \\
\hline
Accuracy & \multicolumn{4}{c}{0.76 (n=45)} \\
Macro     & 0.75 & 0.73 & 0.74 & 45 \\
Weighted & 0.75 & 0.76 & 0.75 & 45 \\
\bottomrule
\end{tabular}
\tablefoot{The metrics shown are averaged over all repetitions to ensure robustness against training set variations.}

\vspace{1em}
\begin{center}
\parbox{0.45\textwidth}{
\centering
\textit{Confusion matrix (averaged):}\\[0.5em]
\[
\quad \quad \quad \quad \quad \quad \quad    \begin{bmatrix}
\text{TN} & \text{FP} \\
\text{FN} & \text{TP} \\
\end{bmatrix} 
\quad = \quad
\begin{bmatrix}
11 & 7 \\
4 & 23 \\
\end{bmatrix}
\]
}
\end{center}
\end{table}
\vspace{0.5em}

\begin{table}[t]
  \centering
\renewcommand{\arraystretch}{1.15}
\setlength{\tabcolsep}{7pt} 
\caption{
Optimal thresholds ($t_{\mathrm{opt}}$) and associated statistics for the main discriminating variables in predicting the resolved nature of the circumstellar envelope.}
\label{tab:predict}
\begin{tabular}{lccccc}
\toprule
\hline
Variable & Med\_unres & Med\_res & $t_{\mathrm{opt}}$ & Dir \\
\midrule
L (L$_\odot$)     & 3098 & 6141 & 4681 & $>$ \\
Av (mag)      & 0.01    & 0.01    & 0.02    & $<$ \\
$\rm E_{IR}$   & 1.80    & 1.81    & 1.42    & $>$ \\
$\rm LIR/L_\star$   & 0.004    & 0.011    & 0.007    & $>$ \\
Ctr      & 0.02    & 0.03    & 0.04    & $>$ \\
$\rm R_{\star}$ (mas)  & 5    & 14   & 9    & $>$ \\
b/a     & 0.89    & 0.86    & 0.97    & $<$ \\
\bottomrule
\end{tabular}

\tablefoot{The \textit{Dir} column indicates the direction of the decision rule: ``$>$'' means that the probability of being resolved increases when the variable exceeds the threshold $t_{\mathrm{opt}}$, while ``$<$'' indicates that the probability is higher below this threshold. The \textit{Med\_unres}  and \textit{Med\_res}  columns give the median of the variable for  unresolved and resolved objects, respectively. The variables listed in this table account for a cumulative Gini importance of 54.81\% in the Random Forest classifier.
}
\end{table} 

\paragraph{\textit{Model stability and consistency at the object level}.}
To assess the robustness of the results with respect to the choice of the training set, we repeated the stratified 5-fold cross-validation 30 times, so that each star is evaluated across a large number of independent training configurations. The relatively small dispersion of the global metrics (standard deviations of order $\sim$0.03) demonstrates that the performance of the model is stable and not driven by a particular random split of the data.

\paragraph{\textit{Interpretation of systematic misclassifications}.}
More importantly, by tracking the classification outcome for each individual star across all repetitions, we identify a limited subset of objects (12 out of 45, i.e.\ $\sim$27\%) that are systematically misclassified, with error rates exceeding 50\%. These objects, summarized in Table~\ref{tab:problematic_objects}, can be separated according to their true morphological class.

This object-level analysis is based on the full set of 30 repetitions of the stratified 5-fold cross-validation and therefore differs from the averaged confusion matrix, which reflects the mean behaviour over a single 5-fold split and yields 11 misclassified objects.

Among the stars whose envelopes are intrinsically unresolved but are frequently misclassified as resolved, we identify S~Lep, R~Aql, RT~Vir, R~Peg, BK~Vir, Bet~Gru, T~Cet, RX~Lep, and R~Hor. Conversely, three stars with intrinsically resolved envelopes SW~Col, V~Hya, and W~Aql are predominantly misclassified as unresolved. This clear partition demonstrates that the same objects are repeatedly affected across independent training subsets, indicating that the misclassifications are not driven by statistical noise or by the limited size of the training set, but rather by intrinsic stellar properties.

A detailed inspection of these problematic objects reveals two complementary regimes. In the case of unresolved stars misclassified as resolved, several objects exhibit high luminosities, large infrared excesses, or relatively large angular stellar radii, placing them close to the resolved population in the multidimensional feature space. These properties mimic the typical signatures learned by the classifier for resolved envelopes and naturally lead to confusion.

In contrast, resolved envelopes misclassified as unresolved tend to display global photometric or geometric parameters that overlap with those of the unresolved population, despite their clearly resolved morphology. This behavior is particularly well illustrated by SW~Col, which is systematically misclassified despite exhibiting a strongly bipolar morphology. Its extreme elongation parameters make it morphologically unambiguous in direct imaging, yet its integrated photometric properties remain partially degenerate with those of unresolved stars.

An additional potential source of bias concerns the availability of a dedicated PSF observation for each target. Among the 12 systematically misclassified objects, 6 lack a dedicated PSF corresponding to 50\% of the problematic subset, while the remaining 6 objects (50\%) do have a dedicated PSF. This indicates that systematic misclassifications occur both with and without a dedicated PSF. Consequently, these errors cannot be primarily attributed to uncertainties in the PSF estimation, but rather reflect intrinsic ambiguities in the stellar properties themselves.

Overall, this analysis demonstrates that increasing the number of training repetitions would not fundamentally alter the conclusions: the same stars remain systematically misclassified, and the associated uncertainties are dominated by intrinsic stellar diversity and variability rather than by the choice of training set. Identifying and characterizing these objects, as summarized in Table~\ref{tab:problematic_objects}, is therefore a key outcome of the method, as it defines a well-motivated subset of targets for future observational and modelling efforts aimed at understanding the physical origin of their atypical properties. This object-level analysis complements the global performance metrics and reinforces the reliability and interpretability of the Random Forest approach adopted in this work.

\begin{table*}[!ht]
\centering
\caption{
Summary of the stars systematically misclassified by the Random Forest classifier
(error rate $>50\%$ across repeated cross-validation runs).
}

\label{tab:problematic_objects}
\renewcommand{\arraystretch}{1.2}
\setlength{\tabcolsep}{8pt}

\begin{tabular}{lccccccr}
\toprule\hline
Name &
True class &
Error rate &
$L$ &
$E_{IR}$ &
$R_\star$ &
Dedicated PSF &
Comment \\
&
&
(\%) &
($L_\odot$) &
 &
(mas) &
(Yes/No) &
 \\
\midrule

SW~Col  & Resolved & 100 & 1059  & 2.45 & 3  & No  & strong asymmetry; low $L$ \\
V~Hya   & Resolved & 73  & 52698 & 6.18 & 6  & Yes & extreme $L$ + extreme $E_{IR}$ \\
W~Aql   & Resolved & 77  & 3698  & 1.30 & 10 & No  & low $L$ \\

\midrule

S~Lep   & Unresolved & 77  & 3144  & 2.65 & 8  & No  & $E_{IR}$ close to resolved \\
R~Aql   & Unresolved & 100 & 32170 & 3.28 & 5  & Yes & high $L$ + high $E_{IR}$ \\
RT~Vir  & Unresolved & 100 & 2243  & 1.99 & 17 & No  & large $R_\star$; overlap regime \\
R~Peg   & Unresolved & 90  & 4525  & 1.75 & 8  & Yes & red colors; intermediate $E_{IR}$ \\
BK~Vir  & Unresolved & 90  & 3683  & 1.73 & 10 & No  & moderate $E_{IR}$; elongation \\
Bet~Gru & Unresolved & 87  & 3052  & 1.22 & 17 & Yes & very large $R_\star$ \\
T~Cet   & Unresolved & 73  & 7694  & 1.62 & 8  & No  & overlap in ($L$, $E_{IR}$) \\
RX~Lep  & Unresolved & 87  & 4122  & 1.31 & 10 & Yes & $R_\star$ near resolved regime \\
R~Hor   & Unresolved & 100 & 25564 & 5.76 & 25 & Yes & extreme $L$ + extreme $E_{IR}$ \\

\bottomrule
\end{tabular}
\tablefoot{For each object, we report a subset of physical parameters that dominate the classification and misclassifications processes.
The parameters $L$, $E_{IR}$, and $R_\star$ were retained because they are (i) among the most discriminant variables identified by the PCA, KS test, and
Random Forest feature importance, and (ii) physically and observationally linked to the detectability and spatial
resolvability of circumstellar envelopes with SPHERE.
The error rate is defined as the fraction of individual cross-validation realizations (out of 30 repetitions of the stratified 5-fold cross-validation) in which a given star is misclassified.
The last column indicates whether a dedicated PSF observation was available for the considered target.}
\end{table*}

\newpage
\section{Stellar observation and derived parameters} 

\begin{table*}[!ht]
\centering
\caption{Stellar parameters used to train the prediction model.}
\label{tab:morpho_param}
\resizebox{\textwidth}{!}{\begin{tabular}{cccccccccccccccc}
\toprule \hline
Name &        L &   D &   Av & E$_\mathrm{IR}$ &  $\mathrm{LIR/L_\star}$ &  Ctr & R$_\star$ &    Mv &    Mj &    Mk &   J-K &  b/a &    e &    $\epsilon$ & Res\_stat \\
- & ($L_\odot$) & (pc) & (mag) & - & - & - & (mas) & (mag) & (mag) & (mag) & (mag) & - & - & - & - \\ 
\midrule
  Chi Cyg & 12715 & 181 & 0.02 & 1.60 &  0.0078 & 0.02 &    23 &  6.96 &  0.17 & -1.70 &  1.86 & 0.96 & 0.29 & 0.04 &     yes  \\
    Z Peg &   752 & 201 & 0.01 & 1.63 &  0.0041 & 0.04 &    15 &  9.15 &  2.22 &  0.92 &  1.31 & 0.99 & 0.13 & 0.01 &      no  \\
    Y Pav & 11213 & 403 & 0.02 & 2.18 &  0.0053 & 0.04 &    13 &  6.70 &  0.00 &  0.27 & -0.27 & 0.90 & 0.44 & 0.10 &     yes  \\
    S Lep &  3144 & 203 & 0.01 & 2.65 &  0.0053 & 0.01 &     8 &  7.00 &  0.80 &  0.00 &  0.80 & 0.82 & 0.57 & 0.18 &      no  \\
    R Aql & 32170 & 422 & 0.24 & 3.28 &  0.0241 & 0.01 &     5 &  7.73 &  0.68 & -0.83 &  1.51 & 0.95 & 0.31 & 0.05 &      no  \\
    U Her &  4836 & 235 & 0.02 & 1.96 &  0.0159 & 0.01 &    25 &  8.43 &  0.86 & -0.63 &  1.49 & 0.82 & 0.58 & 0.18 &     yes  \\
    R Crt & 10371 & 261 & 0.01 & 1.81 &  0.0129 & 0.03 &     8 &  9.07 &  0.17 & -1.40 &  1.58 & 0.80 & 0.60 & 0.20 &     yes  \\
   CW Cnc &  2312 & 244 & 0.00 & 1.84 &  0.0029 & 0.08 &     5 &  9.27 &  1.42 &  0.11 &  1.31 & 0.91 & 0.42 & 0.09 &      no  \\
    S Pav &  6083 & 184 & 0.01 & 1.58 &  0.0127 & 0.01 &    15 &  8.07 & -0.20 & -1.62 &  1.42 & 0.90 & 0.44 & 0.10 &     yes  \\
    R Dor &  4681 &  55 & 0.00 & 1.46 &  0.0108 & 0.08 &    47 &  5.86 &  0.00 &  0.00 &  0.00 & 0.99 & 0.14 & 0.01 &     yes  \\
    U Del & 16561 & 481 & 0.04 & 2.76 &  0.0165 & 0.08 &     5 &  7.00 &  1.03 & -0.21 &  1.24 & 0.81 & 0.58 & 0.19 &     yes  \\
    W Peg &  7639 & 294 & 0.04 & 1.64 &  0.0088 & 0.10 &    11 &  9.51 &  1.14 & -0.16 &  1.30 & 0.84 & 0.54 & 0.16 &     yes  \\
   SW Col &  1059 & 199 & 0.00 & 2.45 &  0.0050 & 0.01 &     3 &  5.94 &  2.12 &  1.06 &  1.06 & 0.46 & 0.89 & 0.54 &     yes  \\
Pi.01 Gru &  7970 & 163 & 0.00 & 1.77 &  0.0096 & 0.01 &    17 &  6.38 & -0.66 & -2.11 &  1.45 & 0.85 & 0.53 & 0.15 &     yes  \\
    W Hya &  9944 & 104 & 0.01 & 1.71 &  0.0079 & 0.01 &    40 &  7.65 & -1.74 & -3.22 &  1.48 & 0.99 & 0.16 & 0.01 &     yes  \\
Alpha Her & 15009 & 110 & 0.02 & 1.42 &  0.0018 & 0.01 &    25 &  3.51 & -2.30 & -3.51 &  1.21 & 1.00 & 0.10 & 0.00 &     yes  \\
    Z Eri &  2637 & 256 & 0.01 & 1.35 &  0.0011 & 0.01 &     5 &  6.90 &  1.59 &  0.00 &  1.59 & 0.96 & 0.29 & 0.04 &      no  \\
    R Leo &  2093 &  71 & 0.00 & 1.59 &  0.0213 & 0.04 &    30 &  0.00 & -0.72 & -2.29 &  1.57 & 0.78 & 0.62 & 0.22 &     yes  \\
   RT Vir &  2243 & 136 & 0.00 & 1.99 &  0.0097 & 0.01 &    17 &  8.72 &  0.36 & -1.06 &  1.42 & 0.86 & 0.52 & 0.14 &     yes  \\
    R Peg &  4525 & 287 & 0.03 & 1.75 &  0.0088 & 0.02 &     8 &  8.64 &  1.84 & -0.05 &  1.89 & 0.91 & 0.42 & 0.09 &      no  \\
   BW Oct &  3993 & 250 & 0.05 & 2.06 &  0.0118 & 0.06 &     8 &  8.85 &  1.08 &  0.00 &  1.08 & 0.75 & 0.66 & 0.25 &     yes  \\
    R Scl &  6141 & 266 & 0.01 & 8.80 &  0.0206 & 0.02 &    15 &  7.10 &  1.44 & -0.20 &  1.64 & 0.99 & 0.11 & 0.01 &     yes  \\
   L2 Pup &  1651 &  64 & 0.00 & 5.03 &  0.0655 & 0.03 &    13 &  4.66 & -1.06 & -2.31 &  1.25 & 0.88 & 0.47 & 0.12 &     yes  \\
    Y Scl &  1054 & 165 & 0.00 & 2.40 &  0.0117 & 0.07 &     7 &  8.59 &  1.57 &  0.00 &  1.57 & 0.89 & 0.46 & 0.11 &     yes  \\
     Mira &  7760 &  92 & 0.00 & 6.48 &  0.0735 & 0.04 &    18 &  0.00 & -0.73 & -2.21 &  1.48 & 0.77 & 0.63 & 0.23 &     yes  \\
   SS Lep &  6732 & 279 & 0.01 & 7.13 &  0.0033 & 0.09 &     3 &  4.99 &  2.90 &  1.67 &  1.23 & 0.75 & 0.66 & 0.25 &      no  \\
    T Mic &  8679 & 211 & 0.03 & 1.33 &  0.0033 & 0.02 &    14 &  7.68 & -0.36 & -1.65 &  1.29 & 0.86 & 0.51 & 0.14 &     yes  \\
    R Hya &  5824 & 124 & 0.01 & 1.66 &  0.0044 & 0.01 &    29 &  0.00 &  0.00 &  0.00 &  0.00 & 0.93 & 0.37 & 0.07 &     yes  \\
   BK Vir &  3683 & 181 & 0.00 & 1.73 &  0.0132 & 0.02 &    10 &  8.36 &  0.54 & -0.73 &  1.27 & 0.72 & 0.69 & 0.28 &     yes  \\
   DZ Aqr &  2717 & 297 & 0.02 & 1.93 &  0.0161 & 0.02 &     4 &  8.98 &  1.85 &  0.56 &  1.28 & 0.77 & 0.64 & 0.23 &      no  \\
   AK Hya &  2069 & 156 & 0.00 & 2.05 &  0.0107 & 0.04 &     7 &  6.81 &  0.64 &  0.00 &  0.64 & 0.99 & 0.12 & 0.01 &     yes  \\
   SW Vir &  7634 & 143 & 0.01 & 1.62 &  0.0070 & 0.07 &    19 &  7.18 & -0.36 &  0.00 & -0.36 & 0.83 & 0.55 & 0.17 &     yes  \\
  Bet Gru &  3052 &  54 & 0.00 & 1.22 &  0.0006 & 0.02 &    17 &  2.30 & -1.97 & -3.32 &  1.35 & 0.94 & 0.35 & 0.06 &      no  \\
    T Cet &  7694 & 270 & 0.01 & 1.62 &  0.0043 & 0.03 &     8 &  5.90 &  0.50 & -0.81 &  1.30 & 0.83 & 0.55 & 0.17 &      no  \\
  Psi Phe &  1162 & 105 & 0.00 & 1.30 &  0.0004 & 0.01 &     5 &  4.61 &  0.45 &  0.00 &  0.45 & 0.78 & 0.62 & 0.22 &      no  \\
V1943 Sgr &  7403 & 197 & 0.04 & 1.46 &  0.0054 & 0.04 &    12 &  8.17 & -0.08 &  0.00 & -0.08 & 0.97 & 0.23 & 0.03 &     yes  \\
  Ups Cet &   548 &  90 & 0.00 & 1.30 &  0.0060 & 0.01 &     3 &  4.18 &  1.00 &  0.10 &  0.90 & 0.83 & 0.56 & 0.17 &      no  \\
   RX Lep &  4122 & 149 & 0.01 & 1.31 &  0.0019 & 0.03 &    10 &  5.84 & -0.16 & -1.40 &  1.24 & 0.91 & 0.42 & 0.09 &      no  \\
    R Hor & 25564 & 210 & 0.00 & 5.76 &  0.0064 & 0.04 &    25 &  7.34 &  0.72 & -0.61 &  1.33 & 0.90 & 0.43 & 0.10 &      no  \\
   GY Aql &  2526 & 223 & 0.10 & 4.49 &  0.0648 & 0.01 &    22 & 12.95 &  1.29 & -0.28 &  1.57 & 0.78 & 0.62 & 0.22 &     yes  \\
    V Hya & 52698 & 694 & 0.01 & 6.18 &  0.0298 & 0.05 &     6 &  8.06 &  1.65 & -0.69 &  2.34 & 0.92 & 0.40 & 0.08 &     yes  \\
   AC Cet &   835 & 289 & 0.01 & 2.20 &  0.0058 & 0.01 &     3 &  8.04 &  3.11 &  1.85 &  1.26 & 0.89 & 0.45 & 0.11 &      no  \\
    W Aql &  3698 & 372 & 0.22 & 1.30 &  0.0010 & 0.04 &    10 &  0.00 &  0.00 &  0.00 &  0.00 & 0.78 & 0.63 & 0.22 &     yes  \\
   SV Aqr &  4525 & 430 & 0.01 & 2.08 &  0.0045 & 0.01 &     4 & 10.00 &  2.24 &  0.95 &  1.29 & 0.71 & 0.71 & 0.29 &      no  \\
    V PsA &  7861 & 304 & 0.01 & 2.09 &  0.0108 & 0.04 &     9 &  8.95 &  0.75 &  0.00 &  0.75 & 0.94 & 0.34 & 0.06 &     yes  \\
\bottomrule
\end{tabular}
}
\tablefoot{All displayed quantities are used for training, including the magnitudes ($V$, $J$, $K$, $J-K$). The distance adopted is that of the Hipparcos catalogue, with the exception of the three stars (W~Aql, SV~Aqr and V~PsA) without Hipparcos IDs for which the Gaia DR3 distance was used. Note that Gaia distances present large uncertainties in the parallaxes for most AGB stars, as mentioned by~\cite{beguin2024}. Therefore, we favored Hipparcos distances whenever possible in this study. Luminosity ($L$), and effective temperature ($T_{\mathrm{eff}}$). Photometry sources: Hipparcos/Tycho (visible), SDSS and DENIS (visible and near-IR), 2MASS (near-IR), and MSX, AKARI, IRAS, WISE (thermal IR). Shape parameters: axial ratio $b/a$, eccentricity $e=\sqrt{1-(b/a)^2}$, ellipticity $\epsilon=1-b/a$, and envelope resolution statute $Res\_stat$ . Columns shown: $L$ ($L_\odot$), $D$ (pc), $A_V$ (mag), $E_{\mathrm{IR}}$, $L_{\mathrm{IR}}/L_\star$, $Ctr$, $R_{\mathrm{mas}}$, $M_V$, $M_J$, $M_K$, $J-K$, $b/a$, $e$, $\epsilon$, $Res\_stat$.}
\end{table*}

\begin{table*}[t]\centering
\caption{Targets observation log with ZIMPOL.}
\label{tab:log}
\begin{tabular}{llccccccc}
\toprule\hline
Date & Time & Target & Program ID & Dedicated PSF & Filter 1 & Filter 2 & Seeing & Airmass \\
\midrule
2015-09-22 & 06:00:05 & Y Pav & 095.D-0309(B) & No & V & N\_R & 0.8 & 1.6 \\
2015-09-23 & 06:33:06 & R Hor & 095.D-0309(B) & Yes & V & N\_R & 0.8 & 1.6 \\
            & 05:22:59 & R Scl & 095.D-0309(B) & Yes & V & N\_R & 0.8 & 1.6 \\
            & 04:32:28 & Y Scl & 095.D-0309(B) & Yes & V & N\_R & 0.8 & 1.6 \\
2015-12-19 & 06:56:44 & R Crt & 096.D-0930(E) & Yes & N\_R & N\_R & 1.0 & 1.5 \\
2016-02-19 & 02:00:31 & SW Col & 096.D-0252(A) & No & V & N\_R & 0.8 & 1.6 \\
2016-03-08 & 07:53:40 & W Hya & 096.D-0930(B) & No & Cnt820 & Cnt748 & 1.0 & 1.5 \\
2016-03-09 & 07:35:37 & SW Vir & 096.D-0930(C) & No & N\_R & N\_R & 1.0 & 1.5 \\
2016-04-22 & 01:37:40 & V Hya & 097.D-0516(A) & Yes & V & N\_R & 0.8 & 1.6 \\
2016-04-30 & 06:08:35 & Alpha Her & 097.D-0352(A) & Yes & V & Cnt748 & 0.8 & 1.5 \\
2016-06-30 & 05:27:44 & Chi Cyg & 097.D-0352(A) & Yes & V & Cnt748 & 0.8 & 2.2 \\
2016-07-22 & 10:13:16 & Z Eri & 097.D-0516(B) & No & V & N\_R & 1.4 & 1.6 \\
            & 07:33:48 & R Peg & 097.D-0516(B) & Yes & V & N\_R & 1.4 & 1.6 \\
            & 06:48:41 & BW Oct & 097.D-0516(B) & Yes & V & N\_R & 1.4 & 1.6 \\
            & 09:09:48 & AC Cet & 097.D-0516(B) & Yes & V & N\_R & 1.4 & 1.6 \\
            & 07:12:00 & DZ Aqr & 097.D-0516(B) & Yes & V & N\_R & 1.4 & 1.6 \\
            & 08:40:12 & Z Peg & 097.D-0516(B) & Yes & V & N\_R & 1.4 & 1.6 \\
            & 08:13:34 & W Peg & 097.D-0516(B) & Yes & V & N\_R & 1.4 & 1.6 \\
2016-07-29 & 23:20:37 & RT Vir & 097.D-0443(A) & No & Cnt820 & Cnt820 & 1.2 & 1.9 \\
2016-10-04 & 06:33:30 & RX Lep & 098.D-0206(C) & Yes & CntHa & CntHa & 1.0 & 1.5 \\
2016-10-12 & 02:07:05 & Beta Gru & 098.D-0200(A) & Yes & V & N\_R & 1.4 & 1.6 \\
            & 01:05:09 & T Mic & 098.D-0200(A) & Yes & V & N\_R & 1.4 & 1.6 \\
2016-11-09 & 04:13:02 & R Dor & 098.D-0731(A) & No & Cnt820 & Cnt748 & 1.2 & 1.8 \\
2016-12-07 & 06:06:15 & AK Hya & 098.D-0206(D) & No & N\_R & N\_R & 1.0 & 1.5 \\
2016-12-14 & 08:47:36 & R Leo & 097.D-0352(A) & No & V & Cnt748 & 0.8 & 1.5 \\
2017-03-07 & 04:08:12 & BK Vir & 097.D-0443(A) & No & Cnt820 & Cnt820 & 1.2 & 1.9 \\
2017-07-31 & 07:44:43 & T Cet & 099.D-0600(A) & No & CntHa & B\_Ha & 0.8 & 1.2 \\
2017-11-27 & 03:03:26 & Mira Ceti & 0100.D-0737(B) & Yes & Cnt820 & Cnt748 & 1.2 & 1.3 \\
2019-07-05 & 05:09:14 & U Del & 0103.D-0772(A) & No & CntHa & CntHa & 2.0 & 2.0 \\
2019-07-08 & 01:55:17 & U Her & 0103.D-0772(A) & No & VBB & VBB & 2.0 & 2.0 \\
            & 09:20:26 & $\pi^1$ Gru & 0103.D-0772(A) & No & Cnt820 & Cnt748 & 2.0 & 2.0 \\
2019-07-09 & 04:53:58 & W Aql & 0103.D-0772(A) & No & VBB & VBB & 2.0 & 2.0 \\
            & 08:51:53 & V PsA & 0103.D-0772(A) & No & N\_R & N\_R & 2.0 & 2.0 \\
2019-07-27 & 01:35:53 & R Hya & 0103.D-0772(A) & No & CntHa & CntHa & 2.0 & 2.0 \\
2019-07-30 & 00:53:26 & R Aql & 0103.D-0772(A) & Yes & N\_R & N\_R & 2.0 & 2.0 \\
2019-09-24 & 03:38:22 & S Pav & 0103.D-0772(A) & No & N\_R & N\_R & 2.0 & 2.0 \\
2019-09-28 & 03:18:41 & GY Aql & 0103.D-0772(A) & Yes & VBB & VBB & 2.0 & 2.0 \\
2019-09-29 & 03:02:18 & SV Aqr & 0103.D-0772(A) & Yes & VBB & VBB & 2.0 & 2.0 \\
2019-10-16 & 03:26:11 & Ups Cet & 0104.D-0240(A) & Yes & V & N\_R & 2.0 & 2.8 \\
            & 01:42:05 & V1943 Sgr & 0104.D-0240(A) & Yes & V & N\_R & 2.0 & 2.8 \\
            & 02:46:51 & Psi Phe & 0104.D-0240(A) & Yes & V & N\_R & 2.0 & 2.8 \\
2020-02-29 & 01:51:51 & S Lep & 0104.D-0240(B) & No & I\_PRIM & I\_PRIM & 2.0 & 2.8 \\
            & 00:10:45 & SS Lep & 0104.D-0240(B) & Yes & I\_PRIM & I\_PRIM & 2.0 & 2.8 \\
            & 00:48:30 & L$_2$ Pup & 0104.D-0240(B) & No & V & N\_R & 2.0 & 2.8 \\
            & 02:20:39 & CW Cnc & 0104.D-0240(B) & No & I\_PRIM & I\_PRIM & 2.0 & 2.8 \\
\bottomrule
\end{tabular}

\end{table*}

\end{appendix}
\end{document}